\documentstyle[aps,prb,amsfonts,epsf,special]{revtex}
\begin{document}

{
\title{Simple Bosonization Solution of the 
2-channel Kondo Model: I. Analytical
Calculation of Finite-Size Crossover Spectrum}
%II. Calculation of Non-Fermi-Liquid Electron Green's Functions    
\author{${}^{1,2}$Gergely Zar\'and and ${}^3$Jan von Delft}
\address{${}^1$Institute of Physics, Technical University of
Budapest, H 1521 Budafoki \'ut 8., Budapest, Hungary \\
${}^2$International School for Advanced Studies, I-34014,
Trieste, Italy \\
${}^3$Institut f\"ur Theoretische Festk\"orperphysik, 
Universit\"at   Karlsruhe, 76128 Karlsruhe, Germany}
\date{Submitted to Phys. Rev. B on December 10, 1998}
\maketitle 

\begin{abstract}
We present in  detail a simple, exact solution of the
 anisotropic 2-channel Kondo (2CK) model at its Toulouse point. 
 We reduce the model to a quadratic resonant-level model
by generalizing the bosonization-refermionization approach of Emery
 and Kivelson to finite system size,
but improve their method in two ways: firstly, 
we construct all boson fields and Klein factors explicitly in terms
of the model's original fermion operators $c_{k \sigma j}$,
and secondly we clarify explicitly how the Klein factors needed
when refermionizing act on the original Fock space. 
This enables us to explicitly follow the
adiabatic evolution of the 2CK model's free-fermion states
to its exact eigenstates, found by 
simply diagonalizing the resonant-level 
model for arbitrary  magnetic fields and spin-flip coupling strengths.
In this way we obtain an {\em analytic}\/
description of  the cross-over from the free to the non-Fermi-liquid
fixed point. At the latter, it is remarkably simple to 
recover the conformal field theory results for
the finite-size spectrum (implying a direct proof of Affleck
and Ludwig's fusion hypothesis). By analyzing the finite-size spectrum,
we directly obtain the operator content of the 2CK fixed
point and the dimension of various relevant and irrelevant perturbations.  
Our method goes beyond previous conformal field theory
results, since it works for arbitrary magnetic fields
and can easily be generalized to include various symmetry-breaking
perturbations, and to study the crossover to other fixed points produced
by these. Furthermore it  establishes instructive connections between 
different renormalization group schemes such as poor man's scaling, 
Anderson-Yuval type scaling, the numerical renormalization group and 
finite-size scaling.
\end{abstract}
%\vskip0.0pc]
}
\narrowtext

\section{Introduction}
Quantum impurity models displaying non-Fermi-liquid (NFL) behavior
have attracted substantial interest during the past few years.
These models have the common property that 
their exact elementary excitations are not free-electron like
and cannot be described using Fermi-liquid theory.
 Such single-impurity 
models have been proposed as relevant for certain properties of
heavy-fermion alloys \cite{Cox87,SML91,AT91} 
and high-$T_c$ superconductors.\cite{EK92,GVR93} 
They also emerge in the tunneling-impurity Kondo
problem,\cite{VZ83,MG86,ZZ94,MF95,Zarand97}   and 
infinite-dimensional strongly-correlated lattice models
 can be mapped onto such models as well.\cite{infdim} 
All these models possess regimes in  which
the physical quantities are non-analytic
(logarithmic or power-law) functions of 
parameters such as temperature or magnetic field. 

The two-channel Kondo (2CK) model, introduced in 1980
by  Nozi\'eres and Blandin,\cite{NB80}   is one of the simplest and 
most-studied  quantum impurity models with NFL behavior. In
this model two channels of spinful conduction electrons interact with a
single spin $1/2$ impurity via a local antiferromagnetic
exchange interaction.  
In contrast to the  single-channel Kondo (1CK) model, 
which has a stable infinite-coupling fixed point 
at which the conduction electrons 
screen the impurity spin completely by forming a spin $0$ complex, 
 in the two-channel case they
{\em overscreen}\/ the impurity spin at infinite coupling, leaving
a non-trivial residual spin object,
so that the 2CK model's infinite-coupling fixed point 
becomes unstable.  A stable fixed point exists at intermediate coupling 
strength, which leads to 
the appearance of a non-zero
residual entropy and to non-analytical behavior for various
physical quantities.
Such non-analytical behavior was directly observed,
for example, in  anomalous conductance
signals of metallic nanoconstrictions  containing 2-state tunneling 
systems,\cite{RLvDB94,KSvK96,ULB96,vDZZ97,vDetal97} 
which are perhaps the most convincing realizations
of 2CK physics found experimentally. 

The two-channel Kondo model has been studied theoretically by an impressive
number of different methods, which are comprehensively reviewed in
Ref.~\onlinecite{CZ95}.  These include approximate methods such as the
multiplicative \cite{MG86,NB80,Multipl} and the path-integral
\cite{VZZ88,GFN96} renormalization group approaches and slave-boson
methods;\cite{CR93,HKH94,Cox?}  effective models such as the so-called
compactified model \cite{ColmIoffeTsvel,CS95-97,ZHB97} which is partially
equivalent to the 2CK model; the numerical renormalization group
(NRG);\cite{CLN80,PC91,AL92b} and exact methods, such as the Bethe
Ansatz,\cite{AD84,WT85,AJ96} conformal field theory (CFT)
\cite{AL92b,Aff90,AL91a,AL91b,AL93,AL94,Lud94a} and abelian
bosonization.\cite{EK92,SG94,kotliarsi,ML95,Ye} Excepting abelian
bosonization, however, the price for using powerful numerical or exact methods
has hitherto always been a very high degree of technical sophistication and a
lack of physical transparency.  The Bethe Ansatz provides an analytical
solution of the model, allowing for the calculation of the cross-over from
Fermi-liquid to non-Fermi-liquid behavior of the thermodynamical quantities.
It is, however, rather involved, and is unable to calculate dynamical
correlation functions.  With the numerical renormalization group technique,
which likewise is able to describe cross-over behavior, one can obtain
thermodynamical properties, carry out a finite-size analysis of the model,
investigate the effect of various perturbations such as different electronic
and impurity magnetic fields and channel anisotropy, and in principle also
calculate dynamical local correlation functions of the impurity. However, this
method, though powerful, is approximate by construction, requires considerable
numerical prowess, is physically not very transparent and is not well-suited
to calculate dynamical properties of the conduction electrons.  Finally, the
elegant conformal field theory solution of Affleck and Ludwig (AL) focuses
exclusively on the NFL regime in the vicinity of $T=0$ fixed point.  By
exploiting its symmetry properties to the full, it provides the finite-size
spectrum of the model, all thermodynamical quantities and furthermore all
dynamical correlation functions.  However, the CFT solution relies crucially
on the so-called fusion hypothesis that can only be verified {\em a
  posterio}\/ by comparing the CFT results with other exact methods.
Moreover, it cannot be used to calculate the cross-over behavior, and
requires, of course, extensive knowledge of the technical subject of boundary
conformal field theory.

A major advance towards finding a {\em simple and transparent}\/ 
exact solution of the 2CK model was achieved by Emery and Kivelson 
(EK)\cite{EK92} with the rather simple technique of 1-dimensional abelian
bosonization (pedagogically reviewed in Ref.~\onlinecite{jvdschoeller}). 
Using bosonization and refermionization, EK showed that along a certain line
in parameter space, known as the Toulouse ``point'' or Emery-Kivelson
line, the anisotropical 2CK model can be mapped exactly onto a {\em
quadratic}\/ resonant-level model, which can be solved straightforwardly by
diagonalization. 
Since spin anisotropy is known to be irrelevant for the multichannel
Kondo model, this yielded new insight also about 
the generic behavior of the isotropic 2CK model.
Though their approach works only in the vicinity of the EK  line, 
the latter connects the Fermi-liquid and non-Fermi-liquid regimes,
so that EK's method captures both the model's NFL behavior and
 the  cross-over from the free to the NFL fixed point.
EK calculated a number of thermodynamic and impurity properties and
some electron correlation functions, and related the NFL behavior to
the fact that, remarkably, only ``one half'' 
of the impurity's degrees of freedom (a Majorana fermion instead
of a proper complex fermion) couple to the electrons. 

In the present work, which is an extended version of a previous
publication,\cite{vDZF}  we generalize EK's bosonization technique to
{\em finite system sizes.}\/ For this purpose
two important modifications are needed: \\
(i) While EK use the field-theoretical approach to bosonization in
which the bosonization relation $\psi_{\alpha j} \simeq F_{\alpha j}
e^{-i \phi_{\alpha j}}$ is used merely as a formal correspondence, we
use the more careful and explicit {\em constructive}\/ bosonization
procedure of Haldane \cite{Haldane81}. 
In the latter approach both the boson fields
$\phi_{\alpha j}$ and Klein factors $F_{\alpha j}$ are constructed
{\em explicitly}\/ from the original $\psi_{\alpha j}$ operators, so
that the bosonization
formula becomes an operator identity in Fock space. \\
(ii) Since EK were interested mainly in impurity properties, they did
not need to discuss at all the Klein factors $F_{\alpha j}$ [which
lower the number of ${\alpha j}$-electrons by one and ensure proper
anticommutation relations for the $\psi_{\alpha j}$'s].  These Klein
factors, however, are essential for quantities like the finite-size
spectrum or various electron correlation functions. Therefore it is
crucial to specify how the Klein factors for the refermionized
operators act on the Fock space.  As we shall see, these new Klein
factors are only well defined on a suitably {\em enlarged} Fock space that
also contains unphysical states, which must be discarded at the end
using  certain {\em gluing conditions}.

With these modifications, EK's bosonization approach enables us 
by straightforward diagonalization of the quadratic resonant-level 
model   (i) for the first time to
analytically trace the cross-over of the 2CK model's 
 finite-size spectrum from the FL
to the NFL fixed point, at which we reproduce the fixed-point
spectrum previously found by CFT using a certain
fusion hypothesis;
   (ii) to construct the eigenstates of the 2CK model 
corresponding  to this crossover spectrum explicitly; 
(iii) and to extract the 
operator content of the NFL fixed point and determine the dimensions
of different relevant and irrelevant operators. Since our
method works also in the presence of an arbitrary magnetic field
(unlike CFT), we
can also (iv) investigate how a finite magnetic field destroys
the NFL spectrum for the low-energy excitations of the model and
restores the FL properties.  (v) Furthermore, 
our finite-size bosonization approach 
can easily be related to various popular renormalization group
methods; it therefore not only provides a useful bridge between them,
but can potentially be used as a pedagogical tool for 
{\em analytically}\/  illustrating their main ideas.

In a future publication\cite{ZvD}  we shall show that this method
furthermore allows one 
(vi) to construct very easily the scattering states of the
model;  (vii) to prove explicitly the validity
of the bosonic description of the NFL fixed point
Maldacena and Ludwig,\cite{ML95}
(viii) to determine the fixed point boundary conditions at the impurity site for
the different currents and fields in a very straightforward way; and
(ix) to calculate with ease all correlation functions at and around
the NFL fixed point.
This implies that all CFT results
can be derived from first principles using the bosonization approach.

The paper is organized as follows.  
In Section~\ref{model} we define the 2CK
model to be studied.
For completeness, and since the proper use of Klein factors is essential,
Section~\ref{bosonizationbasics} briefly reviews the ``constructive''
(operator identity-based) approach to finite-size bosonization used throughout
this paper.  The Emery-Kivelson mapping onto a resonant-level model is
discussed in Section~\ref{EKmap}, using our novel, more explicit formulation
of refermionization within a suitably extended Fock space. The solution of the
resonant level model and the construction of the non-Fermi liquid spectrum
using generalized gluing conditions is presented in Section~\ref{fsize}.  
In Section~\ref{s:RGs} the results of our finite-size calculations are
compared with and interpreted in terms of various RG procedures.  Finally, in
Section~\ref{s:Concl} we summarize our conclusions
and compare our method to a few others. 

The centerpiece of our work, and indeed the prerequisite for all of our
results, is the uncommonly careful and detailed finite-size formulation of the
Emery-Kivelson mapping in the main text.  Technicalities not related to this
mapping are relegated to the Appendices: Appendix~\ref{app:realsp} discusses
in some detail matters related to the choice of an ultraviolet cutoff, and
also gives the often-used position-space definition of the 2CK model, to
facilitate comparison with our momentum-space version.  The construction of
the extended Fock space needed for refermionization is discussed in
Appendix~\ref{Jan'sapp}, and the technical details used to diagonalize the
resonant-level model and to calculate several of its properties are given in
Appendix~\ref{app:2ck}.  Finally, for pedagogical reasons and for the sake of
completeness, in Appendix~\ref{app:1ck} we use our finite-size bosonization
method to solve the 1-channel Kondo model as well.

\section{Definition of the model}
\label{model}

\subsection{Hamiltonian in Momentum Space}

Throughout the main part of this
 paper we shall use the standard 2CK
Hamiltonian in momentum space (its position-space 
representation is given in Appendix~\ref{app:realsp}).
We consider a magnetic impurity with spin 1/2
placed at the origin of a 
sphere of radius $R=L/2$, filled with
two  species of free, spinful conduction electrons, labeled by a spin index 
$\alpha = (\uparrow, \downarrow) = (+,-)$ and a channel or flavor
index $j=(1,2)=(+,-)$.
We assume that 
the interaction between the impurity and the conduction electron
is sufficiently short-ranged
that it involves only  $s$-wave conduction electrons,
whose kinetic energy can be written as 
($v_F = \hbar = 1$)
\begin{equation}
H_0 = \sum_{k \alpha j}    k 
  : \! c_{k \alpha j}^\dagger  c_{k \alpha j} \! : \; ,
\label{hkinetic}
\end{equation}
The operator  $c^\dagger_{k \alpha j} $ creates an 
$s$-wave conduction electron of species $(\alpha j)$ with radial 
momentum $k \equiv p 
- p_F$ relative to the Fermi momentum $p_F$, and the dispersion has
been linearized around the Fermi energy $\varepsilon_F$:
$\varepsilon_k-\varepsilon_F \approx k $.  The symbol $:\phantom{n}:$
in Eq.~(\ref{hkinetic}) denotes normal ordering with respect to the
free Fermi sea or ``vacuum state'' $|\vec 0\rangle_0$ (the reason for
this notation will become clear in the next section), defined by 
\begin{mathletters}
\label{eq:defnormorder}
\begin{eqnarray}
c_{k\alpha j} |\vec 0\rangle_0 \equiv 0 &
\quad \mbox{for} \quad  & k>0, 
\\
c_{k\alpha j}^\dagger |\vec 0\rangle_0 \equiv 0 &
\quad \mbox{for} \quad &  k\le 0.
\end{eqnarray}
\end{mathletters}
The $c_{k \alpha j}$'s obey
standard anticommutation relations,
\begin{equation}
        \{c_{k\alpha j},c_{k'\alpha' j'}^\dagger\}= \delta_{kk'}
        \delta_{\alpha\alpha'}\delta_{jj'}\; ,
\end{equation}
where due to  radial 
momentum quantization in the spherical box, 
the values taken on by $k$ are quantized:
\begin{equation}
        k= {\textstyle \frac {2\pi} L} 
        (n_k - P_0/2)\;,\quad n_k\in {\mathbb Z}\;.      
\label{k_free} 
\end{equation} 
Here  $P_0 = 0$  or $1$,  since 
at zero temperature  the chemical 
potential (and hence $p_F$) must either coincide with a degenerate level
($P_0=0$)  or lie midway between two of them, respectively
($P_0=1$). Evidently the level spacing in both cases is 
\begin{equation}
\Delta_L={\textstyle 2\pi\over L}.
\end{equation}

Since the $s$-wave conduction electrons form an effectively one-dimensional 
system they can also be described by a one dimensional chiral field, defined
as\cite{AL91b}
\begin{eqnarray}
\label{define-psi}
\psi_{\alpha j}(x) \equiv 
        {\textstyle \sqrt{{2\pi\over L}} }
        \sum_{n_k \in {\mathbb Z}} e^{- ikx}c_{k\alpha j} \; , 
        \phantom{nn} \bigl( x \in 
\bigl[- {\textstyle \frac L2}, {\textstyle \frac L2} \bigr] \bigr)
\;, \\
        \{\psi_{\alpha j}(x), \psi^\dagger_{\alpha^\prime 
        j^\prime}(x^\prime)\} =
        \delta_{\alpha \alpha^\prime} \delta_{jj^\prime} 2\pi 
        \delta(x-x^\prime)\; .
\label{anticom}
\end{eqnarray}
In the continuum limit $L\to\infty$, the $x>0$ 
and $x<0$ portions of $\psi_{\alpha j}(x)$ can be 
associated with the incoming and outgoing scattering states, respectively. 
Note that for
$P_0 = 0$ or $1$ the fields $\psi_{\alpha j}(x)$ 
have periodic  
or antiperiodic  boundary  conditions at $x = \pm L/2$, respectively, 
hence $P_0$ will be called the ``periodicity parameter''.

We assume a short-ranged anisotropic exchange interaction 
between the impurity spin and the $s$-wave conduction electron 
spin density at the origin, i.e.\ a 
 Kondo interaction of the form
\begin{equation}
        H_{\rm int} = \Delta_L
        \sum_{\scriptstyle \mu ,k,k^\prime 
        \atop \scriptstyle \alpha,\alpha^\prime,j} 
        \lambda_\mu S_\mu : \!c^\dagger_{k\alpha j} ({\textstyle \frac 1 2}
        \sigma^\mu_{\alpha\alpha^\prime}) c_{k^\prime\alpha^\prime j} \! : \; .
\label{H_int1D}
\end{equation}
Here the $S_\mu$ ($\mu = x,y,z$) are the impurity spin operators, with
  $S_z$ eigenvalues $(\Uparrow,\Downarrow) = ({1 \over 2}, - {1\over 2})$, and
  the $ \lambda_\mu $'s denote dimensionless couplings:
  $\lambda_z $ generates different phase shifts for
  spin-up and spin-down conduction electrons, while $\lambda_x
 \equiv  \lambda_y \equiv \lambda_\perp $ 
describe spin-flip scattering off the impurity.  Finally,
the effect of a finite magnetic field is described by
\begin{equation}
H_h = h_i S_z + h_e \hat {\cal N}_s  , 
\label{Hh}
\end{equation}
where $h_i$ and $h_e$ denote the  magnetic fields
acting on the impurity and conduction electron spins,
respectively, and $\hat{\cal N}_s$ (to be defined slightly below) denotes 
the total spin of the conduction
electrons.

Finally, note that we have taken all sums $\sum_k$ 
  over fermion momenta above
  to be unbounded, since the constructive bosonization scheme we
  intend to use requires an unbounded fermion momentum spectrum.
  We thus have effectively taken the fermion bandwidth, say
  $D$, to be infinite, but will reintroduce
  an ultraviolet cutoff when defining the boson fields in
  Eq.~(\ref{bosonfieldsa}) below.

\section{Bosonization Basics}
\label{bosonizationbasics}

The key to diagonalizing  the  Hamiltonian is to find
the relevant quantum numbers of the problem and to bosonize
the Hamiltonian carefully. 
While the technique of bosonization is
widely used in the literature, the  so-called
Klein factors mentioned in the introduction are often
neglected or not treated with sufficient care. 
However, it has recently been emphasized by several authors
\cite{kotliarsi,jvdschoeller,FabGogcomm,Furusaki} that 
these Klein factors are very important in some situations,
one of which is the calculation of the  finite-size
spectrum.  In the present Section we therefore discuss our 
bosonization approach in somewhat more detail than  usual,
formulating it as a set of {\it operator identities in Fock space},
and emphasizing in particular the proper use of Klein factors
to ladder between states with different particle numbers in
Fock space. (An elementary, pedagogical and detailed introduction
to the bosonization scheme used here,
which is based on that of Haldane,\cite{Haldane81} may be found in
Ref.~\onlinecite{jvdschoeller}.)

\subsection{Bosonization Ingredients}

To characterize the electronic states, we start by
introducing the number operators 
\begin{equation}
\hat N_{\alpha j} \equiv
    \sum_{k} \!\! : \! c_{k \alpha j}^\dagger  c_{k \alpha j} \! :
   \; ,
\label{N_alphaj}
\end{equation}
which count the number of electrons in channel $(\alpha j)$ with respect 
to the free electron reference ground state $| \vec 0 \rangle_0$. The 
non-unique eigenstates of $\hat N_{\alpha j}$
will generically  be denoted by 
$| \vec N \rangle \equiv | N_{\uparrow 1} \rangle \otimes 
| N_{\downarrow 1} \rangle \otimes | N_{\uparrow 2} \rangle \otimes 
| N_{\downarrow 2} \rangle$, where the  $N_{\alpha j}$'s can
be arbitrary integers, i.e.\ $\vec N \in {\mathbb Z}^4$.  

Next, we define bosonic electron-hole creators by
\begin{equation}
b^\dagger_{q \alpha j} \equiv 
               \frac{i}{ \sqrt{n_q}}
               \sum_{n_k \in Z} 
        c^\dagger_{k+q \alpha j} c_{k \alpha j} \; , 
        \quad \; (q = 2 \pi n_q /L > 0) \; ,
\label{e-hole}
\end{equation}
where the $n_q$ are positive integers. The operators $b^\dagger_{q\alpha
j}$ create ``density excitations'' with momentum $q$ in channel 
$\alpha j$, satisfy standard bosonic commutation relations, and 
commute with the $\hat N_{\alpha j}$'s:
\begin{eqnarray}
\label{bqcom}
\mbox{[} b_{q \alpha j}, b^\dagger_{q' \alpha' j'} \mbox{]} &=& 
  \delta_{q q'} \delta_{\alpha \alpha'} \delta_{jj'} \;,  \\
 \mbox{[}b_{q \alpha j} , {\hat N}_{\alpha' j'} \mbox{]} 
 &=&   0 \nonumber \; . 
\end{eqnarray}

Among all states $|\vec N\rangle$ with given $\vec N$, 
there is a unique state, to be denoted by 
$|\vec N\rangle_0$, that contains {\em no holes}\/
and thus has the defining property
\begin{equation}
b_{q  \alpha  j} |{\vec N} \rangle_0 = 0  \phantom{nnn}(\mbox{for any  
$q>0,\alpha,j$} )\;.
\label{|N>_0}
\end{equation}
We shall call it the ``$\vec N$-particle ground state'', since in the
absence of the interaction term~(\ref{H_int1D}), no $|\vec N \rangle$
has a lower energy than $|{\vec N} \rangle_0$;
likewise, no $|{\vec N}
\rangle_0$ has a lower energy than the ``vacuum state'' $| \vec
0\rangle_0$ defined in Eq.~(\ref{eq:defnormorder}). Note, though, that 
if $P_0 = 0$, the states $c_{0 \alpha
j} |\vec 0 \rangle_0$ are degenerate with $|\vec 0 \rangle_0$,
  because then $c_{0 \alpha j}$ removes a zero-energy electron.
 [In the conformal field theory
literature the states $|\vec N\rangle_0$ are sometimes referred
to   as $[U(1)]^{4}$ primary states, since the quantum numbers ${\hat 
N}_{\alpha j}$ are just the charges associated with the $U(1)$ gauge 
transformations of the fields $\psi_{\alpha j}\to \psi_{\alpha j}e^{i 
\delta_{\alpha j}}$.]
It can be proven \cite{Haldane81,jvdschoeller}
that any $\vec N$-electron state $|{\vec N}\rangle$ can 
be written as $|{\vec N}\rangle = f(b^\dagger) |{\vec N}\rangle_0$,
i.e.\ 
by acting on the ${\vec N}$-electron ground state
with an appropriate function of electron-hole operators. 

Next, we define bosonic  fields by 
\begin{eqnarray}
\label{bosonfieldsa}
\phi_{\alpha j} (x) & \equiv &   \sum_{q > 0}
         \frac{- 1}{ \sqrt{n_q}}
           \left( e^{-i q x} b_{q {\alpha j}} 
          +  e^{i q x} b^\dagger_{q {\alpha j}} \right) 
          e^{-a q/2} \; .
\end{eqnarray}
Here $a \sim 1/p_F$ is a short-distance cutoff; it is introduced to
cure any ultraviolet divergences the theory may have acquired by
taking the fermion bandwidth $D$ to be infinite. It is well-known,
however, that within this {\em bosonization cutoff scheme}\/ the
coupling constants have different meanings than for other standard
regularization schemes using a finite fermion bandwidth, and that the
relations between coupling constants in different regularization
schemes can be found by requiring that they yield the same phase
shifts.  For the sake of completeness, we discuss this and other
cutoff related matters in some detail in Appendix~\ref{Pedantic}.

It is easy to prove that 
the fields $\partial_x\phi_{\alpha j} (x)$ are
canonically conjugate to the $\phi_{\alpha j} (x)$'s,
in that \begin{equation}
\label{[phi,dxphi]}
\mbox{[} \phi_{\alpha j} (x), \partial_{x'} \phi_{\alpha'
j'}(x') \mbox{]}
= 2\pi i  (\delta_a(x-x') - 1/L)\; \delta_{\alpha\alpha'}
\delta_{jj'}\; ,
\end{equation}
where $ \delta_a (x-x')$ is a smeared delta function:
\begin{equation}
\label{delta_a}
\delta_a (x-x') = {a / \pi \over (x - x')^2 + a^2}\;.
\end{equation}

As final bosonization ingredient, 
we need the so-called Klein factors $F_{\alpha j}$, which 
ladder between states with different
$N_{\alpha j}$'s (which no function
containing only $b^\dagger_{q \alpha j}$ can accomplish,
since these conserve $N_{\alpha j}$).
By definition, the $F_{\alpha j}$'s are required to 
satisfy the following relations: 
\begin{mathletters}
\label{Fajcomms}
\begin{eqnarray}
        \mbox{[} 
        F_{\alpha j}, {\hat N}_{\alpha^\prime j^\prime}
        \mbox{]} 
&=& 
        \delta_{\alpha\alpha^\prime}
        \delta_{jj^\prime} F_{\alpha j} \;, 
\label{prop2}
\\
        \mbox{[}
        F_{\alpha j}, b_{q \alpha^\prime j^\prime}
        \mbox{]} 
&=&
        \mbox{[} F_{\alpha j}, 
        b^\dagger_{q 
        \alpha^\prime j^\prime} \mbox{]} 
        =0\;, 
\label{prop1}
\\
        F_{\alpha j} F^\dagger_{\alpha j}
        &=& F^\dagger_{\alpha j} F_{\alpha j} = 1 \; ,
\label{prop3} 
\\
        \{ F_{\alpha j}, F^\dagger_{\alpha^\prime j^\prime}\} &=& 2 \, 
        \delta_{\alpha \alpha'} \delta_{j j'} \;
\label{prop4}
\\
        \{ F_{\alpha j}, F_{\alpha^\prime j^\prime}\} &=& 0 
        \qquad \mbox{for}\; (\alpha j ) \neq (\alpha' j') \; .
\label{prop5}
\end{eqnarray}
\end{mathletters}
These relations imply that when  $F_{\alpha j}$ 
is applied to a state $|{\vec N}\rangle = f(b^\dagger) |\vec N \rangle_0$,
it commutes past  $f(b^\dagger)$ [by (\ref{prop1})],
and then removes [by (\ref{prop2})] an $(\alpha j)$ electron from
the topmost filled level of $|\vec N \rangle_0$,
namely $n_{k,\alpha j} = N_{\alpha j}$; to be explicit, 
$F_{\alpha j} |\vec N \rangle
= f(b^\dagger) c_{ N_{\alpha j} \alpha j} |\vec N \rangle_0$. 
Thus $F_{\alpha j}$  decreases 
the electron number in channel $\{\alpha j\}$ by one, $N_{\alpha j} \to 
N_{\alpha j} -1$, without creating particle-hole excitations. 
Similarly, $F^\dagger_{\alpha j}$ adds 
a single $(\alpha j)$ electron. 
 As shown  in Refs.~\onlinecite{Haldane81} or~\onlinecite{jvdschoeller}, 
the   construction $F_{\alpha j} = a^{1/2}
\psi_{\alpha j}  (0)  e^{i \phi_{\alpha j} (0)}$,
which explicitly expresses $F_{\alpha j}$ in
terms of the
fermion operators  $c_{k \alpha j}$, has
all the desired properties.

\subsection{Bosonization Identities}
Having introduced the Klein factors $F_{\alpha j}$ and the 
boson fields $\phi_{\alpha j}$,  we are ready to bosonize, i.e.\ to rewrite 
expressions involving the fermion operators $c_{k \alpha j}$
in terms of the bosonization ingredients defined above.
In our notation, the standard bosonization identities
\cite{Haldane81} for the fermion field, density 
and kinetic energy 
take the following  forms:\cite{jvdschoeller}
\begin{eqnarray}
&&\psi_{\alpha j} (x) 
=     F_{\alpha j}  a^{-1/2} 
    e^{-i (\hat N_{\alpha j} - P_0/2) 2 \pi x /L} \, 
    e^{- i \phi_{\alpha j} (x) } \; , 
\label{fermiboson} \\
&&{1\over 2\pi} : \! \psi^\dagger_{\alpha j} (x) \psi_{\alpha j} (x)  \! : 
= {1\over 2\pi}\partial_x \phi_{\alpha j} (x) 
  + \hat N_{\alpha j} / L \; , 
\label{density}
\\
\label{H0a} 
&&H_0 
        = 
        \sum_{\alpha j} 
        {\Delta_L \over 2}
             {\hat N}_{\alpha j} (\hat N_{\alpha j} + 1 -  P_0)
        +
        \sum_{\scriptstyle \alpha j\atop \scriptstyle q>0} 
        q \; b^\dagger_{q \alpha j} b_{q \alpha j} \; .
\end{eqnarray}

Several comments are in order: Firstly, in the limit $a \to 0$
  Eqs.~(\ref{fermiboson}) to (\ref{H0a}) are not mere formal correspondences
  between the fermionic and bosonic expressions, but hold as rigorous {\it
    operator identities in Fock space}\/.  For $a \neq 0$, (\ref{fermiboson})
  and (\ref{density}) are not rigorously exact, but instead should be viewed
  as conveniently regularized redefinitions of the fermion fields and
  densities (as discussed in Appendix~\ref{app-bos-cutoff}).  Next, in
Eq.~(\ref{fermiboson}) for $\psi_{\alpha j}$, the Klein factors $F_{\alpha j}$
play a twofold role: firstly, by Eq.~(\ref{prop2}) they ensure that the
right-hand side of Eq.~(\ref{fermiboson}) acting on any state indeed does
lower the number of $\alpha j$-electrons by one, just as $\psi_{\alpha j}$
does; and secondly, by Eqs.~(\ref{prop4}) and (\ref{prop5}) they ensure that
fields with different ($\alpha j$)'s do have the proper anticommutation
relations (\ref{anticom}).  In contrast, Eq.~(\ref{density}) for the density
operator contains no Klein factors [because of (\ref{prop3})].  Finally, in
Eqs.~(\ref{H0a}) for the kinetic energy, the first $\Delta_L$ term is just
${}_0 \! \langle \vec N | H_0 | \vec N \rangle_0$, the energy of the $\vec
N$-particle ground state $|\vec N \rangle_0$ relative to $|{\vec 0}\rangle_0$.
Since the Klein factors do not commute with this term, they evidently cannot
be neglected when calculating the full model's finite-size spectrum, for which
all terms of order $\Delta_L$ must be retained.  The second term of
(\ref{H0a}) describes the energy of electron-hole excitations relative to $|
\vec 0 \rangle_0$.  [Its form can be obtained by observing that the commutator
$[b_{q \alpha j}, H_0]$ is the same when calculated in terms of $c_{k \alpha
  j}$'s using (\ref{hkinetic}) and (\ref{e-hole}), or in terms of $b_{q \alpha
  j}$'s using (\ref{bqcom}) and (\ref{H0a}).]

\section{Mapping onto Resonant-Level Model}
\label{EKmap}
In this section we map the 2CK model onto
a resonant level model, using a finite-size
version of the strategy invented by Emery and Kivelson:
using bosonization and refermionization,
we make a unitary transformation to a more
convenient basis, in which the Hamiltonian
is quadratic for a certain choice of parameters.

\subsection{Conserved Quantum Numbers}

The quantum numbers $N_{\alpha j}$ of 
Eq.~(\ref{N_alphaj}) are conserved under the action of
$H_0$, $H_h$ and $H_z$ (the $\lambda_z$ term of $H_{\rm int} \equiv 
H_z + H_\perp$),   but  fluctuate under the action of
the spin-flip interaction $H_\perp$ (the $ \lambda_\perp$ term). 
On the other
hand, the total charge and flavor of the conduction electrons is 
obviously conserved by  all terms in the Hamiltonian,
including   $H_\perp$. 
Therefore it is natural to introduce the following new quantum numbers:
\begin{equation}
\label{2ck:transformationN}
{\textstyle
\left( \begin{array}{c}
        \hat {\cal N}_c \\ \hat  {\cal N}_s \\ 
        \hat {\cal N}_f \\ \hat {\cal N}_x 
        \end{array} \right)}
\equiv
{1\over 2} 
{\textstyle     
        \left( \begin{array}{cccc}
        1 & \phantom{-}1 & \phantom{-}1 & \phantom{-}1 \\
        1 & -1  & \phantom{-}1  & -1 \\
        1 &  \phantom{-}1 & -1  & -1 \\
        1 & -1  & -1 & \phantom{-}1 
        \end{array} \right)
\left( \begin{array}{c}
        \hat N_{\uparrow 1} \\ \hat N_{\downarrow 1} 
        \\ \hat N_{\uparrow 2} \\ \hat N_{\downarrow 2}
        \end{array} \right) \; ,
}
\end{equation}
where $2\hat {\cal N}_c$, $\hat  {\cal N}_s$, and $\hat {\cal N}_f$
denote the total charge, spin, and flavor of the conduction electrons, and 
$\hat {\cal N}_x $ measures the spin difference between channels $1$ and 
$2$. 
Clearly, any conduction electron state
$| \vec N \rangle$ can equally well be labeled by the 
corresponding quantum numbers $ \vec {\cal N} \equiv({\cal N}_c,{\cal 
N}_s,{\cal N}_f,{\cal N}_x)$. 
However, whereas the  $N_{\alpha j}$'s take arbitrary independent
integer values,  the $\vec {\cal N}$'s
generated by  Eq.~(\ref{2ck:transformationN}) (with $\vec N \in {\mathbb Z}^4$)
can easily be shown to satisfy the following two constraints, 
to  be called the 
{\it free gluing conditions}: 
\begin{mathletters}
\label{gluingall}
\begin{eqnarray}
\vec {\cal N} &\in &({\mathbb Z} + P/2)^4 \; , 
\label{2ck:Nyint} \label{gluinga}
\\
 {\cal N}_c  \pm  {\cal N}_f &=&  
( {\cal N}_s   \pm  {\cal N}_x)\,  \mbox{mod}\, 2 \; ,
\label{2ck:GC} \label{gluingb}
\end{eqnarray}
\end{mathletters}
where the {\em parity index}\/ $P$ equals 0 or 1 if the total number of
electrons is even or odd, respectively.  Eq.~(\ref{2ck:Nyint}) formalizes the
obvious fact that the addition or removal of one $\alpha j$ electron to or
from the system necessarily changes {\em each}\/ of the ${\cal N}_y$'s by $\pm
1/2$, so that they are either all integers or all half-integers.
Eq.~(\ref{2ck:GC}) selects out from the set of $\vec {\cal N}$ of the form
(\ref{2ck:Nyint}) those that correspond in the old basis to $\vec N \in
{\mathbb Z}^4$, i.e.\ to physical states (an $\vec {\cal N}$ of the form
(\ref{2ck:Nyint}) that violates (\ref{2ck:GC}) would correspond to $\vec N \in
({\mathbb Z}+1/2)^4$, which does not exist in the physical Fock space).

\begin{figure}
\epsfxsize=8.5cm
\epsfbox{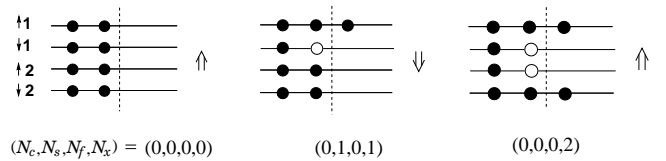}%{nxfluct.eps}
\vskip0.7truecm
\caption[Spin-flips]{
  \label{nxfluctuates}
Under a succession of spin flips,
${\cal N}_s$ fluctuates {\em mildly}\/ between
$S_T \mp 1/2$ (here $S_T = 1/2$);
in contrast, ${\cal N}_x$ fluctuates {\em wildly,}\/
since it can acquire {\em any}\/ value consistent 
with the gluing conditions (\ref{gluingall}).
The dotted line represents the reference  energy 0
up to which the free Fermi sea is filled for $P_0 = 1$,
the filled and empty circles represent filled and 
empty single-particle states with energy $k$. }
\end{figure}

The new basis has two major advantages: firstly,
$ {\cal N}_c$ and 
$ {\cal N}_f$ are conserved quantum numbers;
and secondly, the quantum number 
{\em ${\cal N}_s$ fluctuates only ``mildly''}\/ 
between the values $S_T \mp 1/2$, since the total spin,
\begin{equation}
\label{2ck:tildeNS}
S_T \equiv  {\cal N}_s + S_z \; , 
\end{equation}
is conserved. 
In contrast,  {\em the  quantum number
${\cal N}_x$ fluctuates ``wildly'',}\/  
because an appropriate succession of spin-flips 
can produce {\em any} ${\cal N}_x$  that
satisfies (\ref{2ck:GC}), as illustrated in Fig.~\ref{nxfluctuates}.
{\em This wildly fluctuating quantum number will be seen
below to be at the heart of the 2CK model's NFL behavior.}\/  
In revealing contrast, the 1CK model, which shows no NFL behavior,
 lacks such a wildly
fluctuating quantum number (see  Appendix~\ref{app:1ck}).

Since $S_T$, ${\cal N}_c$ and 
${\cal N}_f$ are conserved, the Fock space ${\cal F}_{\rm phys}$ 
of all physical states can evidently be divided 
as follows into subspaces invariant under the action of $H$:
\begin{eqnarray}
{\cal F}_{\rm phys} &= &\sum_{\oplus' S_T, {\cal N}_c, {\cal
N}_f} {\cal S}_{\rm phys} (S_T,{\cal N}_c, {\cal N}_f)\;, 
\label{2ck:subfock}
\end{eqnarray}
\begin{eqnarray}
{\cal S}_{\rm phys} (S_T,{\cal N}_c,{\cal N}_f) & = &
\sum_{\oplus' {\cal N}_x} \Bigl\{ | {\cal N}_{c}, S_T-1/2, {\cal N}_f, 
{\cal N}_x; 
\Uparrow \rangle \nonumber \\
  && \hspace{-5mm} \oplus  
  |{\cal N}_{c}, S_T + 1/2, {\cal N}_{f}, {\cal N}_x
  \! + \! 1 ; \Downarrow \rangle \Bigr\} .
  \label{2ck:S}
\end{eqnarray}
In both equations the prime on the sum indicates a restriction to those
${\cal N}_y$'s that satisfy the free gluing
conditions~(\ref{gluingall}).
To  diagonalize the Hamiltonian for
given $S_T$, ${\cal N}_{c}$ and ${\cal N}_{f}$, 
it evidently suffices to restrict one's attention
to the corresponding subspace ${\cal S}_{\rm phys}
(S_T, {\cal N}_{c}, {\cal N}_{f})$.

\subsection{Emery-Kivelson transformation}
\label{sec:EK}

Following Emery and Kivelson, we now introduce, in analogy to  Eq.~(\ref{2ck:transformationN}),
new electron-hole operators and boson fields via the transformations,
\begin{equation}
\left.
\begin{array}{l}
b_{q y} \equiv \sum_{\alpha j} R_{y,\alpha j} b_{q \alpha j} \\
\varphi_{y} \equiv \sum_{\alpha j} R_{y,\alpha j} \phi_{\alpha j}
\label{2ck:varphidef}
\end{array}
\right\}
\; \mbox{\phantom{nn}($y=c,s,f,x$)}\;,
\label{2ck:transformationb}
\end{equation}
where $R_{y,\alpha j}$ is the unitary matrix in (\ref{2ck:transformationN}).
These obey relations analogous to (\ref{bqcom}) and (\ref{[phi,dxphi]}), with
$\alpha j \to y$.  Moreover, we define $|\vec {\cal N}\rangle_0$, the
$\vec{\cal N}$-particle vacuum state, to satisfy $b_{qy} |\vec {\cal N}
\rangle_0 = 0$, as in (\ref{|N>_0}).  If $\vec {\cal N}$ and $\vec N$ are
related by (\ref{2ck:transformationN}), then the states $|\vec {\cal
  N}\rangle_0 $ and $|\vec N \rangle_0$ are equal up to an unimportant phase
(see Appendix~\ref{Jan'sapp}), because both have the same $\hat N_{\alpha j}$
and $\hat {\cal N}_y$ eigenvalues and both are annihilated by all $b_{q \alpha
  j}$'s and $b_{qy}$'s.

Using the quantum numbers ${\hat {\cal N}}_y$ and the bosonic fields 
$\varphi_y(x)$, the Hamiltonian takes a transparent form. The free
electron  part $H_0$ of (\ref{H0a}) can be written  as
\begin{equation}
\label{2ck:H0y}
 H_0 = \Delta_L \Bigl[
        \hat {\cal N}_c (1 - P_0)  +
         \sum_y  \hat {\cal N}_y^2 /2 \Bigr] 
        + \sum_{ y, \, q>0} q \; b^\dagger_{q y} b_{q y} \;,
\end{equation}
while Eqs.~(\ref{density}) and (\ref{fermiboson}) are used to
obtain, 
respectively, 
\begin{eqnarray}  
\label{2ck:bosonh}
  H_z &= & \lambda_z   
        \left[ \partial_x \varphi_s (0) + \Delta_L \hat {\cal N}_s
        \right] S_z \, , 
\\
  H_\perp  &=& 
   {\lambda_\perp \over 2a} \biggl[ 
  e^{- i \varphi_s (0)} S_+ ( F^\dagger_{\downarrow 1} F_{\uparrow 1} 
        e^{- i  \varphi_x (0) } \nonumber \\
&+& F^\dagger_{\downarrow 2} F_{\uparrow 2} e^{i  \varphi_x 
(0) 
}) + \mbox{h.c.} \biggr] \; . 
\label{2ck:H_perp}
\end{eqnarray}
Eqs.~(\ref{2ck:H0y}) to (\ref {2ck:H_perp}) 
and (\ref{Hh}) constitute the bosonized form
of the Hamiltonian for the anisotropic 2CK model,
{\em up to and including terms of order $\Delta_L$}.

Next we simplify  $H_z$, which, being diagonal
in spin indices, merely causes a phase shift in the
spin sector. This phase shift can be obtained explicitly
using a unitary transformation (due to EK) parameterized by a real number
$\gamma$, whose value will be determined below:
\begin{eqnarray}
&&H\to H' = U H U^\dagger \;,\\
&&U \equiv e^{i \gamma S_z \varphi_s  (0)} \;.
\label{1ck:ektr}
\end{eqnarray}
Under this transformation the impurity spin, 
the spin-diagonal part of the Hamiltonian,
 the spin boson field and the fermion fields transform as follows
(using e.g.\ the  operator identities of 
Appendix~C of Ref.~\onlinecite{jvdschoeller}):
\begin{eqnarray}
&& \phantom{H_0 + }
S_\pm \; \to \;U S_\pm U^\dagger = e^{\pm i\gamma \varphi_s(0)}
S_\pm \;, 
\label{S^pm'}
\\
&&
H_0 + H_z \; \to 
        \;H_0 + (\lambda_z - \gamma) 
        \partial_x\varphi_s(0) S_z \nonumber \\
&& \phantom{H_0 + H_z \; \to} + \lambda_z  \Delta_L 
        {\hat {\cal N}}_s\; S_z + 
\gamma^2 \mbox{[} 1/(4 a)  - \pi / (4 L) \mbox{]} \; , 
\label{2ck:H_z'}
\\
\label{varphiphase}
&& \varphi_s(x) \; \to \; \varphi_s(x) - 2 \gamma S_z\; {\rm
arctan}(x/a)\;, \quad (|x| \ll L) \; . \label{varphi_s'} 
\\
 \label{eq:EK-psi}
& & \psi_{\alpha j} (x) \to  \psi_{\alpha j} (x)
e^{i \alpha \gamma S_z  \arctan (x/a)} \;, \quad (|x| \ll L) \; .
\end{eqnarray}
Eq.~(\ref{2ck:H_z'}) is most easily derived
in the momentum-space representation [using (\ref{bqcom}), 
(\ref{bosonfieldsa}) and (\ref{H0a}), see Section~7 of
Ref.~\onlinecite{jvdschoeller}];
on the other hand, since  in Eq.~(\ref{varphi_s'}) 
we only give the $|x| \ll L$ limit in which order $1/L$ terms
can be neglected,  this equation is
easier to derive in the position-space
representation (by first evaluating  $U \partial_x \varphi_s (x) U^{-1}$
using (\ref{[phi,dxphi]}) and (\ref{delta_a}), 
then integrating). 
Eq.~(\ref{eq:EK-psi}) follows from (\ref{varphi_s'}),
since $\psi_{\alpha j} \propto e^{- i \alpha  \varphi_s /2 }$
[by (\ref{fermiboson}) and (\ref{2ck:varphidef})].

Recalling [from (\ref{density})] that $\partial_x\varphi_s(x) / 2\pi$
contributes to the conduction electron spin density, we note by
differentiating (\ref{varphi_s'}) that the EK transformation produces a change
in the spin density of $-2\gamma S_z \pi \delta_a(x)/2\pi$; intuitively
speaking, it ties a spin of $- \gamma S_z $ from the conduction band to the
impurity spin $S_z$ at the origin.

To eliminate the $S_z \partial_x \varphi_s$  term in (\ref{2ck:H_z'}), 
we now choose $\gamma \equiv \lambda_z$; then 
the spin-flip-independent part of the
Hamilton takes the form 
\begin{eqnarray}
H'(\lambda_\perp = 0)  &=& \lambda_z \Delta_L 
        \hat {\cal N}_s S_z + 
\sum_y   \Delta_L  \hat {\cal N}_y^2 / 2  \nonumber \\
\label{2ck:lambda=0}
        &+& \sum_{ y, \, q>0} q \; b^\dagger_{q y} b_{q y} + 
        H_h + \mbox{const}\; ,
\end{eqnarray}
and  $H'_\perp$ contains the
factors $e^{\pm i (1- \lambda_z) \varphi_s (0)}$. 
These factors are simply equal to 1 
at the special line $\lambda_z=1$,
the so-called {\it Emery-Kivelson line},
at which   $H'_\perp$ simplifies to
\begin{equation} 
H'_\perp  = 
        { \lambda_\perp \over 2a} \biggl[ 
S_+ ( F^\dagger_{\downarrow 1} F_{\uparrow 1} e^{- i \varphi_x (0)} 
+ F^\dagger_{\downarrow 2} F_{\uparrow 2} 
e^{i  \varphi_x (0) }) + 
\mbox{h.c.}  
\biggr] .
\label{HEK}
\end{equation}
We shall henceforth focus on the case $\lambda_z = 1$,
which will enable us to diagonalize the model exactly by refermionization.
Deviations from the EK line will  be shown in Section~\ref{ss:finsize} to be irrelevant,
by  taking $\gamma = 1$ 
but $\lambda_z = 1 + \delta \lambda_z$,  and doing perturbation
theory in 
\begin{equation}
\label{dev:EKline}
\delta H'_z = \delta \lambda_z [
        \partial_x\varphi_s(0)  \, + \, \Delta_L {\hat {\cal N}}_s ] S_z \; . 
\end{equation}

Before proceeding, 
it is instructive to interpret the significance  of the EK line using 
the standard heuristic language 
of Nozi\`eres and Blandin. They argued
that in the {\em strong-coupling limit}, 
an anti-ferromagnetically-coupled impurity
will ``capture'' one spin 1/2 from {\em both}\/ the $j=1$ and $2$
channels (capturing a spin from just one channel
would break channel symmetry), 
i.e.\ it will tie  a spin of $- 2 S_z$ from the conduction band to 
its own $S_z$, which yields an ``overscreened effective spin of $-S_z$''. 
However, since the latter is again anti-ferromagnetically coupled
to the remaining conduction electrons, 
this strong-coupling fixed point is unstable in the RG sense,
just as the weak-coupling fixed point. Thus, they argued
that there must be a  NFL fixed point at intermediate
coupling which is stable (for $h_e = h_i = 0$).

Now, a crucial property of the EK line is that it contains this
NFL intermediate-coupling fixed point. A heuristic way
to see this it to note that 
on the EK line, the impurity spin {\em is}\/ in fact ``perfectly
screened'': the spin $- \gamma S_z$ from the conduction band
[mentioned after~(\ref{varphiphase})], that is tied to the impurity
by the EK transformation, is equal to $-S_z$ if $\gamma= \lambda_z =1$, thus
precisely ``canceling'' the impurity's spin $S_z$.  Thus, on the EK line
the impurity ``captures'' exactly one spin 1/2 from the conduction
band to form a ``perfectly screened singlet'' with 
{\em zero}\/ total spin (which is the heuristic reason why
the EK line is stable), 
but it does so {\em without}\/ breaking channel symmetry, since $\partial_x
\varphi_s$ is constructed in equal amounts 
from all four species of $\phi_{\alpha j}$ fields.

Of course, there are also more rigorous ways of seeing that the NFL fixed
point lies on the EK line: Firstly, for $\lambda_z =1 $ 
it follows from (\ref{eq:EK-psi})
 that the phase shift $\delta$ of the outgoing relative to the incoming fields, 
defined by 
$  \psi_{ \alpha j} (0^-) \equiv
 e^{i 2 \delta} \psi_{\alpha j} (0^+)  $ (with $|0^\pm| \gg a$),
is $|\delta|  = \pi / 4$, 
which is just the value known for the NFL fixed point from
other approaches.\cite{VZZ88,AL92b}  Secondly, we shall deduce in
Section~\ref{ss:finsize} from an analysis of 
the finite-size spectrum that the leading
irrelevant operators (with dimensions $1/2$) vanish exclusively  {\em
  along}\/  this line, but not away from it. 
Since  the presence or absence of the
leading irrelevant operators strongly
influences the low-temperature properties of the model 
(such as its critical exponents)\cite{EK92,SG94}, 
and since these must stay invariant under any RG transformation,
one concludes that the Emery-Kivelson
line must be  be stable under RG transformations.

\subsection{Refermionization}

\subsubsection{Definition of New Klein Factors}
\label{sec:newKlein}

The most nontrivial step in the solution of the model is the proper treatment
of Klein factors when refermionizing the transformed Hamiltonian. In their
original treatment Emery and Kivelson did not discuss Klein factors at all and
simply identified $e^{-i\varphi_x(x)}/\sqrt{a}$ as a new pseudofermion field
$\psi_x (x)$. Although this procedure happened to be adequate for their
purposes, the proper consideration of the Klein factors and gluing conditions
is essential, as already emphasized in the introduction, for giving a rigorous
solution of the problem and obtaining the finite-size spectrum. Some other
authors tried to improve the Emery-Kivelson procedure by representing the
Klein factors by $F_{\alpha j} \sim e^{-i\Theta_{\alpha j}}$, where
$\Theta_{\alpha j}$ is a ``phase operator conjugate to ${\hat N}_{\alpha
  j}$'', and added these to the bosonic fields $\phi_{\alpha j}$ before
making the linear transformation (\ref{2ck:varphidef}). This procedure is
problematic, however,  since then the $e^{-i\varphi_y(0)}$'s contain factors
such as $e^{-i\Theta_{\alpha j}/2}$, which are ill-defined (for a more
detailed discussion of this point, see Appendix D.2 of
Ref.~\onlinecite{jvdschoeller}).

A novel, rigorous way of dealing with Klein factors when
refermionizing was presented in Ref.~\onlinecite{vDZF}
(and adapted in Ref.~\onlinecite{jvdschoeller} to treat
an impurity in a Luttinger liquid):
We introduce a new set of ladder operators
${\cal F}^\dagger_y$ and ${\cal F}_y$ $(y = c,s,f,x)$
to raise or lower the new  quantum numbers ${\cal N}_y$
by $\pm 1$, 
with, by definition, the following  properties
[in analogy to Eqs.~(\ref{Fajcomms})]: 
\begin{mathletters}
\label{Fajcomms-y}
\begin{eqnarray}
\mbox{[}{\cal F}_y, {\hat {\cal N}}_{y'} \mbox{]} 
&= & \delta_{yy'} {\cal F}_y \;, 
\label{propy2}
\\
\mbox{[} {\cal F}_y, b_{q y'} \mbox{]}
 & = & \mbox{[} {\cal F}_y, b^\dagger_{q y'} \mbox{]} = 0\;, 
\label{propy1}
\\
     {\cal F}_{y} {\cal F}^\dagger_{y}
        &= & {\cal F}^\dagger_{y} {\cal F}_{y} = 1 \; ,
\label{propy3} 
\\
        \{ {\cal F}_{y}, {\cal F}^\dagger_{y'}\} & = & 2\, 
     \delta_{y y'} \; ,
\\
     \{ {\cal F}_{y}, {\cal F}_{y'}\} & = & 0 \; 
\quad \mbox{for}\; y \neq y'\; .
\label{propy4}
\end{eqnarray} 
\end{mathletters}
Now, note that the action of any one of the new Klein factors ${\cal F}_y$ or
${\cal F}_y^\dagger$ respects the first of the free gluing conditions,
(\ref{gluinga}), but not the second, (\ref{gluingb}).  More generally,
(\ref{gluingb}) is respected only by products of an {\em even}\/ number of new
Klein factors, but violated by products of an {\em odd}\/ number of them. This
implies that the physical Fock space ${\cal F}_{\rm phys}$ of all $|\vec {\cal N}
\rangle$ satisfying both (\ref{gluinga}) and (\ref{2ck:GC}) is closed under
the action of even but not of odd products of new Klein factors. For example,
let $|\psi\rangle_{\rm phys}$ be in ${\cal F}_{\rm phys}$, then ${\cal F}^\dagger_y
{\cal F}_{y'} | \psi\rangle_{\rm phys}$ is too, but ${\cal F}_y |
\psi\rangle_{\rm phys}$ violates (\ref{2ck:GC}) and hence is an unphysical state.
{\em The action of arbitrary combinations of new
Klein factors thus generates an  extended Fock space}\/ ${\cal
  F}_{\rm ext}$, which contains  ${\cal F}_{\rm phys}$ as a subspace
and is spanned by the set of {\em all}\/ $|\vec {\cal N}
\rangle$ satisfying (\ref{2ck:Nyint}), including unphysical states violating
(\ref{2ck:GC}). To demonstrate  that ${\cal F}_{\rm phys}$ can indeed
be so embedded in  ${\cal F}_{\rm ext}$, we explicitly construct
a set of basis  states for the latter in Appendix~\ref{Jan'sapp}.

Since {\em odd}\/ products of ${\cal F}_y$'s lead out of ${\cal F}_{\rm
  phys}$, they {\it cannot} be expressed in terms of the original Klein
factors $F_{\alpha j}$, which leave ${\cal F}_{\rm phys}$ invariant.  However,
and this is crucial, the Hamiltonian contains only {\em even}\/ products of
old Klein factors. 
Now, any ``diagonal''
combination $F^\dagger_{\alpha j} F_{\alpha j} = 1$; and any ``off-diagonal''
combination $F^\dagger_{\alpha j} F_{\alpha' j'}$ 
or  $F^\dagger_{\alpha j} F^\dagger_{\alpha' j'}$ acting on any state $| \vec
N\rangle$ {\em just changes two of its $N_{\alpha j}$ quantum numbers.}\/
Using (\ref{2ck:transformationN}) to read off the corresponding changes in
${\cal N}_s$, ${\cal N}_f$ and ${\cal N}_x$, 
we can thus make the following identifications between {\em
  pairs}\/ of the old and new Klein factors:
\begin{mathletters}
   \label{newFs} \label{Kleinchange}
\begin{eqnarray}
  \label{newFsA} 
 {\cal F}^\dagger_x {\cal F}^\dagger_s 
\equiv
 F_{\uparrow 1}^\dagger F_{\downarrow 1} 
\; , & \qquad & 
 {\cal F}_x {\cal F}^\dagger_s \equiv
 F_{\uparrow 2}^\dagger F_{\downarrow 2} 
  \; , 
\\
{\cal F}^\dagger_x {\cal F}^\dagger_f \equiv 
 F_{\uparrow 1}^\dagger F_{\uparrow 2} 
\; , & \qquad & 
 {\cal F}^\dagger_c {\cal F}^\dagger_s =
 F_{\uparrow 1}^\dagger F_{\uparrow 2}^\dagger \; .
\end{eqnarray}
\end{mathletters}
Each of these relations involves an arbitrary choice of phase, whose
consequences for the basis states $| \vec {\cal N} \rangle$ are discussed in
Appendix~\ref{Jan'sapp}.  These choices uniquely fix the phases of all other
similar bilinear relations between old and new Klein factors, which can be
found from composing (and conjugating) the above four, e.g.\ ${\cal F}_s^\dag
{\cal F}_f^\dag = - ( {\cal F}_x {\cal F}_s^\dag) ({\cal F}_x^\dag {\cal
  F}_f^\dag) = F^\dag_{\uparrow 1} F_{\downarrow 2}$.  Since the relations
(\ref{Kleinchange}) by construction respect (\ref{2ck:transformationN}) (as
can be verified by acting on any $|\vec N \rangle$), they, and all similar
bilinear relations derived from them, also respect both of the free gluing
conditions (\ref{gluingall}).  The relations involving ${\cal F}_c$ and ${\cal
  F}_f$ are not needed for the present 2CK model, but are included for
completeness; for example, ${\cal F}_f$ would be needed for models involving
``flavor-flip'' processes.

We can thus replace the Klein factor pairs occurring in Eq.~(\ref{HEK})
by the ones in Eq.~(\ref{newFsA}):
\begin{eqnarray}
H_\perp^\prime &=& 
        {\lambda_\perp \over 2a} \biggl[ 
   S_+ F_s ( F_x  e^{- i  \varphi_x (0) } + F^\dagger_x  e^{i \varphi_x 
(0) }) + \mbox{h.c.} \biggr]\;
\label{Hperpref}
\end{eqnarray}
The only consequence of this change is that we now work in the extended Fock
space ${\cal F}_{\rm ext}$, and will diagonalize $H'$ not in the physical
invariant subspace ${\cal S}_{\rm phys} (S_T, {\cal N}_{c},{\cal N}_{f})$ of
(\ref{2ck:S}), but in the corresponding extended subspace ${\cal S}_{\rm ext}
(S_T, {\cal N}_{c}, {\cal N}_{f})$, given by an equation similar to
(\ref{2ck:S}), but where the $\oplus' {\cal N}_x$ sum now is restricted only
to satisfy (\ref{2ck:Nyint}), not also (\ref{2ck:GC}). At the end of the
calculation we shall then use the gluing condition (\ref{2ck:GC}), satisfied
only by the physical states in ${\cal S}_{\rm ext}$ but not by its additional
unphysical states, to identify and discard the latter (see
Section~\ref{sec:generalEstate}).  This approach is completely analogous to
the use of gluing conditions in AL's CFT solution of the 2CK model. It is also
in some sense analogous to Abrikosov's pseudofermion
technique\cite{Abrikosov}; there an impurity-spin operator is represented in
terms of pseudofermion operators acting in an enlarged Hilbert space, which
contains not only the physical one-pseudofermion states, but also unphysical
many- or no-pseudofermion states that are projected away at the end of the
calculation.

\subsubsection{Pseudofermions and Refermionized Hamiltonian}

We now note that $H'_\perp$ of (\ref{Hperpref}) 
can be written in a form {\em quadratic}\/ in fermionic variables, 
\begin{eqnarray} 
H_\perp^\prime &=& 
{\lambda_\perp\over
2\sqrt{a}} \left( \psi_x(0) + \psi_x^\dagger(0) \right) (c_d - c_d^\dagger ) 
\;,
\label{Halfpsicouples}
\end{eqnarray}
by defining a local pseudofermion $c_d$ and a pseudofermion
field  $\psi_x (x)$ by
the following refermionization relations:
\begin{eqnarray}
c_d \, && \equiv F_s^\dagger S_-  \;, \qquad
c^\dagger_d c_d = S_z + 1/2 \; , 
\label{definec_d}
\end{eqnarray}
\vspace*{-4mm}
\begin{mathletters}
\label{newpsixgeneral}
\begin{eqnarray}
\psi_x(x) \,  && \equiv F_x  a^{-1/2}   e^{-i (\hat {\cal N}_x - 1/2 )
        2 \pi x /L} \, e^{- i \varphi_x (x) }
\label{newpsix}
\\
        && \equiv { \sqrt{2\pi \over  L}}
        \sum_{\bar k} e^{-i \bar k 
x}c_{\bar k x} \;.
\label{cbarkdef}
\end{eqnarray}
\end{mathletters}
where Eq.~(\ref{cbarkdef}) defines the $c_{\bar k x}$ as 
Fourier coefficients of the field $\psi_x (x)$.
For reasons discussed below, the field $\psi_x$ in Eq.~(\ref{newpsix})
has been defined in such a way that its boundary condition 
at $\pm L/2$  is $P$-dependent,
since ${\cal N}_x \in {\mathbb Z} + P/2$
and $\varphi_x(x)$ is a periodic function. 
Thus the quantized $\bar k$ momenta in the Fourier
expansion (\ref{cbarkdef}) must have the form
\begin{equation}
     \bar k = \Delta_L  [n_{\bar k} \!-\! (1 \! - \! P)/2 ]\;,
\qquad (n_{\bar k} \in {\mathbb Z}) 
\label{bark}
\end{equation}
(i.e.\ the periodicity parameter
$P_0$ of (\ref{k_free}) here equals $1-P$).

The new pseudofermions were constructed in such a way that they
satisfy the following 
commutation/anticommutation relations 
\begin{eqnarray}
\{c_{\bar k x},c^\dagger_{\bar k' x}\}  &=& \delta_{\bar k \bar
k'}\;,\\ 
\{c_d,c_d^\dagger\} &=& 1\;, \\
\{c_d,c_{\bar k x}^\dagger\} &=& \{c_d,c_{\bar k x}\} = 
0 \;,\\
\mbox{[}c_d , {\hat {\cal N}}_s\mbox{]} &=& c_d \;,
\label{[N_s,c_d].2ck}
\end{eqnarray}
which follow directly from the properties of $\varphi_x$ 
[by analogy to the relations of Section~\ref{bosonizationbasics}]
and Eqs.~(\ref{Fajcomms-y}). 
Note that  $c_d$ lowers the impurity
spin, raises the total electron spin ${\hat {\cal N}}_s$
and hence conserves the total spin $S_T$, whereas
$\psi_x$ conserves each of the impurity, electron and total spins.

Next we have to relate the number operator for the new
$x$-pseudofermions to the quantum number ${\cal N}_x$.
This requires defining a free reference ground state,
say $|0 \rangle_{{\cal S}_{\rm ext}}$,
in the extended subspace ${\cal S}_{\rm ext}$,  
with respect to which the number of pseudofermions
are counted. In analogy to  (\ref{eq:defnormorder}),
we define  $|0 \rangle_{{\cal S}_{\rm ext}}$ by
\begin{mathletters}
\label{eq:defnormorderbark}
 \begin{eqnarray}
c_{\bar k x} | 0\rangle_{{\cal S}_{\rm ext}} \equiv 0 &
\quad \mbox{for} \quad & \bar k>0, 
\\
c_{\bar k x}^\dagger | 0\rangle_{{\cal S}_{\rm ext}} \equiv 0 
& \quad \mbox{for} \quad &
\bar k\le 0 ,
\\
c_{d} | 0\rangle_{{\cal S}_{\rm ext}}  \equiv 0 &
\quad \mbox{for} \quad & \varepsilon_d >0, 
\qquad \mbox{i.e.} \; \; n_d^{(0)} \equiv 0 , \\
c_{d}^\dagger | 0\rangle_{{\cal S}_{\rm ext}} \equiv 0 
& \quad \mbox{for} \quad &
\varepsilon_d \le 0 ,
\qquad \mbox{i.e.} \; \; n_d^{(0)} \equiv 1 .
\end{eqnarray}
\end{mathletters}
Here $\varepsilon_d$, whose value will be derived below [see
(\ref{Hquadratic})], is the energy associated with the $c_d$
pseudofermion, and $n_d^{(0)}$ denotes its occupation number in the
reference ground state $|0\rangle_{{\cal S}_{\rm ext}}$.
Using the symbol $:\; :$ to henceforth
denote normal ordering of the pseudofermions w.r.t.\ $|0\rangle_{{\cal
    S}_{\rm ext}}$, we have $: \! c^\dagger_d c_d \! : \, =
c_d^\dagger c_d - {n_d^{(0)}}$.  Furthermore, we define the number
operator for the $x$-pseudofermions by ${\hat {\bar N}}_{x} \equiv
\sum_{\bar k} : \! c^\dagger _{\bar k x} c_{\bar k x} \! :$.  Then
equations (\ref{newpsixgeneral}), (\ref{bark}) and
(\ref{eq:defnormorderbark}) together imply that
\begin{eqnarray}
{\hat {\bar N}}_{x} &= &  
  {\hat {\cal N}}_x -  P/ 2 
\label{xN_k}
\end{eqnarray}
holds as an operator identity. This 
can be seen intuitively by noting that 
$\psi_x \sim {\cal F}_x \sim c_{\bar k x}$
[by (\ref{newpsixgeneral})], hence 
the application
of $\psi_x$ (or $\psi_x^\dagger$) to a state decreases (or increases)
both ${\cal N}_x$ and $\bar N_x$ by one. 
These two numbers can thus
differ only by a constant, which must ensure  that $\bar N_x$ 
is  an integer.  Our definition of $| 0\rangle_{{\cal S}_{\rm ext}} $ 
effectively fixes this constant to be $P/2$,
by setting   $\bar N_x = 0$ for ${\cal N}_x = P/2$. 
(This can be verified  rigorously by checking\cite{a-to-0} that 
\mbox{$\lim_{x_0 \to 0} \lim_{a \to 0} \int_{-L/2}^{L/2} 
{dx \over 2 \pi}  [\psi_x^\dagger(x +x_0 )\psi_x(x) - {1 \over x_0}]$},
when evaluated 
using either (\ref{newpsix}) or (\ref{cbarkdef}), yields the
right- or left-hand sides of (\ref{xN_k}), respectively. 
Similarly, Eq.~(\ref{kinxferm}) below can be proven by evaluating 
\mbox{$\lim_{x_0 \to 0} \lim_{a \to 0}
\int_{-L/2}^{L/2} {dx \over 2 \pi} [\psi_x^\dagger(x +x_0 ) i \partial_x
\psi_x(x) - {1 \over x_0^2}]$}.)

We are now ready to refermionize the Hamiltonian $H'$. 
The kinetic energy of the $\bar k$ pseudofermions obeys 
the following operator identity:
\begin{equation}
\label{kinxferm}
\sum_{\bar k} \bar k : \! c^\dagger_{\bar k x}
c_{\bar k x} \! : \, 
= {\Delta_L \over 2}
        {\hat {\bar {N}}}_x ({\hat {\bar {N}}}_x + P) 
        + \sum_q q\, b_{qx}^\dagger b_{qx}\;.
\end{equation}
This follows by analogy with (\ref{hkinetic}) and 
(\ref{H0a}), with $\hat {\bar  N}_x $ and $1-P$ instead of 
$\hat N_{\alpha j}$ and  $P_0$.  Now note that 
${\hat {\bar {N}}}_x ({\hat {\bar {N}}}_x + P) 
= \hat {\cal N}_x^2  - P/4$, i.e.\ (\ref{kinxferm})
does {\em not}\/ contain a term linear in ${\hat {\cal N}}_x$.
Actually, the choice of the phase $e^{-i (\hat {\cal N}_x - 1/2)2 \pi x /L}$
in our refermionization Ansatz (\ref{newpsix})  for $\psi_x(x)$
was made specifically to achieve this.
Hence (\ref{kinxferm}) can be directly used to represent the
kinetic energy of the $x$-sector in Eq.~(\ref{2ck:H0y})
in terms of $c_{\bar k x}$ fermions:
\begin{mathletters}
\begin{eqnarray}
\label{barkH0a}
H_{x0} &=& \Delta_L  \hat {\cal N}_x^2 / 2 
        + \sum_{q>0} q \; b^\dagger_{qx} b_{qx} \\
 &=& \sum_{\bar k} \bar k  : \!  c^\dagger_{\bar k x} c_{\bar kx} \! :
        + \Delta_L P / 8 \;.
\label{barkH0}
\end{eqnarray}
\end{mathletters}
As a check, note that this equation also follows from the following
observations: firstly,  the equation of motion for the field $\psi_x(x)$,
expressed as (\ref{newpsix}) or (\ref{cbarkdef}), 
is the same when calculated using (\ref{barkH0a}) or
(\ref{barkH0}), respectively,
and therefore the latter two expressions
 can differ only by a constant; and secondly, this
 constant can be determined to be $\Delta_L P/8$, by requiring 
the free ground state energies for $|0\rangle_{{\cal S}_{\rm ext}}$ 
given by the two expressions to be the same.

Finally, in the subspace 
${\cal S}_{\rm phys}$ 
[of (\ref{2ck:S})] and hence also in ${\cal S}_{\rm ext}$,
we can use (\ref{2ck:tildeNS}) and (\ref{definec_d}) 
 to  express $\hat {\cal N}_s S_z$  and $\hat {\cal N}_s^2$
in terms of $c^\dagger_d c_d$. Thus, 
the EK-transformed 2CK Hamiltonian of Eqs.~(\ref{2ck:lambda=0}) and
(\ref{HEK}) takes the form 
\begin{eqnarray}
\label{fullH'}
H' &=& 
H_{csf} + H_x + E_G + const. \; , 
\\
H_{csf} & = & \sum_{ \scriptstyle c,s,f} \sum_{\scriptstyle\, q>0} q 
\;b^\dagger_{q y} b_{qy} \; ,
\label{H_csf}
\\
H_x &=& \varepsilon_d : \! c_d^\dagger c_d \! :
+ \sum_{\bar k} \bar k 
: \! c^\dagger_{\bar k  x} c_{\bar kx} \! :
\nonumber \\
&&+ 
    \sqrt{ \Delta_L  \Gamma } \sum_{\bar k} 
(c^\dagger_{{\bar k} x} \! + c_{{\bar k} x}) (c_d - c_d^\dagger) \; , 
\label{Hquadratic}
\\
\nonumber 
E_G &=& 
 \Delta_L \mbox{$\Bigl[ $} 
{\cal N}_c (1 - P_0)  +
  ( {\cal N}_c^2 +     
   {\cal N}_f^2 +  S_T^2 - 1/4)/2 
\\
&+& P/8         \mbox{$\Bigr] $}
   + \varepsilon_d ({n_d^{(0)}} - 1/2)  +  S_T h_e  \, . 
\label{EGfree}
\end{eqnarray}
The charge, spin and flavor degrees of freedom in
$H_{csf}$ evidently decouple completely. 
$H_x$ in  (\ref{Hquadratic})
has the form of a quadratic resonant level model
whose ``resonant level'' has energy $\varepsilon_d$
and width $\Gamma$, 
where  $\varepsilon_d \equiv  h_i - h_e$
is the energy cost   for  an impurity spin-flip,
and $\Gamma =  \lambda^2_\perp /4 a$, 
which will be identified below as 
the  Kondo temperature.

 $E_G$ is the ``free ground state energy'' of the subspace ${\cal S}_{\rm
    ext}$ in the presence of magnetic fields.  Its $S_T h_e $ term implies
  that the magnetic fields do {\em not}\/ enter only in the combination $h_i -
  h_e$ of $\varepsilon_d$, thus the role of the magnetic field $h_e$ applied
  to the conduction electrons is somewhat different from that of the local
  field $h_i$.  Note though, that for $h_e=2n\Delta_L$ (with $n\in {\mathbb Z}$) the
  $S_T h_e$ term can formally be absorbed (up to a total energy shift) by
  introducing a ``new total spin'' $S_T^\prime = S_T + 2n$, since then
$\Delta_L S^2_T/2 + S_T h_e =
\Delta_L S_T^{\prime 2} /2 
- 2 n^2 \Delta_L$.
Now, since the construction of the complete finite-size spectrum involves
enumerating all possible values of $S_T$, and since the generalized gluing
condition (\ref{GGC}) to be derived below is invariant under $S_T \to S_T +
2n$, the finite-size spectrum for $h_e=2n\Delta_L$ 
and a local field $h_i$ (so that $\varepsilon_d = h_i - 2 n \Delta_L$)
will be identical to that for $h_e = 0$ and a local field 
of $h_i - 2n\Delta_L$ (so that $\varepsilon_d$ is unchanged). The physical
origin of this ``periodicity'' is that as $h_e$ increases, at each value $2
n\Delta_L$ a ``level crossing'' occurs in which the free-electron ground state
changes from, say, $| {\cal N}_c, {\cal N}_s, {\cal N}_f, {\cal N}_x
\rangle_0$ to a new one differing from it {\em only}\/ in the spin quantum
number, namely $| {\cal N}_c, {\cal N}_s -2, {\cal N}_f, {\cal N}_x
\rangle_0$, by flipping the topmost spin-up electrons in both channels $j = 1$
and 2 to spin down.

For general values  $h_e \neq 2 n \Delta_L$, 
there is no such symmetry (essentially since electron-hole symmetry 
in the spin sector is lost), and the corresponding finite-size
spectrum differs from that at the periodicity points
in that some additional splitting of states occurs.\cite{eh} 
For simplicity we henceforth set $h_e = 0$ and
consider only a local magnetic field, with  $\varepsilon_d \equiv h_i$,
but the more general case $h_e \neq 0$ can be treated
completely analogously.

\section{Finite-Size Spectrum of 2CK Model}
\label{fsize}

EK studied the resonant level model $H_x$ of (\ref{Hquadratic})
in the continuum limit $L \to \infty$. They mainly
analyzed its {\em impurity}\/ properties,  showing that
these have NFL behavior because 
 ``half of the impurity'', namely  $c_d + c_d^\dagger$,
decouples.  By keeping $L$ finite, one
can extend their analysis to include
also the NFL behavior of electron properties.
In this section, we illustrate this by
diagonalizing $H_x$ at finite $L$ and
constructing its  finite-size spectrum in terms
of its exact eigenexcitations.

\subsection{Diagonalization of $H_x$}

Since $H_{csf}$ is trivial, we just have to diagonalize the resonant
level part $H_x$ in the extended subspace
${\cal S}_{\rm ext} (S_T, {\cal N}_c, {\cal N}_f)$, 
which is straightforward in principle, since $H_x$
is quadratic. However, care has to be exercised, in particular
regarding normal ordering: the change in ground state energy due to
the interaction turns out to be of order $- \Gamma $, and the
sub-leading (state-dependent) contributions of order $\Delta_L$
relative to this energy have to be extracted carefully when
constructing the finite-size spectrum.

\begin{figure}
\epsfxsize=8.5cm
\epsfbox{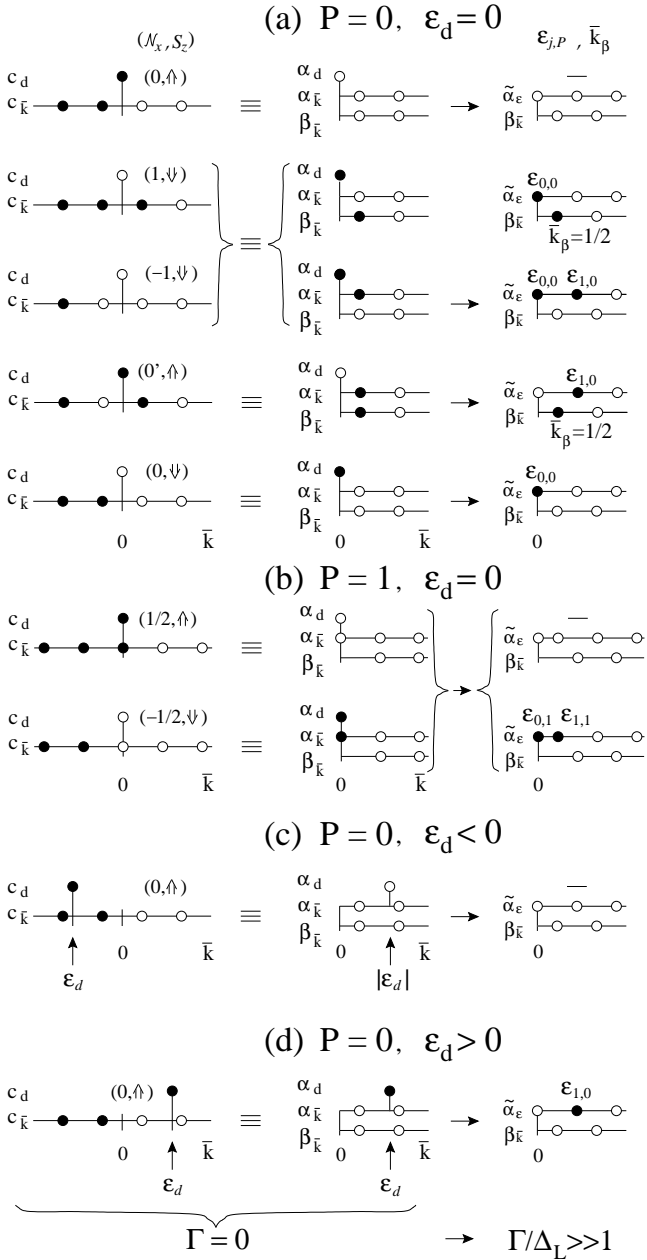}%{manystatesnew.eps}
\vskip0.7truecm
\caption[many-body-states]{\label{manystate} The left and middle columns
show, for various combinations 
of $({\cal N}_x, S_z)$,  some representative
free many-body states [eigenstates of $H'(\lambda_\perp = 0)$
of (\ref{2ck:lambda=0})] 
in a general physical subspace
${\cal S}_{\rm phys} (S_T, {\cal N}_c, {\cal N}_f)$, 
constructed in terms of both $c_n$'s and  $\alpha_n$'s,
thus illustrating
the transformation (\ref{ddeeffalphabeta}) between
these operators. The braces between these two columns
indicate that the states to their 
left and right are not in one-to-one correspondence,
but linear combinations of each other.
The right column 
shows some of the  exact physical many-body eigenstates $|\tilde E\rangle$ 
of the full $H'$ of (\ref{generalEstate})
at $\Gamma / \Delta_L \gg 1$.
Each $|\tilde E \rangle$ is labeled by the excitation
energies $\varepsilon_{j,P}$ and $\bar k_\beta$
of its occupied single-particle states. 
When $\Gamma$ is turned off to  0, each
state in the right column  reduces to the free state 
in the same row in the middle column,  unless they are separated by braces, in
which case it reduces to a linear combination
of the two degenerate free states grouped within  the braces to its left.
The first four rows in (a) and the two rows in (b)
correspond, in that order, 
to the first four $S_T = 1/2$ states 
and the two $S_T=0$ states listed in 
Table~\protect{\ref{2ckbigtable}}; the fifth  row in (a) 
is the spin-reversed  partner of the first row in (a),
illustrating how the two-fold degeneracy guaranteed by spin-reversal 
symmetry comes about due to the presence or absence
of a $\varepsilon_{0,P} = 0 $ excitation;
(c) and (d) illustrate the case $\varepsilon_d \neq 0$
relevant for non-zero magnetic fields. 
} 
\end{figure}

As first step, we define
 new fermionic excitations,
illustrated in  Fig.~\ref{manystate}, whose energies
are strictly non-negative, 
\begin{mathletters}
\label{ddeeffalphabeta}
\begin{eqnarray}
&&\left.
\begin{array}{l}
\alpha_{\bar k} \equiv 
\phantom{-i\,} (c_{\bar k x} + c^\dagger_{- \bar k 
x}) / \sqrt 2\\
\beta_{\bar k} \equiv -i\,  (c_{\bar k x} - c^\dagger_{- \bar k 
x}) / \sqrt 2 
\end{array}
\right\} \mbox{  \mbox{\phantom{nnn} for} $\bar k >0$,} 
\label{defalphabeta} \\
&&\alpha_0 \equiv c^\dagger_{0x} \mbox{\phantom{nnn} for 
$\bar k = 0$ if $P=1$,} 
\label{ehtransf} \\
&&\alpha_d \equiv 
\left\{
\begin{array}{lll}
c_d         & \mbox{\phantom{nnn} for } & \varepsilon_d > 0\; \\
c^\dagger_d & \mbox{\phantom{nnn} for } & \varepsilon_d \le 0\;
\end{array}
\right. \; , 
\label{defalphad}
\end{eqnarray}
\end{mathletters}
and which have the virtue that the $\beta_{\bar k}$ excitations
decouple completely from the impurity:\hfill\\
\begin{eqnarray}
H_x &=& \sum_{\bar k \ge 0}\bar k \alpha_{\bar k}^\dagger
\alpha_{\bar k} +\sum_{\bar k > 0}\bar k \beta_{\bar k}^\dagger
\beta_{\bar k} + |\varepsilon_d| \alpha_d^\dagger \alpha_d 
\nonumber \\
&+& \sum_{\bar k\ge 0}V_{\bar k} \bigl(\alpha^\dagger_{\bar k} +
\alpha_{\bar k}\bigr)\bigl(\alpha_d -\alpha_d^\dagger\bigr)\;.
\label{Hxeh}
\end{eqnarray}
Here the possible $\bar k$-values are 
given by Eq.~(\ref{bark}),  and 
the hybridization amplitudes $V_{\bar k}$  by
\begin{equation}
V_0 \equiv {V_{\bar k\ne 0} / \sqrt{2}} \equiv 
e^{i\pi\, {n_d^{(0)}}} \sqrt{\Gamma \Delta_L }\; .
\label{V_k}
\end{equation}
Note that in (\ref{ddeeffalphabeta}) we purposefully defined 
$\alpha_n^\dagger$ and $\beta_{\bar k}^\dagger $ such that
the free
reference ground state $|0\rangle_{{\cal S}_{\rm ext}}$,  by
(\ref{eq:defnormorderbark}), contains {\em no}\/ 
$\alpha_n^\dagger$ or $\beta_{\bar k}^\dagger $ excitations, 
i.e.\ 
\begin{equation}
\label{referenceS}
\alpha_d|0\rangle_{{\cal S}_{\rm ext}} = \alpha_{\bar k}
|0\rangle_{{\cal S}_{\rm ext}}
= \beta_{\bar k}|0\rangle_{{\cal S}_{\rm ext}} =  0\; , 
\label{|0>_S}
\end{equation}
as illustrated in the middle entries of
the first rows of Figs.~\ref{manystate}(a), 
\ref{manystate}(b) and \ref{manystate}(c). 
Note too that $\alpha_d^\dagger |0\rangle_{{\cal S}_{\rm ext}} $
is degenerate with  $|0\rangle_{{\cal S}_{\rm ext}} $  if 
$\varepsilon_d = 0$, as is 
$\alpha_0^\dagger |0\rangle_{{\cal S}_{\rm ext}} $ in the odd electron
sector, $P=1$.

Since the Hamiltonian Eq.~(\ref {Hxeh}) is quadratic, it can
be brought into the diagonal form 
\begin{equation}
H_x = \sum_{\bar k>0} \bar k \; \beta^\dagger_{\bar
k}\beta_{\bar k}  + \sum_{\varepsilon \ge 0} \varepsilon \;
\tilde \alpha^\dagger_\varepsilon \tilde \alpha_\varepsilon
+ \delta E_G \; ,
\label{Hdiagfer}
\end{equation}
describing a ground state energy shift $\delta E_G$ 
and non-negative-energy excitations relative to 
a reference state 
$| \tilde 0 \rangle_{{\cal S}_{\rm ext}}$,
an exact ground state of $H'$ 
 in  ${{\cal S}_{\rm ext}}$, defined by
\begin{equation}
\label{NFLreferenceS} 
\tilde \alpha_{\varepsilon} | \tilde 0 \rangle_{{\cal S}_{\rm ext}}
=  \beta_{\bar k} | \tilde 0 \rangle_{{\cal S}_{\rm ext}}  \equiv  0 \; .
\end{equation}
This diagonalization can be 
accomplished by a Bogoliubov transformation of the form
\begin{equation}
\tilde \alpha^\dagger_{\varepsilon} 
   =    \sum_{ n\in\{\bar k, d\}}  \sum_{\nu = \pm} 
     B_{  \varepsilon n \nu }
   (\alpha_n^\dagger + \nu \alpha_n) /2  \; ,
\label{2ck:bogolubov}
\end{equation}
where the new operators $\tilde \alpha_\varepsilon$ are
required to satisfy:
\begin{eqnarray}
\label{Heisenberg}
&&[H_x,\tilde \alpha^\dagger_{\varepsilon} ]= \varepsilon \;
\tilde \alpha^\dagger_{\varepsilon} \;,\\
\label{tildealphaunit}
&&\{\tilde \alpha^\dagger_{\varepsilon} , \tilde
\alpha_{\varepsilon'} \} = \delta_{\varepsilon\varepsilon'}\;.
\end{eqnarray}
Eqs.~(\ref{Heisenberg}) and (\ref{tildealphaunit})
yield  a  closed system of equations for the  
eigenenergies $\varepsilon$ and 
the coefficients $B_{\varepsilon n \nu}$.
In Appendix~\ref{app:2ck}, we solve them explicitly
(by transforming to conveniently chosen Majorana fermions),
with the following results.

The excitation energies $\varepsilon$  are the non-negative 
roots of the transcendental equation
\begin{equation}
{ \varepsilon  4 \pi \Gamma \over
   \varepsilon^2 - \varepsilon_d^2}  = 
   -  \cot \pi \left( {\varepsilon / \Delta_L } - P/ 2  \right)
     \; ,
\label{2ck:eigenenergies}
\end{equation}
and the ground state energy shift is 
\begin{equation}
\delta E_G = \;  
{|\varepsilon_d| \over 2} +  \sum_{\bar k \ge 0} {\bar k \over 2}  
\: - \: \sum_{\varepsilon \ge 0} {\varepsilon \over 2}  \; .
\label{2ck:EG}
\end{equation}
For $\varepsilon > 0$, the coefficients
$B_{\varepsilon n \nu}$ are given by 
\begin{mathletters}
\label{allBeqs}
\begin{eqnarray}
B_{ \varepsilon d +}  = \varrho(\varepsilon) \, |\varepsilon_d| \; , 
&\qquad&
 B_{ \varepsilon d -}  =\varrho(\varepsilon) \, \varepsilon \; , 
\label{B_ed-} 
\\ 
B_{ \varepsilon \bar k +}  = \varrho(\varepsilon) \, 
{2 V_{\bar k} \, \varepsilon^2 \over \varepsilon^2- {\bar k}^2} \; , 
\label{B_ek+}
& \qquad &
B_{ \varepsilon \bar k -}  = \varrho(\varepsilon) \, 
{2 V_{\bar k} \, \varepsilon \, {\bar k} \over \varepsilon^2- {\bar k}^2} \; ,
\end{eqnarray}
\end{mathletters}
where the normalization factor $\varrho(\varepsilon)$ is 
\begin{equation}
\varrho(\varepsilon)= \left[ { 2 \Delta_L \Gamma} \over {
    {1 \over 4} ( \varepsilon^2 \! - \varepsilon_d^2)^2 \! 
   +  \!  \Delta_L \Gamma (  \varepsilon^2 \! + \varepsilon_d^2)
  + \! 4 \pi^2   \Gamma^2   \varepsilon^2 }\right]^{1/2}\;.
\label{varrho}
\end{equation}
For $\varepsilon = 0$, 
the coefficients $B_{0 n \nu}$ must be considered separately and
are given in Appendix~\ref{2ck:zeromodes}.

Eqs.~(\ref{H_csf}), (\ref{EGfree}), (\ref{Hdiagfer}) and
(\ref{2ck:eigenenergies}) to (\ref{varrho}), together with the gluing
conditions (\ref{GGC}) discussed in the next subsection, constitute a
complete, analytic solution of the 2CK model along the EK line.

\subsection{Evolution of Excitation Energies}
\label{sec:evolution}

\begin{figure}
\epsfxsize=8.5cm
\epsfbox{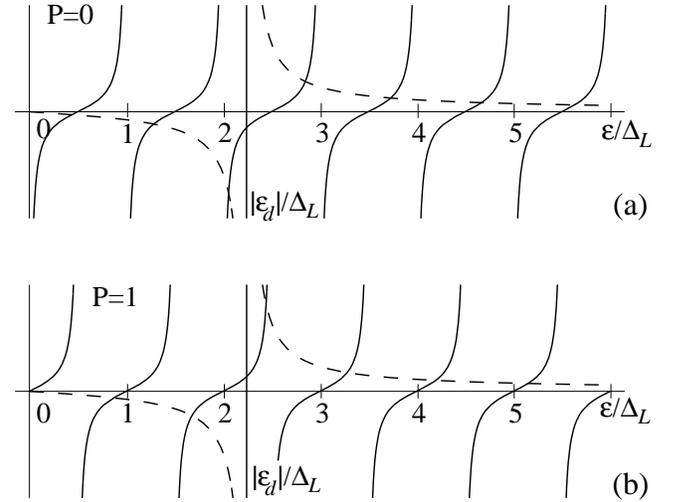}%{graphical-final.eps}
\vskip0.7truecm
\caption[Graphical-solution]{ \label{fig:graphical}
Graphical solution of
the eigenvalue equation (\protect\ref{2ck:eigenenergies})
for (a)  $P=0$  and (b) $P=1$. The vertical solid line
marks the position of $|\varepsilon_d|$.
Dashed and solid lines
represent the left- and right-hand sides of (\protect\ref{2ck:eigenenergies});
their intersections give the allowed eigenvalues $\varepsilon$. 
The ``amplitude'' of the  dashed lines is proportional to 
$\Gamma/\Delta_L$; if this increases from 0, the $\varepsilon$'s
thus shift away from their free values 
$|\varepsilon_d|$ or $\bar k = \Delta_L [n_{\bar k} -
(1-P)/2]$.} 
\end{figure}

The eigenvalue equation (\ref{2ck:eigenenergies})
is a central ingredient of our analytical solution,
since it yields the exact excitation energies $\varepsilon$ of $H_x$, 
and also allows one to explicitly identify the various crossover scales
of the problem. 
Let the label $j = 0, 1, 2, \dots$ enumerate,  in increasing order,
the solutions $\varepsilon_{j,P}$ of  (\ref{2ck:eigenenergies})
in a sector with parity $P$. 
Their smooth evolution as functions of $\Gamma$ and 
$|\varepsilon_d|$
can readily be understood by a graphical analysis of
Eq.~(\ref{2ck:eigenenergies}), see Fig.~\ref{fig:graphical}, and is shown in
Figs.~\ref{onepscaling}(a) and \ref{onepscaling}(b) for  $P=0$ and
1, respectively. All but the lowest-lying $(j=0$) solutions
can be parameterized as
\begin{mathletters}
\label{2ck-ejp}
\begin{eqnarray}
  \label{2ck:epsappr}
  \varepsilon_{j,P} &=& \Delta_L \left[ j - 
{\textstyle {1\over 2}  - {P \over 2}}  + \delta_{j,P} \right],
\qquad j = 1, 2, 3, \dots ,
\end{eqnarray}
where $\delta_{j,P} \in [0,1]$ is the shift of $\varepsilon_{j,P}/\Delta_L$
from its $\Gamma = \varepsilon_d = 0$ value and is determined selfconsistently
 [from (\ref{2ck:eigenenergies})] by
\begin{eqnarray}
\label{delta-shift}
 \delta_{j,P} &=&  {1 \over 2} + {1 \over \pi}
\arctan {1 \over 4 \pi} \left[ {T_h \over \varepsilon_{j,P} }
 - {\varepsilon_{j,P} \over \Gamma } \right] \, ,
\end{eqnarray}
\end{mathletters}
with $T_h \equiv \varepsilon_d^2 / \Gamma$. The lowest-lying modes
are given by
\begin{mathletters}
\label{eq:zeromodes}
\begin{eqnarray}
  {\varepsilon_{0,0} \over \Delta_L} &=& \left\{
      \begin{array}{ll}
        0 & \quad \mbox{for} \; \varepsilon_d = 0 , \\
        (-1/2 + \delta_{0,0} ) \in (0,1/2\mbox{]} 
        & \quad \mbox{for} \; \varepsilon_d \neq 0 ,
        \end{array} \right.
\\
  \varepsilon_{0,1} &=& 0 \qquad \qquad \mbox{for all} \;
 \Gamma, \varepsilon_d \; 
\end{eqnarray}
\end{mathletters}
(see also Appendix~\ref{2ck:zeromodes}). 

Eq.~(\ref{delta-shift}) shows very nicely that 
$\Gamma$ and $T_h$ are cross-over scales:
Firstly,  in the absence of magnetic fields, 
i.e.\ for $|\varepsilon_d| =  |h_i| = T_h =  0$,
the spectral regime {\em below 
 $\Gamma$ is strongly perturbed}\/ [$\delta_{j,P} \simeq 1/2$
for $\varepsilon_{j,P} \ll \Gamma$], whereas 
{\em above $\Gamma$
it is only weakly perturbed}\/
[$\delta_{j,P} \simeq 0$
for $\varepsilon_{j,P} \gg \Gamma$].
This also follows directly from a graphical analysis of 
the eigenvalue equation (\ref{2ck:eigenenergies}).
 It is thus natural to identify the crossover scale $\Gamma$ with the
Kondo temperature, $T_K \simeq \Gamma$.

Secondly, in the presence of a local magnetic field,
$T_h = {h_i^2}/\Gamma > 0$ furnishes 
another crossover scale. When 
considering the $T_h$-induced
shifts in $\delta_{j,P}$ relative to
their values for $T_h=0$,
 several cases can be distinguished
[by direct inspection of (\ref{delta-shift})]:

(i) For  $T_h \ll \Delta_L$, 
i.e.\ for $|{h_i}|$ 
much smaller than a crossover field  
$h_c \sim \sqrt{\Gamma \Delta_L}$,
none of the $T_h$-induced shifts are strong.

(ii) For $T_h \gg \Delta_L, \Gamma$, the crossover scale $T_h$ divides the
spectrum into two parts: the $T_h$-induced shifts are weak for all levels
with $\varepsilon \gg T_h$, but strong for all those
with $\varepsilon \ll T_h$.

(iii) For $\Gamma \gg T_h \gg \Delta_L$ [a special case of (ii)]
 one can distinguish  three
physically different regimes: the spectrum is
NFL-like (non-uniform level spacings) in the intermediate regime
 $T_h \ll \varepsilon \ll
 \Gamma$, and  Fermi-liquid like
(with uniform level spacing) in the extreme
regimes $\varepsilon \gg \Gamma$
and  $\varepsilon \ll T_h$. 
In the last of these regimes (rightmost part of Fig.~\ref{onepscaling}),
the set of lowest-lying $\varepsilon$'s  is in fact identical
to that for  the {\em free}\/  case $T_h = 0$, $\Gamma =0$
(leftmost part of Fig.~\ref{onepscaling}),  except
that the free case has one more $\varepsilon = 0$ mode,
associated with the impurity level's two-fold degeneracy 
due to spin reversal symmetry
for  $|{h_i}| = 0$  
(which is broken if $T_h \gg \Delta_L$, implying $|{h_i}| \gg \Delta_L$).
 This shows quite beautifully that 
a magnetic field very literally suppresses
the effect of spin-flip scattering for the low-energy
part of the spectrum;  heuristically, this happens,
of course, since  low-energy electrons do not have enough energy
to overcome the Zeeman energy necessary for
a spin flip in a magnetic field. 

\begin{figure}
\epsfxsize=8.5cm
\epsfbox{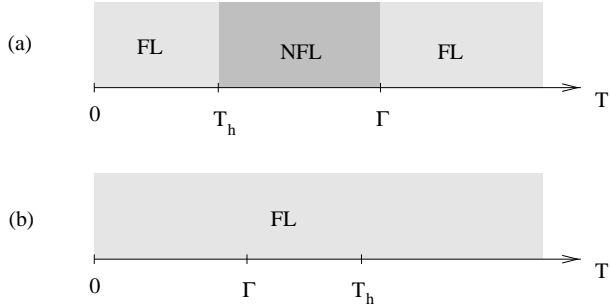}%{nflregions.eps}
\vskip0.7truecm 
\caption[NFL-regions]{
\label{fig:regions}  
Sketch  of the different Fermi liquid and non-Fermi liquid regions for
a finite magnetic field on the EK line. }
\end{figure}

Since at a finite temperature physical quantities are governed mostly by
excitations of energy $\varepsilon \sim T$, they will show NFL behavior for
$\Gamma \gg T \gg T_h$ and Fermi liquid behavior for $T\gg \Gamma$ or $T \ll
T_h$ \cite{AD84,Sacr,EK92,SG94}, as sketched in Fig~\ref{fig:regions}.

\subsection{Generalized Gluing Conditions}
\label{sec:generalEstate}

A general eigenstate of $H_x$ in ${\cal S}_{\rm ext}$ has the form
\begin{equation}
\label{generalEstate}
|\tilde E \rangle \propto 
 \prod_{i=1}^{{\cal N}_{\tilde \alpha}}
 \tilde \alpha^\dagger_{\tilde \varepsilon_i} 
 \prod_{j=1}^{{\cal N}_{\tilde \beta}}
  \beta^\dagger_{\bar k_j} | \tilde 0 \rangle_{{\cal S}_{\rm ext}} \; . 
\end{equation}
However,  as emphasized earlier, of all such 
states only those in the physical
subspace ${{\cal S}_{\rm phys}}$ must be retained,
and all others discarded as being unphysical.
(Recall that we had
to extend ${{\cal S}_{\rm phys}}$ to ${{\cal S}_{\rm ext}}$
to  define the operators $\psi_x$ and
$c_d$.) To identify which $|\tilde E\rangle$
are physical, we shall now derive a {\em generalized gluing
condition}\/ satisfied by them, by
noting that $|\tilde E \rangle$ can be physical
only if 
the state $|E \rangle \equiv \lim_{\Gamma \to 0} | \tilde E \rangle$,
to which it  reduces when $\Gamma$
is adiabatically switched off,  satisfies the
free gluing conditions (\ref{gluingall}). Key to the derivation 
is the fact that although 
 the hybridization interaction
$H'_\perp$ of (\ref{Halfpsicouples}) does not
conserve the number of $\alpha^\dagger_n$ excitations, 
it {\em does conserve the parity}\/  of their number. 

To be explicit, let ${\cal P}_{\tilde E}$ be the 
the parity of the number 
of excitations of $|\tilde E \rangle $ relative to
$| \tilde 0 \rangle_{{\cal S}_{\rm ext}}$:
\begin{equation}
 {\cal P}_{\tilde E}
 \equiv \langle \tilde E | \Bigl[
\sum_{\varepsilon \ge 0} \tilde  \alpha^\dagger_{\varepsilon} \tilde 
\alpha_{\varepsilon} +
\sum_{\bar k > 0 } \beta^\dagger_{\bar k}\beta_{\bar k}
\Bigr] \, \mbox{mod} \, 2 | \tilde E \rangle  \; .
\label{Ahat}
\end{equation}
During the adiabatic switch-off of $\Gamma$, this quantity
of course remains {\em fixed}\/, and hence equals 
${\cal P}_{\tilde E} (\Gamma \to 0)$. This in turn can be
written as 
\begin{eqnarray}
\label{GGC1} \nonumber 
\lefteqn{
\phantom{.} \hspace{-5mm}
{\cal P}_{\tilde E} (\Gamma \to 0) =
\langle E | \Bigl[ \sum_{n = d, \bar k \ge 0}
 \alpha_n^\dagger \alpha_n + \sum_{\bar k > 0 }
\beta^\dagger_{\bar k} \beta_{\bar k} \Bigr] \mbox{mod} \, 2
| E \rangle }
\\
\nonumber
&=& 
\langle E | \left[ \hat {\bar { N}}_x + 
 \alpha_d^\dagger \alpha_d \right] \mbox{mod} \, 2
| E \rangle 
\\
\nonumber 
&= &
\langle E | \! \left[ ( \hat {{\cal N}}_x \!-\! {\textstyle { P \over 2}})  +
\hat {\cal N}_s - S_T - {\textstyle {1 \over 2}} + 
{n_d^{(0)}} \right] 
\mbox{mod} \,  2 | E \rangle \; . 
\end{eqnarray}
The first equation follows because the
hybridization interaction preserves the parity
of the excitation numbers
(in other words, since $\tilde \alpha_\varepsilon^\dagger$
is a linear combination of $\alpha_n^\dagger$ and $\alpha_n$);
the second follows because the $c^\dagger_{\bar k x}$ excitations
counted by $\hat {\bar {\cal N}_x}$ are linear
combinations of $\alpha_{\bar k}$, $\alpha^\dagger_{\bar k}$,
$\beta_{\bar k}$ and $\beta^\dagger_{\bar k}$
[this is illustrated by Fig.~\ref{manystate},
which depicts  how states in the $ c_{\bar k x}$
and  $\alpha_{\bar k}, \beta_{\bar k}$ representations 
are related to each other]; 
and the third follows from 
(\ref{xN_k}) for $\hat {\bar {\cal N}_x}$
and (\ref{defalphad}), (\ref{definec_d}) and (\ref{2ck:tildeNS}) for
$\alpha_d$. Imposing now the condition that
$|E \rangle$ must be in the physical subspace ${\cal S}_{\rm phys}$
and hence must satisfy  the free gluing condition (\ref{2ck:GC}), we  
obtain 
\begin{equation}
 {\cal P}_{\tilde E} = 
        \left\{\begin{array}{ll}
          \mbox{[} {\cal N}_c + {\cal N}_f -  S_T -{\textstyle {(P+1)/2}}
        \mbox{]  mod 2} 
& \quad (\varepsilon_d>0)\;, 
\\
          \mbox{[} {\cal N}_c + {\cal N}_f - S_T - (P-1) /  2 \mbox{] mod  2} 
& \quad (\varepsilon_d\le 0)\; .
     \end{array}\right. 
\label{GGC}
\end{equation}
This {\em generalized gluing condition}\/
specifies which of all the possible 
states in ${\cal S}_{\rm ext}$ are physical, i.e.\ are
in ${\cal S}_{\rm phys}$;
it supplements the free gluing condition (\ref{2ck:Nyint}),
which stipulates that $S_T \pm 1/2 $ must
be integer (half-integer) if ${\cal N}_c$ and ${\cal N}_f$ are
 integer (half-integer). The unphysical states
in ${\cal S}_{\rm ext}$ that do not satisfy (\ref{GGC}) must be discarded
when constructing the finite-size spectrum (and
thus no such states occur in Table~\ref{2ckbigtable} below).

\subsection{Ground State Energy Shift}
\label{groundstateshift}

The form of Eq.~(\ref{2ck:EG}) for 
the change in ground state energy $\delta E_G$ 
suggests that it can be interpreted 
as {\em the dynamical binding energy of the impurity spin,}\/
which results from the impurity-induced energy shifts 
of all the  states in the filled Fermi sea. (The factor
$1/2$ in (\ref {2ck:EG})  reflects 
the fact\cite{EK92,EKrev} 
 that only ``half'' of the $x$-pseudofermion field,
namely  $\psi_x + \psi_x^\dag$,
couples to the impurity [in (\ref{Halfpsicouples})].)
For $\varepsilon_d = 0$, the number of
levels strongly shifted by the interaction
is [by (\ref{delta-shift})]
of order $\Gamma / \Delta_L$, and each of these gets shifted
roughly by $\Delta_L/2$;  we can  thus estimate that $|\delta E_G|$ will
be of order $\Gamma$, which supports 
the heuristic statement that ``the impurity's binding energy
is of order $T_K$''. 

However, since the level shifts 
$ \Delta_L \, \delta_{j,P}$  also have a $P$-dependence
of order $\sim \Delta^2_L/ \Gamma$ [from (\ref{delta-shift})],
the total ground state energy shift $\delta E_G$
will have a $P$-dependence too, of order $ \sim\Delta_L$.
We therefore  write 
\begin{equation}
  \label{eq:deltaEg}
\delta E_G \equiv
 \delta E_G^0 +  P \,  \delta {\cal E}_G^P , 
\end{equation}
 where
the first term is   $P$-independent
and hence gives only an overall energy shift.
In contrast,   $\delta E_G^P$ affects the finite-size spectrum
since it shifts the odd electron states, $P=1$, relative to 
even electron states, $P=0$, 
and hence  must be evaluated with particular care.
This is done in Appendix~\ref{2ck:calcEG},
where we find, for $\Gamma/ \Delta_L \gg 1$, 
\begin{eqnarray}
\delta E_G^P  &=& 
\left\{ \begin{array}{cl}
 - \Delta_L  / 8 &
\quad (T_h = 0) \vspace{1mm} , 
\\
0 &
\quad (T_h \gg  \Delta_L ) ,
\end{array} \right. 
\label{sums}
\label{sumsPP}
 \label{eq:deltaEG-P-h}
\\
\label{EG0h=0}
\delta E_G^0 & \approx & 
\left\{ \begin{array}{cl}
-2 \Gamma \bigl[ \ln(D/4 \pi \Gamma) + 1  
 \bigr] \; , 
& \quad (T_h = 0) \vspace{1mm} ,
\\
- 2 \Gamma \bigl[ \ln (D/|\varepsilon_d|) + 1 
\bigr] \; , 
& \quad (T_h \gg \Delta_L,  \Gamma ) .
\end{array} \right. 
\end{eqnarray} 
Here $D \gg \Gamma, T_h$ is a cutoff needed to regularize the sums in
(\ref{2ck:EG}).  Note that for $T_h = 0$, (\ref{EG0h=0}) is consistent with
the estimate for $\delta E_G$ above if we take $D \simeq 1/a$ and recall that
$\Gamma =\lambda^2_\perp/4a$.  For $T_h \gg \Gamma$, the magnetic field
$|\varepsilon_d|$ takes over as lower energy scale in the logarithm instead of
$\Gamma$.

\subsection{Construction of the Finite-Size Spectrum}
\label{sec:ffs}

Now we are  finally ready  to construct the
finite-size many-body excitation spectrum of the 2CK model.
In doing so, 
we shall generally use calligraphic ${\cal E}$'s 
to denote  dimensionless energies measured in units of $\Delta_L$.
Specifically, we shall construct the 
dimensionless energies
\begin{equation}
  \label{eq:fssdef}
\tilde {\cal E} (L)  = [\tilde E (L) - \tilde E_{\rm min}(L) ]/ \Delta_L ,
\end{equation}
associated with the lowest few
exact many-body eigenstates $| \tilde E \rangle$ of
the full Hamiltonian $H'$ of (\ref{fullH'}),
measured relative to its ground state energy, $\tilde E_{\rm min}$. 
For the sake of
simplicity we only consider the case with periodicity index
$P_0=1$ [see (\ref{k_free})],
for which the $\psi_{\alpha j}$'s have 
anti-periodic boundary conditions. In this
case the free ground state in the electronic sector
is unique, namely $|\vec 0 \rangle_0$,
 which somewhat simplifies the counting of states. (Of
course, one can use the same procedure for $P_0=0$, with similar
results.)

The construction proceeds in three steps: we 
first evolve 
toward the EK line, second evolve along the EK line,
and third turn on a local magnetic field. 
The results are summarized in Fig.~\ref{2ckscaling}
and Tables~\ref{2ckbigtable} and \ref{concise}. 
The caption of  Table~\ref{2ckbigtable}
also summarizes the technical details of the construction.
Here we just state the main ideas:

(i) {\em Phase-shifted Spectrum:---}\/ First we 
study the evolution
of the spectrum {\em toward}\/ the EK line for
$\lambda_z \in [0, 1]$ at $\lambda_\perp = \varepsilon_d = 0$.
Since the impurity has no dynamics for $\lambda_\perp = 0$,
the spectrum is that of a {\em free-electron Fermi liquid
with a $S_z$-dependent phase shift in the spin sector,}\/
given by $H'(\lambda_\perp = 0)$ of (\ref{2ck:lambda=0});
it evolves linearly with increasing $\lambda_z$,
from ${\cal E}_{\rm free}$ at $\lambda_z = 0$ to 
${\cal E}_{\rm phase}$ at $\lambda_z = 1$, see 
Fig.~\ref{2ckscaling}(a).

(ii) {\em Crossover Spectrum:---}\/
Next we study the spectrum's  further evolution
{\em along} the EK line 
for $\Gamma/ \Delta_L \in [0, \infty)$
at $\lambda_z = 1$, $\varepsilon_d = 0$.
To this end we enumerate in Table~\ref{2ckbigtable}
the lowest-lying  physical eigenstates
$|\tilde E \rangle$ of the full Hamiltonian $H'$
in terms of the excitations $\tilde \alpha^\dagger_{\varepsilon_{j,P}}$,
$\beta_{\bar k}^\dagger$ and $b_{qy}^\dagger$
which diagonalize it,
and {\em follow the evolution with increasing $\Gamma/ \Delta_L$
of the excitation energies}\/ 
$\varepsilon_{j,P}$ (shown in Fig.~\ref{onepscaling}),
and of the ground state energy shift $\delta E_G^P$ [see (\ref{sums})]. 
This  yields the crossover shown in Fig.~\protect\ref{2ckscaling}(b) 
from the phase-shifted
to the NFL fixed point spectrum, consisting 
of a set of universal, dimensionless energies defined by 
\begin{equation}
{\cal E}_{\rm NFL}
  \equiv  \lim_{L\to\infty} 
{\tilde E(L; \varepsilon_d = 0, \Gamma) -
\tilde E_{\rm min} (L; \varepsilon_d = 0, \Gamma)
\over \Delta_{L}} \;.
\label{Euniversal}
\end{equation}
Satisfyingly, the spectrum of ${\cal E}_{\rm NFL}$ energies found in
Fig.~\ref{2ckscaling}(b) and Table~\ref{concise}
(degeneracies are given in brackets)
coincides with the ones obtained
in NRG and CFT calculations.\cite{CLN80,Aff90,AL91a,AL91b,AL92b} 
This constitutes a direct and
straightforward analytical proof of the soundness of the latter
approaches. In particular, it  proves the so-called fusion hypothesis
employed by Affleck and Ludwig in their CFT calculation of this
spectrum.\cite{Aff90,AL91a,AL91b}
As is well-known from
CFT,\cite{Cardy} each of the fixed-point values ${\cal E}_{\rm NFL}$ 
can be associated with the scaling dimension of one of the 
operators characterizing the fixed point. 
 The occurrence of ${\cal E}_{\rm NFL}$'s that 
are not simply integers or half-integers is thus 
a very direct sign of NFL physics, since these
correspond to non-fermionic operators. 

Our NFL spectrum  demonstrates explicitly
the well-known fact that the spin anisotropy is
irrelevant at the NFL fixed point, since if we take the
continuum limit $\Delta_L \to 0$ at fixed $\Gamma$, the fixed point
spectrum is evidently reached {\em independently}\/ of the specific value of
$\Gamma$.  Said more formally: the symmetry of our anisotropic
starting Hamiltonian with respect to transformations in the charge,
spin and flavor sectors is $U(1)_c \times U(1)_s \times SU(2)_f $,
i.e.\ in the spin sector it is only invariant under spin rotations
around the $z$ axis; in contrast, Affleck and Ludwig derived the NFL
fixed point spectrum by assuming it to have the complete $U(1)_c \times
SU(2)_s\times SU(2)_f $ symmetry of the free model.  The fact that the
low-energy part ($\varepsilon \ll T_K$) of our NFL fixed point spectrum
coincides with theirs beautifully illustrates how the broken
symmetry of the original model is restored in the vicinity of the NFL
fixed point, and thus proves another central assumption of the CFT
solution of the 2CK model, in agreement with the NRG study of Pang and
Cox.\cite{PC91}

The fact that the exact eigenenergies of $H'$ 
 interpolate smoothly
between their values for $\lambda_\perp=0$ and
$\lambda_\perp\not=0$ [Fig.~\ref{2ckscaling}(b)]
 may  at first seem somewhat surprising, 
because a common way of heuristically  characterizing
a NFL is that its quasiparticles are orthogonal 
to the bare ones of the corresponding free Fermi liquid.
This is referred to as the ``breakdown 
of Landau's quasiparticle construction'',
since in Landau's
picture of a Fermi liquid, the  dressed quasiparticles and the
corresponding bare ones have finite overlap. 
Here, in fact, one can readily check
that ${}_{{\cal S}_{\rm ext}} \langle \tilde 0 | \alpha_{\bar k} \tilde 
\alpha^\dagger_{\varepsilon (\bar k)} | \tilde 0 \rangle_{{\cal S}_{\rm ext}}$
is non-zero (where $\varepsilon (\bar k)$ is the excitation
energy that reduces to $\bar k$ as $\Gamma/\Delta_L \to 0$), 
implying that in the $\alpha$-basis
the system  {\em is}\/ a Fermi liquid
(in accord with the fact that $H_x$ is quadratic, i.e.\ 
``simple''). 
However, this does not contradict the
fact that in the {\em original $c_{k \alpha j}$
basis}\/ the system nevertheless  behaves like a NFL,
since the bosonization-refermionization relation between states
in the $\alpha$   and 
 $c_{k \alpha j}$ bases is very highly non-linear.

 (iii) {\em Crossover due to local magnetic field:---} 
Finally, we turn on
 a local magnetic field, $ \varepsilon_d = {h_i} \neq 0$ 
 at fixed $\lambda_z = 1$
 and $\Gamma / \Delta_L \gg 1$, thus breaking spin reversal symmetry.
 {\em The further evolution of the excitation energies
   $\varepsilon_{j,P}$ as functions of increasing $T_h/ \Delta_L$,}\/
 shown in Fig.~\ref{onepscaling}(b), {\em yields the
   magnetic-field-induced crossover, }\/ shown in
 Fig.~\ref{fig:scaling}(c), from the NFL fixed point energies ${\cal
   E}_{\rm NFL}$ to a set of energies ${\cal E}_{\rm ph}$ corresponding
 to a phase-shifted Fermi liquid fixed point.  For $T_h  / \Delta_L
 \gg 1$, the impurity level evidently becomes empty for all  low-lying
 states, $\langle c_d^\dagger c_d \rangle =0$, i.e. the impurity spin
 is frozen in the state $S_z=\; \Downarrow$. Indeed, the 
 spectrum ${\cal E}_{\rm ph}$ which one recovers is
 precisely the same phase-shifted spectrum as $ {\cal E}_{\rm phase}$
 at the point $\lambda_z=1$ and $\lambda_\perp =0$, apart from a
 degeneracy factor of two, due to the lack of spin reversal symmetry,
 compare Table~\ref{concise}. This shows very nicely that the magnetic
 field ``erases'' all traces of NFL physics for the lowest-lying part
 of the spectrum, because it effectively switches off spin-flip
 scattering.

\subsection{Finite-Size Behavior of Physical Quantities}
\label{physicalffs}

Let us now briefly discuss the finite-size, $T=0$ 
behavior of three physical quantities at the NFL fixed point, 
namely the entropy, susceptibility and 
the fluctuations in ${\cal N}_x$. 

The {\em entropy}\/ of the  ground state at 
$T=0$, $\varepsilon_d = 0$ is evidently simply
ln2 for any $L$, since the ground state is two-fold degenerate
(see Fig.~\ref{fig:scaling}).
This should be contrasted\cite{Rozhkov} with the celebrated
result ${1 \over 2} \ln 2$ that one obtains\cite{AL91d}
when taking the limit $L\to \infty$ before $T \to 0$. 
The difference simply illustrates the triviality that the order
of limits does not commute, since for finite $L$
the system is always gapped. 

The {\em susceptibility}\/
 at $T=0$ due to a local field ${h_i} $ 
is defined by  $\chi = - \frac{\partial^2 \tilde 
E_G} {\partial {h_i^2}}$. Since $\tilde E_G = E_G + \delta E_G$,
we simply have to evaluate 
[by (\ref{EGfree}), (\ref{2ck:EG})] the sum 
$\chi = \frac12 \sum_\varepsilon  \frac{\partial^2 \varepsilon}
{\partial {h_i^2}}$. For ${h_i} = 0$, the summands can be determined 
by differentiating (\ref{2ck:eigenenergies}), giving
\begin{eqnarray}
\chi ({h_i}=0) &=& \sum_{\varepsilon >0 }
{1\over \varepsilon}
{4\pi \Gamma \Delta_L \over 
( \Delta_L 4 \pi \Gamma + \pi [(4\pi \Gamma)^2 + \varepsilon^2])} \\
& \approx & {1\over 4\pi^2 \Gamma} \ln (4 \pi \Gamma / \Delta_L)\; 
\qquad (\mbox{for}\; \Gamma \gg \Delta_L) \; .
\end{eqnarray}
The fact that $\chi ({h_i} = 0) \to \infty$ as $L\to\infty$ is
of course a characteristic sign of 2CK NFL physics:
it illustrates  the instability of the
NFL phase with respect to a local symmetry
breaking.\cite{Sacr} 
At finite temperatures $T$ takes over the role of the infrared cutoff
$\Delta_L$, so that the
susceptibility diverges logarithmically with $T$.\cite{AD84,EK92} 

The {\em fluctuations in $\hat {\cal N}_x$}\/ can be quantified
by calculating 
$\langle \hat  {\cal N}_x^2 \rangle - \langle \hat {\cal N}_x \rangle^2$.
In Appendix~\ref{app:Nxfluctuations}
 this is done at $\varepsilon_d = 0$ for the
physical ground state of ${\cal S}_{\rm phys}$ for both $P=0$ and 1. 
We find that $\langle \hat {\cal N}_x \rangle = 0$ for arbitrary
ratios of  $\Gamma / \Delta_L$, 
showing that the ground state
contains equal amounts of spin from both  flavors $j=1,2$,
as expected from the 2CK model's flavor symmetry.
Furthermore, we find that $\langle \hat {\cal N}_x^2 \rangle = P/4$
for $\Gamma/\Delta \to 0$,  as expected intuitively,
since in this limit the considered ground states are linear
combinations of states with ${\cal N}_x = \pm P /2$
[compare Fig.~\ref{manystate}]. 
In contrast, in the limit $\Gamma /\Delta_L \gg 1$ we find that the
fluctuations diverge logarithmically with system size,
 $\langle \hat {\cal N}_x^2 \rangle \approx {1 \over \pi^2 } \ln {\Gamma L}$.
This illustrates very
vividly how severely the dynamical impurity 
stirs up the original Fermi sea at the NFL fixed point.

\section{Connection to different renormalization group methods}
\label{s:RGs}
In the literature several renormalization group (RG) methods have 
been applied to the multichannel Kondo model. In this section 
we relate these RG methods to our finite-size bosonization technique,
by showing how the strategies employed by the former
can  be implemented, in an {\em exact
way,}\/ within the latter.

\subsection{High-energy cutoff scaling techniques}
\label{ss:standard}
The most common types of RGs are the ones used in particle physics and 
in  the standard treatment of critical phenomena.
 In these RG procedures, one reduces a high-energy cutoff, say $\tilde D$,
in order to
gradually eliminate some high-energy degrees of freedom,
arguing that they only slightly 
influence the low-energy physics of the system. The change in the 
cutoff must be compensated by  rescaling  the model's 
dimensionless 
coupling constants and masses in order to keep the physical properties 
(different inherent 
energy scales and dressed masses) invariant.
 These kinds of
scaling procedures, which include Anderson's poor man's
scaling,\cite{poorman} the multiplicative RG,\cite{Multipl} 
and the Yuval-Anderson real time RG,\cite{Yuval_And} have 
been widely used in the continuum limit, $L\to\infty$, to study
the multichannel Kondo  model.\cite{NB80,VZ83,Gan_etal,Zarand97,VZZ88}

In our case the high-energy cutoff $\tilde D$ can be identified with the 
ultraviolet cutoff $1/a$ introduced  when defining
the boson fields $\phi_{\alpha j}$ [in
(\ref{bosonfieldsa})], i.e.\  $\tilde D \sim 1/a$. 
Thus, if a dimensionless
coupling constant, say $\lambda$, has the scaling equation
\begin{equation}
\label{generalscalingeq}
  { d \ln \lambda \over d \ln \tilde D} =
  - { d \ln \lambda \over d \ln a} = \gamma (\lambda) \, ,
\end{equation}
then in the weak-coupling regime $\lambda \ll 1$,
its  scaling dimension is $\gamma (0)$, and it
is relevant, marginal or irrelevant for $\gamma (0)<0$,
$=0$ or $>0$, respectively.

Now, along the EK line 
one immediately obtains the scaling equations\cite{Ye}
\begin{eqnarray}
\label{scaling1}
\frac{d\lambda_\perp}{d\ln a} = \frac1 2 \lambda_\perp\;,
\qquad 
\label{scaling2}
\lambda_z \equiv 1 \; .
\end{eqnarray}
The first,
which follows from the requirement of the 
invariance of the Kondo scale   $\Gamma = \lambda_\perp^2/4a$, 
 shows that $\lambda_\perp$
is relevant and grows under bandwidth rescaling, with dimension $-1/2$.
As explained at the end of Section~\ref{sec:EK}, 
the second equation follows  from the fact,
(proven in Subsection~\ref{ss:finsize}),  that the leading 
irrelevant operator is absent only  at the EK line, where $\lambda_z=1$.
Eqs.~(\ref{scaling1}) 
exactly coincide with the ones 
obtained with the Yuval-Anderson technique,\cite{VZZ88}
\begin{mathletters}
\begin{eqnarray}
\frac{d\lambda_\perp}{d\ln a} &=& 
\left(4{\delta\over \pi} - 8 {\delta^2\over \pi^2}\right)  
\lambda_\perp\;,\\
\frac4\pi \frac{d \,  \delta}{d\ln a} & = & 
\left( 1 - {4 \delta \over \pi } \right) \lambda_\perp^2 \; ,
\end{eqnarray}
\end{mathletters}
if  in these the phase shift $\delta = \lambda_z\pi/4$ is replaced by
$\pi/4$ at the EK line,\cite{PC91}
as discussed at the end of Section~\ref{sec:EK}.
Remarkably, while these latter  equations result from rather non-trivial
calculations in terms of the original Hamiltonian, they can
trivially  be derived  after
the EK transformation, which in effect resums all 
appropriate diagrams into a quadratic  form.

As discussed Subsection~\ref{sec:evolution}, in a finite
local magnetic field $\varepsilon_d={h_i}$, a further natural energy
scale appears: $T_h = {h_i^2}/\Gamma$.  For energies below this
scale the magnetic 
field destroys the non-Fermi liquid behavior and a Fermi-liquid
is recovered. 
By requiring the invariance of $T_h$ one
immediately derives that, as long as the high-energy cutoff $1/a$
is much larger than the Kondo scale $\Gamma$, the field ${h_i}$ must be
invariant under the RG transformation: 
\begin{equation}
\label{hi-1}
\frac{d {h_i}}{d\ln a} = 0\;\;\;\; (1/a \gg \Gamma).
\end{equation}

However, once the cutoff is reduced sufficiently so that $1/a \ll
\Gamma$,  the role of $\Gamma$ is taken over by $1/a$,
i.e.\ $T_h$ is now given by ${h_i^2} \, a$, 
and therefore the scaling equation (\ref{hi-1}) must be replaced by
\begin{equation}
\frac{d {h_i}}{d\ln a} = -\frac 12 {h_i} \;\;\;\; (1/a \ll \Gamma) .
\label{eq:dimh}
\label{hi-2}
\end{equation}
Since $h_i$ is dimensionfull, the scaling dimension
of the local field can not be read off directly
from this equation; instead one must consider
the corresponding equation for the 
{\em dimensionless}\/ field  measured in units of the effective bandwidth, 
$\tilde h \equiv h_i a$, namely 
\begin{equation}
\frac{d \tilde h}{d\ln a} = \frac 12 \tilde {h} \;\;\;\; (1/a \ll \Gamma) .
\label{eq:dimlessh}
\end{equation}
Close to the NFL fixed point the local field thus
[by (\ref{generalscalingeq})] has dimension $-1/2$  and 
is relevant: when measured
in units of the effective bandwidth, it grows when
the latter is decreased. (By a similar 
argument, the dimension of the local field
in the regime $1/a \gg \Gamma$ of (\ref{hi-1}) is $-1$.)

Equations (\ref{hi-1}) and (\ref{hi-2})
are in complete  
agreement with those obtained by the Yuval-Anderson technique.\cite{VZZ88}
We remark at this point that perpendicular local magnetic fields $h_{x,y}$
(i.e.\ perturbations of the form $h_x S_x$ or $h_y S_y$)
are known\cite{VZZ88}  to scale differently from ${h_i}=h_z$,
and at the EK line their scaling dimension is 
known to be $-1/2$  even in the region $1/a\gg \Gamma$.

\subsection{Connection to Numerical Renormalization Group}

In this subsection we show that an analysis  of 
our finite-size spectrum as function of $L$ in fact represents 
an analytical version of Wilson's numerical
renormalization group\cite{Wilson} (NRG), which can simply be viewed as a
special type of finite-size scaling.

In Wilson's procedure one divides the Fermi sea into energy shells
using a logarithmic mesh characterized by a parameter $\Lambda >
1$, and then maps the model onto an equivalent one in which 
the impurity is
coupled to the end of an infinite conducting chain, where the hopping
between the sites $n$ and $n+1$ scales as $\Lambda^{-n}$.
The $n$'th site in this chain represents an ``onion-skin'' shell
of conduction electrons,  characterized by
spatial extension $\sim \Lambda^{n/2}$ around the impurity site 
and energy $\sim
\Lambda^{-n}$.
The NRG transformation is then defined
by considering truncated chains of length $N$ with 
Hamiltonian $H_N$, and consists of (i) adding
a new site to the end of the chain, $H_N\to
H_{N+1}$, and (ii)  rescaling the new Hamiltonian by $\Lambda$:
$H_{N+1} \to \Lambda H_{N+1}$. Trivially, step (i)  reduces the mean
level spacing  by a factor of $1/\Lambda$, while step (ii) 
is needed to measure all energies in units of the new mean level
spacing. This strategy is implemented by numerically diagonalizing
$H_{N+1}$ and retaining only the lowest few hundred levels.
One finds that after a number of iterations the 
spectrum of  $H_N$ converges to a fixed, universal 
set of energies, characteristic of some fixed point 
Hamiltonian.\cite{PC91} For the 2CK model this
spectrum has been shown to be identical to the one obtained by
boundary CFT.\cite{AL92b}

The NRG procedure outlined above can easily be interpreted in terms of 
our finite-size calculations. 
Step (i)  corresponds to 
increasing the system size, $L\to  \Lambda L$, (i.e.,
reducing the level spacing, $\Delta_L\to
\Delta_{\Lambda L} = \Delta_L/\Lambda$),
while  step (ii) is equivalent to measuring all energies in units of
$\Delta_{\Lambda L }$. 
Combining both steps, an ``analytical RG step'' thus has the form:
\begin{equation}
{H_x(L,\Gamma,\varepsilon_d)\over \Delta_L} \to
{H_x(\Lambda L,\Gamma,\varepsilon_d)\over \Delta_{\Lambda L}}
= {H_x(L,\Lambda \Gamma,\Lambda \varepsilon_d)\over \Delta_L}\;,
\end{equation}
where the last equality follows identically from Eq.~(\ref{Hquadratic}). For
$\varepsilon_d = 0$ this means that increasing the system size
at fixed $\Gamma$ is equivalent to increasing $\Gamma$ at fixed
$L$, emphasizing once more that  in this case the spectrum depends only on
$\Gamma/\Delta_L$.
Therefore the ``spectral flow'' as function of $\Gamma/\Delta_L$ 
in Fig.~\ref{2ckscaling} can be viewed as the
analytical version of an NRG spectrum as a function of iteration number.

\subsection{Finite-Size Scaling}
\label{ss:finsize}

It  is also straightforward to implement 
Wilson's prescription for extracting the exact 
scaling exponent of a perturbation around
the fixed point, say $\delta \lambda \, \hat O$, 
 from its effect on the finite-size spectrum:
In general, it causes the dimensionless  energy $\tilde {\cal E} (L)$
[of (\ref{eq:fssdef})]
 (calculated at a  finite, non-zero $\Delta_L \ll \Gamma$)
to differ from its universal fixed point value ${\cal E}_{\rm NFL}$ 
[of (\ref{Euniversal})] by an amount $\delta \tilde {\cal E} (L)$,
whose leading  asymptotic behavior for $L \to \infty$ is 
\begin{equation}
\delta \tilde {\cal E} (L) \equiv \tilde {\cal E} (L)
- {\cal E}_{\rm NFL} \sim (\delta \lambda / L^{\gamma})^n\;,
\label{deltauniversal}
\end{equation}
where $n\ge 1$ is some integer. 
By definition, 
$\gamma$ is the scaling dimension of the operator $\hat O$,
which is relevant for $\gamma < 0$, marginal for $\gamma = 0$ 
 and irrelevant for $\gamma > 0$.
Thus deviations from the universal spectrum are
characteristic of the  operator content of the fixed point. 

We first consider the situation  {\em on}\/ the EK line
(i.e.\ for $\lambda_z = 1$),  and close to the NFL fixed 
point, where  $\Delta_L/\Gamma$ and $T_h / \Delta_L$ are
are both  $\ll 1$ ({\em at}\/ NFL fixed point they are both 0).
The leading deviations 
$\varepsilon_{j,P}/\Delta_L - (\varepsilon_{j,P}/\Delta_L)_{\rm NFL}$ 
of the  dimensionless single-particle excitation eigenenergies 
from their NFL fixed point values are then given 
[from (\ref{2ck-ejp})] by  
\begin{equation}
 \label{eq:Wilsonillustrate}
\delta_{j,P} - (\delta_{j,P})_{\rm NFL} 
 =    {1 \over 4 \pi^2} 
    \left[ {T_h \over   \Delta_L (j - P/2)} 
- { \Delta_L (j - P/2 ) \over  \Gamma } \right]  
\end{equation}
(for $j \ge 1$; for $j=0$, Eq.~(\ref{e00-energy}) yields
the same conclusions). The leading dependence on the
local magnetic field via $ T_h/\Delta_L$  is evidently
$[{h_i} L^{1/2}]^2 $, which grows
as $L \to \infty$. This shows that a local magnetic field
has dimension $ \gamma_{{h_i}} = -1/2$ and is relevant:
for an arbitrarily small ${h_i}$,
 there exists a system size $L$ above which
the lowest part of the spectrum and the ground state properties of the
model are drastically affected, namely when
$\Delta_L \lesssim T_h$; in the language
of section \ref{sec:evolution}, this occurs as soon as 
the crossover
field below which the
spectrum is unaffected, namely $h_c = \sqrt {\Gamma \Delta_L}$,
becomes smaller than $|{h_i}|$ as $L$ is increased. 
The dimension $\gamma_{{h_i}} = -1/2$ also follows from the
$L$-dependence of $h_c$, and agrees with the conclusions
of our bandwidth rescaling arguments [see (\ref{eq:dimlessh})].

In the absence of magnetic fields, the leading term
in (\ref{eq:Wilsonillustrate}) vanishes with increasing $L$ as
$(\Gamma L)^{-1}$, 
implying that the least irrelevant irrelevant operator {\em on}\/
the EK line has dimension $\gamma_{EK} =1$. Thus, we 
conclude that the leading irrelevant 
operators with dimension $\gamma=1/2$ that were found in the CFT
treatment \cite{AL93} are absent {\em on}\/ the EK line,
in agreement with Refs.\onlinecite{EK92,SG94}.

Now let us move away from the EK line by taking
$\lambda_z = 1 + \delta \lambda_z$, 
and do perturbation theory in $\delta \lambda_z$, i.e.\
in $\delta H'_z$ of (\ref{dev:EKline}).
Then the operators with dimension $\gamma=1/2$ just mentioned
immediately show up: As shown in 
detail in Appendix~\ref{app:pertEK}, 
we find that the ``zero mode'' term 
$\delta \lambda_z \Delta_L \hat {\cal N}_s S_z$ of 
(\ref{dev:EKline}) (which does not occur 
in the continuum limit considered in Ref.~\onlinecite{EK92}), 
affects the  spectrum
already in  first order in $\delta \lambda_z$:
in the absence  of magnetic fields,
the first excited states (with ${\cal E}_{\rm NFL} = 1/8$)
are shifted relative to the doubly degenerate
ground states (with ${\cal E}_{\rm NFL}  = 0$) 
by an amount 
\begin{eqnarray}
\delta \tilde {\cal E} (L)
 \simeq
-  {\textstyle { 1 \over 4}}
 \delta \lambda_z  \bigl(1 + 4\pi^2 
{\Gamma/  \Delta_L}\bigr)^{-1/2} \sim L^{-1/2} \; . 
\end{eqnarray}
This implies  that 
the leading irrelevant operator away from the EK line 
has dimension  $1/2$.   

In the presence of a local
magnetic field $\varepsilon_d = {h_i} $, one finds
in the continuum limit $\Delta_L \ll \Gamma, h_i$ 
that the ground state degeneracy is split by an amount
\begin{eqnarray}
  \label{eq:splitgsdegenhh}
  \delta \tilde {\cal E} (L) = 
\left\{
\begin{array}{ll} 
{\displaystyle 
  {\delta \lambda_z \over 2 \pi^2 }  { | {h_i} | \over \Gamma} 
  \ln  { |{h_i}|  \over 4 \pi \Gamma }  }
& \quad (\Delta_L \ll h_i \ll \Gamma)  \; \vspace{2mm}, 
\\
{\displaystyle 
 {\delta \lambda_z \over 2 } \left( 1 - {4 \Gamma \over |{h_i}|} \right) }
& \quad (\Delta_L \ll \Gamma \ll h_i ) \; . 
\end{array} \right. 
\end{eqnarray}
This shows that the magnetic-field behavior
{\em along}\/ the EK line is not completely
generic, since 
it misses  this part of the $h_i$-dependence
of the magnetic-field-induced crossover. 
Note that the 
$ |{h_i} | / \Gamma \ln | {h_i}| / \Gamma$
behavior that occurs for a local magnetic
field of intermediate strength is consistent with the conclusions 
of the NRG studies of Ref.~\onlinecite{AL92b}
for the $h_i$-dependence of a certain
phase shift that can be used to characterize the NRG
spectra.

Finally, we would like to comment here on the identification of the Kondo
scale $T_K$.  In Section~\ref{sec:evolution} we showed  that
the crossover scale below which the finite-size spectrum  
takes its fixed-point form  (at $h_i = 0$) was $\Gamma$, and hence
concluded that $T_K \simeq \Gamma$. This differs from the
suggestion of Sengupta and Georges\cite{SG94}
that the Kondo scale in the anisotropic 2CK  model close to the EK
line is not $\Gamma$ but rather $\Gamma /(\delta \lambda_z)^2$.
This scale emerged naturally in their calculation of the
total susceptibility enhancement due to the impurity, which yielded
$\chi_{\rm imp}\sim (\delta \lambda_z)^2/\Gamma \ln(\Gamma/T)$
(at $h_i = 0$).  However, the factor $(\delta \lambda_z)^2$
only expresses the fact that the amplitudes of the leading
irrelevant operators vanish {\em on}\/ the EK line, so that the
characteristic logarithmic features appear only in second order in
$\delta \lambda_z$. The fact that the scale above which
these logarithmic features vanish is $T\simeq \Gamma$,
not  $T\simeq \Gamma/( \delta \lambda_z)^2$, 
supports our above conclusion that it is rather
$\Gamma$ that should be identified as the Kondo scale.

\section{Discussion and Conclusions}
\label{s:Concl}
The main general conclusion of our work is that constructive finite-size
bosonization is an unexpectedly powerful tool for investigating quantum
impurity problems. Firstly, for the 2CK model, it enables one to {\em
  analytically calculate by elementary means the crossover 
along the EK-line of the finite-size
  spectrum}\/ ({\em and}\/ the corresponding eigenstates) between the free
Fermi liquid and the NFL fixed point. This crossover had hitherto been
tractable only with the numerical renormalization group, and has been beyond
the reach of all analytical approaches used to study this model.

Secondly, finite-size bosonization {\em can deal without much
additional effort with symmetry-breaking perturbations,}\/ such as a finite
magnetic field (or channel symmetry breaking \cite{CS95-97},
which was not discussed here,
but can be included by a straightforward extension of our methods).
Indeed, it is to be expected that the methods developed
here can fruitfully be applied to a number of related
quantum impurity problems. For example,
an adaption of our finite-size refermionization approach was very
recently used to rigorously  resolve a recent controversy regarding the
tunneling density of states at the site of an impurity in
a Luttinger liquid.\cite{jvdschoeller} 
Other potential applications would be to the 
generalized Kondo models studied by Ye \cite{Ye},  or by
Moustakas and Fisher  \cite{MF95,Dobler}, or by 
Kotliar and Si \cite{kotliarsi}.

Thirdly, finite-size bosonization 
allows one {\em to mimic in an exact way the strategy of standard
RG approaches}\/ such as poor man's bandwidth rescaling and 
finite-size scaling; thus  it should 
be useful also as a pedagogical tool for teaching 
and analytically illustrating  standard RG ideas.

Crucial to the success of our method is that we do not use
field-theoretical bosonization, in which the bosonization relation
just has the status of a formal correspondence, but
Haldane's  {\em constructive}\/ formulation of 
bosonization, in which all operators and fields needed
are constructed explicitly in terms of the initial set of electron
operators $c_{k \alpha j}$ in terms of which the model is defined. This has the
great advantage that Emery and Kivelson's
bosonization-refermionization mapping of the model onto a quadratic
resonant level model can be implemented rigorously, not merely as a
formal correspondence, but as a set of operator identities in Fock
space. To achieve this, however, the Klein factors have to be treated
with due care. Our main technical innovation was
to demonstrate how refermionization can be performed at the same level
of rigor as bosonization, by 
extending the Fock space to include unphysical states,
and identifying and discarding these at the end of the calculation
using a generalized gluing condition. 
However, we wish to emphasize that our
method is truly elementary: in principle it requires nothing 
more than a knowledge of standard second quantization,
since that is all one needs to derive the constructive
bosonization formalism.\cite{jvdschoeller}

Our rigorous implementation of EK's mapping of the 2CK model
onto the resonant level model ensures that
the latter is not merely an
``effective model that captures the essential physics'', but truly {\em
identical to the original model}\/ one starts out with (after the
standard reinterpretation of the meaning of the coupling constants
discussed in Appendix~\ref{sec:redefinecouplings},
to compensate for differences that can arise relative to other methods
due to the use of different regularization schemes).
The fact that we are thus  able to solve
the {\em original}\/ model exactly,  
constitutes a significant advance relative to a number
of alternative approaches that have tried to
analytically access the NFL fixed point, but 
end up doing so using either an ``effectively 
equivalent'' model or some assumptions that are not
proven from first principles. 
Let us briefly mention three of these:

(i) Coleman and coworkers\cite{ColmIoffeTsvel,CS95-97} have proposed a
``pedestrian solution'' of the 2CK model, in which it is argued that
many of its properties can be calculated using a so-called
``compactified model'' involving only a single channel of spinful conduction
electrons.  This model was argued to  represent that part of the 2CK
model that is left over when one ``factorizes out'' the charge and
flavor degrees of freedom.  Indeed, using field-theoretic
bosonization, Schofield showed that there is a formal correspondence
between the compactified model
 and our $H_\perp$ of Eq.~(\ref{2ck:H_perp}) (which
involves only $\varphi_s$ and $\varphi_x$), and that it yields the
same results as the 2CK model for the {\em impurity contribution}\/ to
thermodynamical properties. In this sense, the compactified model can
be viewed as an effective model for calculating impurity properties.
However, as first emphasized by Ye,\cite{Ye} it is {\em not}\/
equivalent to the original 2CK model, since Schofield's arguments
ignored the fact that there are gluing conditions such
as (\ref{gluingall}) between the $c$,$f$
sectors and the $s$,$x$ sectors.  As long as these are ignored, the
compactified model can {\em not}\/ be used to calculate conduction
electron properties, since that requires adding back the contributions
from the charge and flavor channels.  This was attempted by Zhang, Hewson
and Bulla,\cite{ZHB97}   who calculated the fixed-point spectrum of the
compactified model using the NRG, from which they tried to reconstruct
that of the 2CK. However, their construction has an ad hoc character
(it requires knowledge of the answer) and succeeded only partially (it
did not correctly reproduce all degeneracies of the 2CK fixed point
spectrum).

Our constructive bosonization approach allowed us to clarify this issue
completely: it makes precise in what sense the $c$ and $f$ sectors can be
``factorized out'', rigorously yields an appropriate model for the remaining
$s$ and $x$ sectors, emphasizes the gluing conditions between the $c$,$f$ and
$s$,$x$ sectors, and shows how they can be used at the NFL fixed point to
combine the contributions from all four sectors to obtain the NFL fixed point
spectrum.

(ii) Affleck and Ludwig's path-breaking conformal field theory solution of the
model in the NFL regime, though very elegant and highly successful, rests on
two assumptions that can not be proven from within their theory, but have to
be confirmed by comparison with other exact methods: firstly, they assume that
the NFL fixed point has the same $U(1)_c \times U(2)_s \times SU(2)_f $
symmetry as the free model; and secondly, they need to use a certain conformal
fusion hypothesis to obtain the operator content of the NFL fixed point.  In
Section~\ref{fsize}, we analytically proved these assumptions in a simple,
natural, direct manner.  (The second assumption has also been proven by
Sengupta and George using bosonization, but their proof requires extensive
knowledge of CFT, and the assumption that the fixed point is invariant under
modular transformations).

(iii) Maldacena and Ludwig \cite{ML95} have used CFT to show 
that Affleck and Ludwig's CFT solution can be
reformulated in terms of free boson fields 
$\varphi_y (x)$ satisfying certain  asymptotic 
boundary conditions. Ye \cite{Ye} reproduced this result 
using field-theoretic bosonization (in the continuum limit) 
and scaling arguments, which, however, 
have the standard weakness of scaling arguments in
strong-coupling problems: they give the initial
flow of the coupling constants, but become non-rigorous
once the weak-coupling regime is left.

We have shown in Ref.~\onlinecite{vDZF} (and will elaborate this in a future
publication \cite{ZvD}) 
that these results can be reproduced with great ease and much
more rigor by simply taking the continuum limit $L\to \infty$ of our above
finite-size calculation.  In fact, this allows us to fully reproduce all
Affleck and Ludwig's results for electronic correlations functions.

In summary, we believe that our finite-size bosonization approach is the first
straightforward analytical calculation which, starting from first principles
and without any assumptions, yields the crossover of the finite-size spectrum
of the 2CK model from the free to the NFL fixed point and allows a detailed
finite-size scaling analysis of the fixed points.

{\em  Acknowledgments:---} 
We are deeply indebted to  Michele Fabrizio for carefully 
reading  the manuscript and for invaluable help in the early
stages of this work.   We are also grateful to 
Andreas Ludwig and Abraham Schiller for illuminating discussions  
and generous  help, Gabriel Kotliar and Anirvan Sengupta for 
useful suggestions, and Volker Meden for a discussion 
on cutoff-related matters. 
  This research has been supported by SFB195 
of the Deutsche Forschungsgesellschaft and by the Hungarian 
Grants OTKA~T026327 and OTKA~F016604. G. Z. has been  supported by the
Magyary Zolt\'an  Scholarship. 
\widetext
\vspace*{6mm}
\appendix 
\narrowtext

\section{Cutoff-related Matters}
\label{app:realsp}
\label{Pedantic}
\label{sec:redefinecouplings}

In this appendix we discuss in some detail various matters related
to the choice of an ultraviolet cutoff scheme,\cite{Meden} 
since this is a rather subtle matter, 
which can be elucidated more explicitly 
when using constructive rather than the more usual
field-theoretic bosonization formalism. In Section~\ref{app-extend-bandwidth}
we explain the need to reintroduce an ultraviolet cutoff
after removing the fermionic bandwidth cutoff $D$,
and in Section~\ref{app-bos-cutoff} discuss
how this is accomplished by the {\em bosonization cutoff
scheme}\/ used in the main text. 

\subsection{Extending the Bandwidth to Infinity}
\label{app-extend-bandwidth}

On physical grounds,
the  momentum sums in Eqs.~(\ref{hkinetic}) and (\ref{H_int1D})
for $H_0$ and   $H_{\rm int}$ 
must be cut off at some large momenta ($|k| \le D \sim p_F$),
to account for the finite width of the fermion conduction band
and the fact that a realistic impurity potential
always has non-zero range. 
However, the bosonization procedure used by us 
requires a single-particle  Hilbert space with
an  {\em unbounded}\/ fermion  momentum spectrum.\cite{Haldane81} 
To achieve this we  have removed, following Haldane \cite{Haldane81},
the implicit  cutoff  $D$ in Eqs.~(\ref{hkinetic})
and (\ref{H_int1D}), and used   $k \in [- \infty, \infty]$ instead. 
By doing so, we  extended the Hilbert space of single-electron states
to include unphysical ``positron''
states with arbitrarily large negative momenta, but  this should 
not change the low-energy physics of the system, since
by construction they require very high energies ($ > \varepsilon_F)$ 
for their excitation. 

In the resulting single-particle Hilbert
space with an unbounded fermion momentum spectrum, 
the fermion fields defined in (\ref{define-psi}) 
and  the corresponding densities
do not depend on any cutoff parameter.
In this Appendix we shall denote this fact by a superscript ${}^{(0)}$:
\begin{eqnarray}
  \label{eq:psi-0-define}
\psi^{(0)}_{\alpha j}(x) \equiv 
        {\textstyle \sqrt{{2\pi\over L}} }
        \sum_{n_k \in {\mathbb Z}} e^{- ikx}c_{k\alpha j} \; , 
\\
    \label{eq:psi-0-bosonize}
\rho^{(0)}_{\alpha j}(x) \equiv \; :\! 
{1 \over 2 \pi } 
\psi^{(0)}_{\alpha j}(x) \psi^{(0)}_{\alpha j}(x) \! : \; , 
\end{eqnarray}
In this notation, the position-space representation of the
Hamiltonian, given here  to facilitate comparison with field-theoretic
treatments of the 2CK model, reads:
\begin{eqnarray}
\label{H0position}
        H_0 
& = &
        \sum_{\alpha j} \int_{-L/2}^{L/2}{dx\over 2\pi}
        : \! \psi_{\alpha j}^{(0) \dagger} (x) \; 
        i\partial_x \psi_{\alpha j}^{(0)} (x) \! : \;, 
\\
        H_{\rm int}
&=& 
        \sum_{\mu ,\alpha,\alpha^\prime,j}  \lambda_\mu 
        S_\mu : \! \psi^{(0) \dagger}_{\alpha j}(0)  {\frac 1 2}
        \sigma^\mu_{\alpha\alpha^\prime}  
        \psi^{(0)}_{\alpha^\prime j}(0) \! :\;, 
        \label{H_int}
\\
        H_h
&=& 
        h_i S_z + h_e \sum_{\alpha j} 
        \int_{-L/2}^{L/2} {dx\over 2\pi}
        {\frac \alpha 2} : \! \psi_{\alpha j}^{(0) \dagger} (x)  
        \psi_{\alpha j}^{(0)} (x)\! :   \;.
\end{eqnarray}

Now, some physical quantities, such as the  phase shift $\delta$ 
of the outgoing relative to the incoming fields, 
\begin{equation}
  \label{eq:def-phaseshift}
  \psi^{(0)}_{ \alpha j} (0^-) \equiv
 e^{i 2 \delta}  \psi^{(0)}_{\alpha j} (0^+)  ,
\end{equation}
depend explicitly on an ultraviolet cutoff and in fact would  be ill-defined 
without any. Therefore, the decision  to use an infinite fermion band
must always be accompanied by a {\em reintroduction},
in some other fashion, of an ultraviolet cutoff.
Moreover, the precise way in which this is done is well-known
 \cite{EKrev,Noz-Dom,Schotte}
to strongly influence the meaning of the coupling constants in 
$H_{\rm int}$. 
As an example, 
we consider the case of no spin-flip scattering ($\lambda_{\perp}
=0$), and reintroduce an ultraviolet cutoff by 
replacing $\lambda_z$ in Eq.~(\ref{H_int1D}) for $H_{\rm int}$
either\cite{Noz-Dom}
 by the separable  form $\lambda_{z1} e^{-(|k| + |k'|)a/2}$,
or\cite{Schotte} by the nonseparable form $ \lambda_{z2} e^{-|k- k'|a/2} $. 
Choice 1 restricts  both momenta of a fermion scattering process separately
 to a band of width $1/a$, choice
2  only the momentum difference. They imply two different
versions for $H_z$ in position space, namely
\begin{eqnarray}
\nonumber
        H_{z1}
&=& 
        \lambda_{z1}
        S_z \sum_{\alpha j}  
        {\frac \alpha 2}  \int_{-L/2}^{L/2} dx 
         \int_{-L/2}^{L/2} dx'   \,      \delta_{a/2} (x)
       \delta_{a/2} (x')
       \\ & & \times 
        : \! \psi^{(0) \dagger}_{\alpha j}(x)  
        \psi^{(0)}_{\alpha j}(x') \! :\;, 
  \label{eq:Hz1}
\\
\nonumber
        H_{z2}
&=& 
        \lambda_{z2}
        S_z \sum_{\alpha j}  
        {\frac \alpha 2}  \int_{-L/2}^{L/2} dx \, 
       \delta_{a/2} (x)
        : \! \psi^{(0) \dagger}_{\alpha j}(x)  
        \psi^{(0)}_{\alpha j}(x) \! :\;, 
\\
& &
  \label{eq:Hz2}
\end{eqnarray}
as follows by noting that $e^{-|k|a}$ is the Fourier-transform
of the smeared $\delta_{a} (x)$ function of (\ref{delta_a}):
\begin{eqnarray}
  \label{eq:delta-a-Fourier}
  {1 \over L} \sum_{n_k \in Z} 
  e^{-i k x} e^{-|k|a } = \delta_{a} (x) + {\cal O}(1/L^2).
\end{eqnarray}
Eq.~(\ref{eq:Hz1}) shows
that choice 1  separately smears out
both  $\psi^{(0) \dagger}_{\alpha j} (x)$ and $\psi^{(0)}_{\alpha j} (x)$ 
 over a range $a$, i.e.\ corresponds to a zero-range potential
in a finite band,
whereas by (\ref{eq:Hz2})
choice 2 smears out the density $\rho^{(0)}_{\alpha j} (x)$,
i.e.\ corresponds to a finite-range potential in an infinite band.
Consequently, the equations of motions differ:
Choice 1 yields
\begin{eqnarray}
\nonumber
& & i(\partial_t - \partial_x)   \psi^{(0)}_{\alpha j}(t,x)
\\   \label{eq:1-eqm}
& = &   \pi \alpha \lambda_{z1} S_z  \, \delta_{a/2} (x) 
\int_{-L/2}^{L/2} \! dx'  \delta_{a/2} (x') \, 
    \psi^{(0)}_{\alpha j}(t,x')
\\
& \simeq  & \pi \alpha \lambda_{z1} S_z  \, \delta (x) \, 
   \frac{1}{2} \left[ \psi^{(0)}_{\alpha j}(t,0^-) +
    \psi^{(0)}_{\alpha j}(t,0^-) \right] \; , 
\end{eqnarray}
(we took $a\to 0$ in the second line).
This is solved by \cite{Noz-Dom} 
\begin{eqnarray}
  \label{eq:psisolve-1}
   \psi^{(0)}_{\alpha j}(t,x)
 & = & e^{-i k (t+x)} \left[ \theta (x) + \theta(-x) e^{2 i \delta_1} \right] 
   \; ,
\\
\label{eq:phase-shift-1}
 \delta_{1} &=& - \arctan (  \pi \alpha \lambda_{z1} S_z /2) \; , 
\end{eqnarray}
where $\theta (x)$ is a sharp step function, and the phase shift agrees with
that found in the Bethe Ansatz Kondo literature, or in non-1D treatments that
use a finite bandwidth \cite{Hewson}.  In contrast, the equation of motion for
choice 2,
\begin{eqnarray}
  \label{eq:2-eqm}
 i(\partial_t - \partial_x)   \psi^{(0)}_{\alpha j}(t,x)
 =    \pi \alpha \lambda_{z2} S_z  \, 
\delta_{a/2} (x)     \psi^{(0)}_{\alpha j}(t,x) \, , 
\end{eqnarray}
has the solution \cite{Schotte}
\begin{eqnarray}
  \label{eq:psisolve-2}
   \psi^{(0)}_{\alpha j}(t,x)
 & = & e^{-i k (t+x)} 
 e^{i  \alpha \lambda_{z2} S_z \arctan( 2 x /a) } \; , 
\end{eqnarray}
so that (for $a \ll |0^\pm|$) the phase shift of (\ref{eq:def-phaseshift}) is 
\begin{eqnarray}
  \label{eq:phase-shift-2}
 \delta_{2} &=& - \pi \alpha \lambda_{z2} S_z /2 \; .   
\end{eqnarray}
Evidently, regularization schemes 1 and 2 yield different
relations between coupling constant and phase shift.
Since the latter, being a physical quantity,
must have the same value in both schemes, $\delta_1 \equiv \delta_2$, 
we thus conclude that the coupling constants must be related
by $\lambda_{z2} = {4 \over \pi } \arctan ( \pi \lambda_{z1} /4)$. 

Finally, note that after the removal of the fermion band cutoff $D$,
even the free theory ($H_{\rm int} = 0$)
requires the reintroduction of an ultraviolet cutoff: the free 
imaginary-time-ordered  zero-temperature correlator,
\begin{eqnarray}
  \label{eq:psi-0-correlator}
  \langle {\cal T} \psi^{(0)}_{\alpha j} (\tau, x)
 \psi^{(0)\dagger}_{\alpha j} (0,0) \rangle = {1 \over \tau +  i x} \; ,
\end{eqnarray}
has a divergence at $t = x = 0$, which 
is often regularized by the replacement $\tau \to \tau
+ \mbox{sgn} (\tau) a$, where $a \simeq 1/p_F$
(though we reuse the notation $a$ here, this cutoff parameter
in general need not be the same as that used in $H_z$ above).
 Alternatively, one can introduce a bandwidth cutoff into the
definition of the fermion field itself, e.g.\
replace  $ \psi^{(0)}_{\alpha j}(x)$ by
\begin{eqnarray}
\label{define-tilde-psi-a}
\tilde \psi^{(a)}_{\alpha j}(x) & \equiv & 
        {\textstyle \sqrt{{2\pi\over L}} }
        \sum_{n_k \in {\mathbb Z}} e^{- ikx} e^{- |k|a /2} c_{k\alpha j} 
\\
        &=&
   \int_{-L/2}^{L/2} d x'    \delta_{a/2} (x-x')
   \psi^{(0)}_{\alpha j}(x') \;  , 
\end{eqnarray}
i.e.\ by a smeared version of $ \psi^{(0)}_{\alpha j}(x)$,
which results in 
\begin{eqnarray}
  \label{eq:tilde-psi-a-correlator}
  \langle  {\cal T}  \tilde \psi^{(a)}_{\alpha j} (\tau, x)
 \tilde \psi^{(a)\dagger}_{\alpha j} (0,0) \rangle = {1 \over \tau +
 i x + \mbox{sgn}(\tau) a}
\; .
\end{eqnarray}

\subsection{The Bosonization Cutoff Scheme}
\label{app-bos-cutoff}

The {\em bosonization cutoff scheme}\/ used in the main text constitutes
yet another way, {\em alternative}\/ to those just discussed,
 of reintroducing an ultraviolet cutoff 
after removing the fermion bandwidth cutoff $D$: one
bosonizes the theory completely and introduces an ultraviolet  cutoff 
$e^{-q|a|/2}$ in
the {\em boson}\/  fields of Eq.~(\ref{bosonfieldsa}), 
whose $a$-dependence we shall
indicate in this appendix by an explicit superscript ${}^{(a)}$:
\begin{eqnarray}
  \label{eq:bosonfieldsa-a}
  \phi_{\alpha j}^{(a)} (x) & \equiv &   \sum_{q > 0}
         \frac{- 1}{ \sqrt{n_q}}
           \left( e^{-i q x} b_{q {\alpha j}} 
          +  e^{i q x} b^\dagger_{q {\alpha j}} \right) 
          e^{-a q/2} \; .
\end{eqnarray}
Evidently,  $1/a$ can be viewed as ``effective boson bandwidth'' for 
the particle-hole excitations occuring in $\phi^{(a)}$.
In this notation, the bosonization relations 
(\ref{fermiboson})  
and (\ref{density}) used in the main text read
\begin{eqnarray}
&&\psi^{(a)}_{\alpha j} (x) 
=     F_{\alpha j}  \Delta_L^{1/2}
    e^{-i (\hat N_{\alpha j} - P_0/2) 2 \pi x /L} \, 
    : \! e^{- i \phi^{(a)}_{\alpha j} (x) } \! : 
\label{fermiboson-a} \\
&& \rho^{(a)}_{\alpha j} (x) 
= {1\over 2\pi}\partial_x \phi^{(a)}_{\alpha j} (x) 
  + \hat N_{\alpha j} / L \; 
\label{density-a}
\end{eqnarray}
[un-normal-ordering the exponential in (\ref{fermiboson-a}), 
which is possible only for $a \neq 0$, 
yields a factor $(a \Delta_L)^{-1/2}$ (see Eq.~(42) of 
Ref.~\onlinecite{jvdschoeller}) and thus reproduces 
(\ref{fermiboson})].
The superscripts ${}^{(a)}$ emphasize that for $a \neq 0$,
$\psi^{(a)}$ and $\rho^{(a)}$ are {\em not}\/
identically equal to the  original 
$\psi^{(0)}$ and $\rho^{(0)}$  of  (\ref{eq:psi-0-define}) 
of (\ref{eq:psi-0-bosonize}),
which do {\em not}\/ depend on $a$. Instead, 
the rigorously exact bosonization identities for $\psi^{(0)}$ and $\rho^{(0)}$
are the $a=0$  versions of  (\ref{fermiboson-a}) and
(\ref{density-a}) (cf.\ Ref.~\onlinecite{jvdschoeller},
footnote 7). Correspondingly, the exact bosonized position-space version of
$H_0$ depends on $\phi^{(0)}$ (not $\phi^{(a)}$):
\begin{eqnarray}
\nonumber
H_0   & = &  
        \sum_{\alpha j} 
        \Delta_L 
             {\hat N}_{\alpha j} (\hat N_{\alpha j} + \! 1\!  - \! P_0) /2
\\
\label{H0phifields}
 && +
                \sum_{\alpha j}  \int_{-L/2}^{L/2}  
{dx\over 4\pi} : \! 
        \left( \partial_x \phi^{(0)}_{\alpha j} (x)  \right)^2 \! :
         \;.
\end{eqnarray}
 By taking $a \neq 0$ in (\ref{fermiboson-a}) and
(\ref{density-a}) (but not in $H_0$), as we do in the main text, we thus
effectively make the replacement
$\psi^{(0)} \to \psi^{(a)}$, i.e.\ we {\em redefine}\/ the fermion fields 
to be explicitly cutoff dependent
 and thereby modify the ultraviolet behavior of the theory. 
Although this  redefinition is not identical to the replacement
$\psi^{(0)} \to \tilde \psi^{(a)}$ of (\ref{define-tilde-psi-a})
since $\psi^{(a)} \neq \tilde \psi^{(a)}$,
it similarly {\em regularizes the correlation function 
$\langle {\cal T} \psi^{(a)}  \psi^{(a)\dagger} \rangle $,}\/
which turns out \cite{jvdschoeller}
to be given by (\ref{eq:tilde-psi-a-correlator}) too. 
Moreover, it  {\em smears out the density}\/ by $a$,
since  [by (\ref{eq:bosonfieldsa-a}) and (\ref{eq:delta-a-Fourier})]
\begin{eqnarray}
\lefteqn{  \rho^{(a)}_{\alpha j} (x) = \Delta_L
        \sum_{\scriptstyle k,k^\prime}  e^{-|k - k^\prime | a/2} 
        : \!c^\dagger_{k\alpha j} c_{k^\prime\alpha} \! : }
\\
          &&=
   \int_{-L/2}^{L/2} dx' \, 
       \delta_{a/2} (x-x')
        : \! \psi^{(0) \dagger}_{\alpha j}(x')  
        \psi^{(0)}_{\alpha j}(x') \! :\; ,
\label{rho-a-smeared}
\end{eqnarray}
and {\em thus also  regularizes $H_z$,}\/ which depends
on the spin density (cf.\ (\ref{2ck:bosonh}) for $H_z$ in the main text).
In fact, comparison of (\ref{rho-a-smeared}) with $H_{z2}$ of (\ref{eq:Hz2})
shows that our bosonization cutoff scheme 
regularizes $H_z$ precisely according to the choice 2 discussed above. 

It is therefore not surprising that 
the phase shift for $\psi_{\alpha j}^{(a)}$ found at the end
of Section~\ref{sec:EK}  via the EK-transformation 
$U \equiv e^{i \lambda_z S_z \varphi_s^{(a)}  (0)}$,
namely  $|\delta| =  \pi/4$ for $\lambda_z = 1$,
 agrees with 
(\ref{eq:phase-shift-2}) for choice 2. 
(Note that $U$ would be undefined for $a \neq 0$, since its exponential, in
order to be unitary, {\em must}\/ be non-normal-ordered.)  However, if one
examines the phase shift more closely than for $|x| \gg a$, one discovers that
the phase factor $\arctan(2x/a)$ in (\ref{eq:psisolve-2}), obtained by solving
the equation of motion for $\psi^{(0)}_{\alpha j}$, {\em differs}\/ from the
$\arctan(x/a)$ in (\ref{eq:EK-psi}), obtained by EK-transforming
$\psi^{(a)}_{\alpha j}$. This simply illustrates that $\psi^{(0)} \neq
\psi^{(a)}$.  Indeed, if one EK-transforms $\psi^{(0)}$ instead of
$\psi^{(a)}$, one recovers the $\arctan(2x/a)$ of (\ref{eq:psisolve-2}),
either by using $\psi_{\alpha j}^{(0)} (x) \propto e^{-i \alpha
  \varphi_s^{(0)} (x)/2}$ and
\begin{eqnarray}
  \label{eq:EK-a-phi}
 U \varphi^{(0)}_s(x) U^\dagger  =  \varphi_s(x) - 2 \lambda_z S_z\; {\rm
arctan}(2 x/a) \; , 
\end{eqnarray}
or by using the {\em fermionic}\/
definition (\ref{eq:psi-0-define}) for $\psi^{(0)}$ to find
\begin{eqnarray}
  \label{eq:EK-ferm-psi-0}
  [ i \varphi_s^{(a)} (0), \psi_{\alpha j}^{(0)} (x)]
 = \alpha i \arctan (2 x/a) \, \psi_{\alpha j}^{(0)} (x) ,
\end{eqnarray}
together with the fact that if $[A,B] = c B$ 
with $[c,A] = [c,B] = 0$, then $e^A B e^{-A} = B e^c$. 

This example illustrates the subtle difference
between our bosonization cutoff scheme,
which replaces $\psi^{(0)}$ by $\psi^{(a)}$
and thereby modifies $H_z \to H_{z2}$, and the
regularization scheme of choice 2,
which  modifies only $H_z \to H_{z2}$ but does not change
$\psi^{(0)}$. As far as $H_z$ is concerned, 
both schemes can be used with equal merit, but once one
has chosen one of them, one must use it 
consistently throughout. 

For the treatment of $H_\perp$, however, our bosonization cutoff scheme is
distinctly more convenient. To see this, note that the EK-transformed version
of $H_\perp$ differs, depending on whether $H_\perp$ is expressed through
$\psi^{(a)}$ (as in the main text), or through $\psi^{(0)}$. 
In the former case the factor multiplying 
$(\lambda_\perp/2) F^\dagger_{\alpha j} F_{-\alpha j}$
in (\ref{HEK}) is  
\begin{eqnarray}
\nonumber 
\lefteqn{ U \Bigl[ \Delta_L 
: \!  e^{ i \alpha [\varphi_s^{(a)} (0) + j \varphi_x^{(a)} (0)] } \! :
 S_{- \alpha} \Bigr] U^\dagger }
\\ 
&=&  \left(\Delta_L \over a \right)^{1/2}  
: \! e^{ i \alpha j \varphi_x^{(a)} (0) } \! : S_{- \alpha} \; , 
\end{eqnarray}
(for $\lambda_z = 1$), the latter case instead yields
\begin{eqnarray}
\nonumber 
\lefteqn{
U \Bigl[ \Delta_L : \!  
e^{ i \alpha [\varphi_s^{(0)} (0) + j \varphi_x^{(0)} (0)] } \! :
 S_{- \alpha} \Bigr] U^\dagger } 
\\
&=& 2  \left(\Delta_L \over a \right)^{1/2} 
: \!  e^{i \alpha  [ \varphi_s^{(0)} (0)  -
\alpha \varphi_s^{(a)} (0) + j \varphi_x^{(0)} (0)] } \! :  S_{-\alpha} \;  .
\end{eqnarray}
Whereas in the former case the $\varphi_s$-dependence conveniently drops out 
for arbitrary $a$,  in the latter it 
inconveniently drops out
only in the limit $a \to 0$, in which the prefactor $a^{-1/2}$ 
diverges (and moreover the fermion correlation functions are
ultraviolet divergent). Note also that the extra prefactor of 2 in
the latter  case (which stems from normal-ordering
the product of $U S_{-\alpha} U^\dagger =  e^{- i \alpha
  \varphi_s^{(a)} (0)} S_{-\alpha}$ 
and $: \!  e^{i \alpha \varphi_s^{(0)} (0)}  \! :$
instead of 
$: \!  e^{i \alpha \varphi_s^{(a)} (0)}  \! :$), implies that
the coupling constant must be reinterpreted such that 
 $2 \lambda_\perp$ of the latter case corresponds 
to the $\lambda_\perp$ used 
in the former. This illustrates once more 
how sensitively the meaning of the couplings
depends on the choice of regularization scheme. 

\subsection{Point-Splitting vs.\  Normal-Ordering}
In the literature the position-space versions 
of $H_0$, Eqs.~(\ref{H0position}) or (\ref{H0phifields}),  
 are used more frequently than
the momentum-space versions of Eqs.~(\ref{hkinetic})
and (\ref{H0a}), perhaps because the former may seem
more concise. The product of two fields
at the same point is then regularized using the
point-splitting prescription
\begin{eqnarray}
\nonumber 
&& \lim_{x_0 \to 0} \left[ \hat O_1 (x - i x_0 ) \hat O_2 (x) 
- {}_0 \langle \vec 0 | 
 \hat O_1 (x - i x_0) \hat O_2 (x ) | \vec 0 \rangle_0 \right] \; ,
\end{eqnarray}
which in most cases is equal to the normal-ordered product $: \! \! \hat O_1
(x) \hat O_2 (x) \! \! :$, evaluated by normal-ordering the $c_{k \alpha j}$'s
in the Fourier expansions of these operators (see e.g.\ Appendix~G of
Ref.~\onlinecite{jvdschoeller}).  However, when using point-split operators,
great care is required if terms of order $\Delta_L$ are to be treated
correctly.  Since in practice they are more easily dealt with by using
normal-ordering in the momentum-space representation than
point-splitting in the position-space representation, 
we  use the former throughout this paper.

\section{Constructing a Basis for the extended Fock Space}
\label{Jan'sapp}

In Section \ref{sec:newKlein} we transformed from
an old to a new set of quantum numbers, $\vec N \to \vec {\cal N}$,
and embedded the physical Fock space
${\cal F}_{\rm phys}$
in the extended Fock space ${\cal F}_{\rm ext}$
[all  $|\vec {\cal N} \rangle \in {\cal F}_{\rm phys}$
satisfy both the free gluing conditions (\ref{gluinga}) and (\ref{gluingb}),
but only  (\ref{gluinga}) is satisfied by all  $|\vec {\cal
  N} \rangle \in {\cal F}_{\rm ext}$]. 
In this Appendix we show explicitly how such an embedding 
can be accomplished, by constructing a basis
of $\vec {\cal N}$-particle ground states 
$\{ |\vec {\cal N} \rangle_0\} $ 
that spans ${\cal F}_{\rm ext}$, in terms of ordered products
of new Klein factors ${\cal F}_y^\dagger$ acting on two
reference states.  

To begin, we fix the relative phases of the set $\{ | \vec N \rangle_0 \}$
of $\vec N$ particle ground states that span  ${\cal F}_{\rm phys}$, by defining
\begin{equation}
  \label{eq:N=calNphys}
|\vec N \rangle_0   \equiv
  F_{\uparrow 1}^{\dagger N_{\uparrow 1}} 
          F_{\downarrow 1}^{\dagger N_{\downarrow 1}}
          F_{\uparrow 2}^{\dagger N_{\uparrow 2}}
           F_{\downarrow 2}^{\dagger N_{\downarrow 2}} | \vec 0 \rangle_0
\; .
\end{equation}
States with an even
or odd total number of particles have $P =  2 {\cal N}_c \mbox{mod}2=0$
or 1, respectively. 
Clearly, all even or odd states can
be generated, respectively, from the even or odd
reference states $| \vec 0 \rangle_0$ or $|\vec \frac12 \rangle_0
\equiv F^\dagger_{\uparrow 1} | \vec 0 \rangle_0$, defined as
 \begin{eqnarray}
\label{eq:refstates}
  | {\textstyle {\vec P \over  2}} \rangle_0 
  \equiv |\vec {\cal N} = {\textstyle \frac{P}{2},
  \frac{P}{2}, \frac{P}{2},\frac{P}{2}} \rangle_0
  \equiv  |\vec { N} = P,0,0,0\rangle_0, 
\end{eqnarray}
by the application of a product of an {\it even number} of old Klein factors
$F^\dag_{\alpha j}$ or $F_{\alpha j}$.  By using Eqs.~(\ref{Kleinchange}) and
related bilinear relations, this product can be transcribed into a product of
an even number of new Klein factors ${\cal F}^\dag_y$ or ${\cal F}_y$.  The
resulting state evidently is an eigenstate of $\hat {\cal N}_y$, and since
Eqs.~(\ref{Kleinchange}) by construction respect
Eq.~(\ref{2ck:transformationN}), its eigenvalues $\vec {\cal N}$ are related to
$\vec N$ by Eq.~(\ref{2ck:transformationN}).  Therefore
\begin{equation}
 |\vec N\rangle_0 = e^{i \Phi(\vec {\cal N})}\; |\vec {\cal N}\rangle_0
  \; , 
\label{eq:phase}
\end{equation}
where  the $\vec {\cal N}$ particle ground state
$|\vec {\cal N} \rangle_0$ is defined to be 
\begin{eqnarray}
\label{eq:constr}
 | \vec {\cal N} \rangle_0 \equiv
 {\cal F}_c^{\dag {\bar N}_c}{\cal F}_s^{\dag {\bar N}_s} 
 {\cal F}_f^{\dag {\bar N}_f}{\cal F}_x^{\dag {\bar N}_x} 
 |{\textstyle {\vec P \over  2}} \rangle_0 
  \end{eqnarray}
for $\vec {\cal N} \in ({\mathbb Z} + P/2)^4$, 
the integers  $\bar N_y$ are defined
by  $\bar N_y \equiv {\cal N}_y - P/2$, 
and $\sum_y {\bar N}_y $  is, by construction,  an even number.
The phase factor $e^{i \Phi(\vec {\cal N})} = \pm 1$ 
can be determined, if necessary, by explicitly
rearranging the above-mentioned 
even product of new Klein factors into the standard order
of (\ref{eq:constr}). It ensures that the 
action of corresponding pairs of old or new Klein factors 
on the l.h.s.\ or r.h.s.\ of Eq.~(\ref{eq:phase}), respectively, produces
the  same result.

Evidently, the set $\{ | \vec {\cal N} \rangle_0 \}$ of all 
states with  $\sum_y {\bar N}_y = $ even constitutes
a basis for  the physical Fock space ${\cal F}_{\rm phys}$, 
just as  $\{ | \vec  N \rangle_0 \}$ does. 
The unphysical part of the extended Fock space ${\cal F}_{\rm ext}$ can now be
formally constructed by using  the definition (\ref{eq:constr}) 
also for 
integers with $\sum {\bar N}_y = \mbox{odd}$. Note once again that these new
states violate the second free gluing condition (\ref{gluingb}) and are
purely mathematical constructs outside the original Fock space. Then the total
extended Fock space can be formally written as
\begin{eqnarray}
\nonumber
{\cal F}_{\rm ext} = \sum_{\{ {\bar N}_y\} \in {\mathbb Z}^4}
{\cal F}_c^{\dag {\bar N}_c}{\cal F}_s^{\dag {\bar N}_s} 
{\cal F}_f^{\dag {\bar N}_f}{\cal F}_x^{\dag {\bar N}_x} 
(\{| \vec 0 \rangle\} \oplus \{|{\textstyle {\vec 1 \over 2}} \rangle \} ) \;,
\end{eqnarray}
where $\{| \vec 0 \rangle\}$ and $\{|{\textstyle {\vec 1 \over 2}}\rangle\}$
denote the set of all states that can be generated from the reference states
by the action of bosonic excitations $b^\dagger_{q y}$.

Within ${\cal F}_{\rm ext}$,
 which is the natural extension of the
original physical Fock space ${\cal F}_{\rm phys}$, the action of arbitrary (even
{\em and}\/ odd) products of new Klein factors evidently is trivially defined,
and they leave the subspaces generated by $| \vec 0 \rangle_0$ and
$|{\textstyle {\vec 1 \over 2}}\rangle_0$ separately invariant.  Note though,
that it is impossible to reach $| \vec 0 \rangle_0$ from $|{\textstyle {\vec 1
    \over 2}}\rangle_0$ or vice versa using new Klein factors, simply because
these change ${\cal N}_y$ by $\pm 1$, whereas the $\vec {\cal N}$ eigenvalues
in the two subspaces ``differ by $\vec \frac12$''.  However, they {\em are}\/
of course connected by the original Klein factors, e.g.\ $|{\textstyle {\vec 1
    \over 2}}\rangle_0 = F_{\uparrow 1}^\dag | \vec 0 \rangle_0$.  This shows
again that there is no way to express an 
{\em individual}\/ old Klein factor in terms of the
new ones, or vice versa.

\section{Explicit Diagonalization of $H_x$ for 2CK Model}
\label{app:2ck}

In this Appendix we diagonalize the Hamiltonian $H_x$ of
(\ref{Hxeh}) in explicit detail. We also calculate the
ground state energy shift $\delta E_G$, the ${\cal N}_x$ fluctuations
$\langle {\cal N}_x \rangle$, and do perturbation theory about
the EK line. 

\subsection{Introducing Majorana Fermions}

Our aim is to find the unitary transformation that brings the Hamiltonian
$H_x$ of (\ref{Hxeh}) into the diagonal form (\ref{Hdiagfer}), and
to determine the discrete set of eigenenergies $\varepsilon$.  This
transformation will map the original set of orthonormal operators occurring in
(\ref{Hxeh}), $\{ \alpha_n ; n = d, k \ge 0 \}$ (with $ \{ \alpha_n,
\alpha^\dagger_{n'} \} = \delta_{n n'}$), onto the new orthonormal set
occurring in Eqs.~(\ref{Hdiagfer})
to (\ref{tildealphaunit}), $\{ \tilde \alpha_\varepsilon ; \varepsilon \ge
0\}$ (with $ \{ \tilde \alpha_\varepsilon,
 \tilde \alpha^\dagger_{\varepsilon'} \} =
\delta_{\varepsilon \varepsilon'} $); however, the transformation
 does not involve the
$\beta_{\bar k}$'s in (\ref{Hxeh}) at all, since they are completely
decoupled and  ``just go along for the ride'' below.

Since the hybridization term in Eq.~(\ref{Hxeh}) only contains the
combinations $(\alpha^\dagger_{\bar k} + \alpha_{\bar k})$ and $(\alpha_d
-\alpha_d^\dagger)$, {\it ``half of the impurity'', $(\alpha_d +
  \alpha_d^\dagger)$, is completely decoupled from conduction electrons} if
$\varepsilon_d = 0$.  (EK were the first to emphasize that this causes
the model's NFL behavior.)  To exploit this fact,
it is convenient to transform the two sets
of fermions, $\{ \alpha_n \}$ and $\{ \tilde \alpha_\varepsilon \}$, to two
sets of Majorana fermions $\{ \gamma_{n \nu} \}$ and $\{ \tilde
\gamma_{\varepsilon \nu} \}$ ($\nu = \pm$), respectively:
\begin{eqnarray}
\label{gamman}
\left(\!\!\begin{array}{l} 
      \gamma_{n +} \\  \gamma_{n-}
      \end{array}\!\!
\right) 
&\equiv&  {1\over \sqrt{2}}
   \left(\!\! \begin{array}{rr}
                     1 & 1 \\ - i & i
                     \end{array}\!\! \right)
 \left(\!\! \begin{array}{c} 
      \alpha_n \\  \alpha^\dagger_n
      \end{array} \!\! \right), 
    \quad (n = d, k \ge 0), 
\\
\label{gammae}
\left(\!\! \begin{array}{l} 
      \tilde \gamma_{\varepsilon +} \\  \tilde \gamma_{\varepsilon -}
      \end{array}
\!\! \right) 
&\equiv&  {1\over \sqrt{2}}
   \left(\!\! \begin{array}{rr}
                     1 & 1 \\ - i & i
                     \end{array} \!\! \right)
 \left(\!\! \begin{array}{c} 
      \tilde \alpha_\varepsilon \\ \tilde  \alpha^\dagger_\varepsilon
      \end{array} \!\! \right) \;  
    \quad (\varepsilon \ge 0).
\end{eqnarray}
By  construction they are real ($\gamma_{n \nu} = \gamma^\dagger_{n  \nu}$,
$\tilde \gamma_{\varepsilon \nu} = \tilde \gamma^\dagger_{\varepsilon  \nu}$) 
and obey the anticommutation relations
\begin{eqnarray}
\{\gamma_{n \nu} , \gamma_{n' \nu'} \} &=& \delta_{n n'} \delta_{\nu \nu'} \;,
\label{app:anticomm1}
\\
\{\tilde \gamma_{\varepsilon \nu} , \tilde \gamma_{\varepsilon' \nu'} \} 
& = & \delta_
{\varepsilon \varepsilon'} \delta_{\nu \nu'} \;.
\label{app:newanticomm1}
\end{eqnarray}
When rewritten in terms of these Majorana operators, 
the original form (\ref{Hxeh}) 
for  $H_x$ becomes 
\begin{eqnarray}
    H_x &=& 
     \sum_{\bar k > 0 } \bar k (i \gamma_{\bar k +} \gamma_{\bar
      k -}  + 1/2)
    + |\varepsilon_d| (i \gamma_{d+} \gamma_{d-} + 1/2)
   \nonumber \\ &&
   + i \sum_{\bar k \ge 0} 2 V_{\bar k} \,  \gamma_{\bar k +} \gamma_{d-} \, 
   + \sum_{\bar k > 0}\bar k \beta_{\bar k}^\dagger
\beta_{\bar k} \; ,
\label{Hxmaj}
\end{eqnarray}
[with $\bar k$ and $V_{\bar k}$ given by Eqs.~(\ref{bark}) and~(\ref{V_k})],
and the  sought-after diagonalized form (\ref{Hdiagfer})
becomes 
\begin{equation}
    H_x =  \sum_{\varepsilon \ge 0} 
    \varepsilon \; (i \, \tilde \gamma_{\varepsilon  +} 
    \tilde \gamma_{\varepsilon -} + {\textstyle \frac12})
        + \delta E_G \; 
    + \sum_{\bar k > 0} \bar k \beta_{\bar k}^\dagger
    \beta_{\bar k} \, . 
\label{Hdiagmaj}
\end{equation}
To find the transformation that
brings (\ref{Hxmaj}) into the form (\ref{Hdiagmaj}), we make 
a Ansatz for the $\tilde \gamma_{\varepsilon \nu}$
[which by (\ref{gamman}) and (\ref{gammae}) is equivalent to
the Bogoliubov Ansatz~(\ref{2ck:bogolubov})]:
\begin{eqnarray}
\tilde \gamma_{\varepsilon \nu} \equiv
 \sum_{n\in\{d, \bar k\ge 0\}} B_{\varepsilon n \nu} \gamma_{n \nu} 
              \qquad (\nu = \pm)  \; .
\label{Bogolubov}
\end{eqnarray}
It suffices for the Ansatz to be linear, since $ H_x$ is quadratic in $
\gamma_{n \nu}$'s, and for it to be diagonal in the index $\nu$, since both
(\ref{Hxmaj}) and (\ref{Hdiagmaj}) are purely {\em off-diagonal}\/ in
$\nu$.  Since the orthonormality conditions~(\ref{app:newanticomm1}) imply
\begin{equation}
\label{Bnormal}
   \sum_n B_{\varepsilon n \nu} B_{\varepsilon' n \nu}
   = \delta_{\varepsilon \varepsilon'} \;  \qquad (\nu = \pm),
\end{equation}
the $B_{\varepsilon n \nu}$'s are orthogonal matrices [with matrix
indices $(\varepsilon,n)$], so that Eq.~(\ref{Bogolubov}) 
can trivially be inverted:
\begin{eqnarray}
 \gamma_{n \nu} =
 \sum_{\varepsilon \ge 0  } 
 B_{\varepsilon n \nu} \tilde \gamma_{\varepsilon \nu}
 \; ,   \qquad (\nu = \pm)  . 
\label{invertBogolubov}
\end{eqnarray}
We can deduce $\delta E_G$ even without 
having determined the $B_{\varepsilon n \nu}$ yet,
by inserting (\ref{invertBogolubov}) into $H_x$ to transform
(\ref{Hxmaj}) into (\ref{Hdiagmaj}):
since both equations
are off-diagonal in $\nu$,
no diagonal terms $\tilde \gamma_{\varepsilon \nu} 
\tilde \gamma_{\varepsilon \nu}$ (= 1, 
i.e.\ no constants) can arise, so that the constants in both  
equations must be equal; this yields Eq.~(\ref{2ck:EG}) for $\delta E_G$.

\subsection{Determination of $B_{\varepsilon n \nu}$'s and $\varepsilon$'s}
\label{B-and-es}

To determine the coefficients $B_{\varepsilon n \nu}$,
we substitute the Ansatz (\ref{Bogolubov}) it into
the Heisenberg equation
\begin{eqnarray}
 \mbox{[} \tilde \gamma_{\varepsilon \nu}, H_x \mbox{]}
 = \nu \; i 
\;\varepsilon 
\;\tilde \gamma_{\varepsilon - \nu} \qquad (\nu = \pm) \; ,
\label{anticomm}
\end{eqnarray}
[which follows from (\ref{app:newanticomm1}) and (\ref{Hdiagmaj})
and is equivalent to (\ref{Heisenberg})],
evaluate the commutator using (\ref{app:anticomm1})
and (\ref{Hxmaj}), and equate the coefficients of
$\gamma_{n \nu}$. 
This readily yields: 
\begin{mathletters}
\label{app-all}
\begin{eqnarray}
&&\varepsilon B_{\varepsilon\bar k +} = \bar k B_{\varepsilon\bar k -} +
 2 V_{\bar k}B_{\varepsilon d -}\;,
\label{app1}\\
&&\varepsilon B_{\varepsilon\bar k -} = \bar k B_{\varepsilon\bar k +
}\;, \label{app2}\\ 
&&\varepsilon B_{\varepsilon d +} = | \varepsilon_d | B_{\varepsilon d - }
\;, \label{app3}\\ 
&&\varepsilon B_{\varepsilon d -} = | \varepsilon_d | B_{\varepsilon d + }
+ \sum_{\bar k\ge 0} 2 V_{\bar k} B_{\varepsilon \bar k +}\;
\label{app4}.
\end{eqnarray}
\end{mathletters}
We consider the $\varepsilon\neq 0$ and $\varepsilon =0$ solutions 
separately.

\subsubsection{$\varepsilon \neq 0$ Solutions}

For $\varepsilon\not=0$ we write $B_{ \varepsilon d -} \! = \varrho(\varepsilon)
\, \varepsilon $, where $\varrho (\varepsilon)$ is a normalization factor to
be determined below.  Then Eqs.~(\ref{app-all})  
yield Eqs.~(\ref{allBeqs}) 
after some simple algebra. Substituting these
into (\ref{app4}) yields the eigenvalue equation
\begin{equation}
\label{eigen0}
S_1 (\varepsilon) = \varepsilon_d^2 / \varepsilon^2 - 1 \, ,
\end{equation}
where the infinite sum $S_1 (\varepsilon)$ can be 
evaluated as follows:
\begin{eqnarray}
\label{S1a}
  S_1 (\varepsilon) &\equiv &\sum_{\bar k \ge 0} {4 V^2_{\bar k}
    \over \bar k^2 - \varepsilon^2}
\\ 
&=& \sum_{n_{\bar k} \in {\mathbb Z}} 
   {- 4 V_0^2 / \varepsilon \over  \Delta_L (n_{\bar k} + \frac12
     - \frac{P}2)  + \varepsilon} 
\\
  &=& (4 \pi \Gamma / \varepsilon) 
  \tan [\pi \left( {\varepsilon/ \Delta_L} - P/2 \right)] .
\label{trick} \label{tanpi}
\end{eqnarray}
Equating Eqs.~(\ref{eigen0}) and (\ref{tanpi}) gives the 
eigenvalue equation~(\ref{2ck:eigenenergies})
that determines the allowed $\varepsilon$-values. 

The normalization factor $\varrho (\varepsilon)$ can be determined
by writing the  $\varepsilon \! = \! \varepsilon' \neq 0$, $\nu \! = \! +$ 
version of Eq.~(\ref{Bnormal})
 in the form [using Eq.~(\ref{allBeqs})] 
\begin{eqnarray}
\label{S2a}
 & &     \varrho^2 (\varepsilon) \left[ \varepsilon_d^2 + \varepsilon^2 
     S_{2+} (\varepsilon) \right]         = 1
\; ,
\\
\label{S2b}
& &    S_{2+} (\varepsilon) \equiv \sum_{\bar k \ge 0} {4 V^2_{\bar k}
       \varepsilon^2
    \over (\bar k^2 - \varepsilon^2)^2 }\; .
\end{eqnarray}
Noting from Eqs.~(\ref{S1a}) and (\ref{S2b}) that $S_{2+} (\varepsilon) =
{\textstyle \frac12} \varepsilon \partial_\varepsilon S_1 (\varepsilon)$,
evaluating this by reexpressing the derivative of $S_1$=(\ref{trick}) in terms
of $\tan[\,]$, and simplifying  using $\tan[\,] = (\varepsilon_d^2 - \varepsilon^2)/(\pi
\varepsilon \Gamma)$ [by (\ref{trick})=(\ref{eigen0})], one readily finds
Eq.~(\ref{varrho}) for $\varrho(\varepsilon)$.

Note that the eigenvalue equation (\ref{2ck:eigenenergies}) is symmetrical
under the transformation $\varepsilon \to -\varepsilon$ and therefore 
also has negative roots. However, 
 from Eqs.~(\ref{app-all}) 
the corresponding
coefficients are given by $B_{-\varepsilon n \nu} = \nu B_{\varepsilon n \nu}$,
thus the excitations corresponding to $\varepsilon$ and 
$-\varepsilon$ are not independent, but related by 
$
\tilde \alpha_{-\varepsilon} = \tilde \alpha_\varepsilon^\dagger \; . 
\label{dependence}
$
This confirms that only  non-negative eigenvalues 
need to be considered 
[as was intuitively obvious already when writing
down Eq.~(\ref{Hdiagfer})].

\subsubsection{$\varepsilon = 0$ Solutions}
\label{2ck:zeromodes}

There are three situations in which 
a root $\varepsilon_{j,P}$ of 
the eigenvalue equation (\ref{2ck:eigenenergies})
can be equal to zero:

(i) For arbitrary $\Gamma$, the lowest  root in
the $P=0$ sector, $\varepsilon_{0,0}$,
approaches 0 if and only if $\varepsilon_d \to 0$.
For $\varepsilon_d \neq 0$,
its asymptotic behavior close to  0 or $\Delta_L/2$
is as follows:
\begin{mathletters}
\label{e00-energy}
\begin{eqnarray}
\label{eq:e0asympt=0} 
{\varepsilon_{0,0} \over \Delta_L }  &\simeq &
{|\varepsilon_d | \over [\Delta_L^2 + 4 \pi^2 \Gamma
\Delta_L]^{1/2}}  \qquad \mbox{for}\quad 
{\varepsilon_{0,0} \over \Delta_L } \to 0  \; ,
\\
{\varepsilon_{0,0} \over \Delta_L }  & \simeq & { 1 \over 2}  - 
 {2 \Gamma \Delta_L \over 
\varepsilon_d^2  - \Delta_L^2/4} 
\qquad \quad \; \, \mbox{for}\quad 
 {\varepsilon_{0,0} \over \Delta_L } \to {1\over 2}
\; . 
  \label{eq:e0asympt=1/2} 
\end{eqnarray}
\end{mathletters}

For $\varepsilon_d = \varepsilon_{0,0} = 0$, the 
coefficients $B_{\varepsilon_{0,0}=0, n \nu}$ are 
simply the $\varepsilon_d \to 0$
limits of the $B_{\varepsilon_{0,0} n \nu}$ of Eqs.~(\ref{allBeqs}),
 \begin{eqnarray}
  \label{eq:p=0ed=0}
B_{\varepsilon_{0,0}  n +} =  \delta_{n d} \; , 
& \qquad &
B_{\varepsilon_{0,0} n  - } =  {\delta_{n d} - (2 V_{\bar k}/ \bar k) \,
\delta_{n \bar k} 
   \over  [1 + 4 \pi^2 \Gamma/ \Delta_L]^{1/2}},
\end{eqnarray}
reflecting that $\gamma_{d+}$ decouples from
$H_x$ for $\varepsilon_d = 0$.

(ii) The lowest root in the $P=1$ sector, $\varepsilon_{0,1}$,
identically equals 0 for all $\varepsilon_d = 0$ and  $\Gamma$. 
Solving the $\varepsilon =0$, $\Gamma \neq0$ 
versions of (\ref{app-all}) 
{\em directly}\/  (since (\ref{allBeqs}), 
derived for $\varepsilon \neq0$, 
 can {\em not}\/ be used here) gives
 \begin{eqnarray}
 \label{e00}
 B_{\varepsilon_{0,1} n +} = 
{2 V_0 \, \delta_{nd} - |\varepsilon_d| \, \delta_{n 0} 
\over [4 V_0^2 + \varepsilon_d^2]^{1/2}}  \; , 
& \qquad &
  B_{\varepsilon_{0,1} n -} =    \delta_{n 0}  ,
\end{eqnarray}
reflecting that 
$\gamma_{\bar k=0,-}$ decouples from $H_x$.

(iii) The second-lowest root in the $P=1$ sector, $\varepsilon_{1,1}$,
approaches zero if and only if $\Gamma \to 0$ at $\varepsilon_d = 0$
(to be precise, in this limit 
 $\varepsilon_{1,1} \simeq 2 \sqrt {\Gamma \Delta_L}$);
 this reflects the fact that $H_x$ has {\em two}\/  zero modes 
for $\varepsilon_d = \Gamma =0$, namely 
$\alpha_0$ and $\alpha_d$.
Note, though, that in this limit $\tilde \gamma_{\varepsilon_{1,1} \nu}$
[found using (\ref{allBeqs})] 
and $\tilde \gamma_{\varepsilon_{0,1}, \nu}$
do {\em not} reduce simply to 
$\gamma_{d \nu}$ and $\gamma_{k=0,\nu}$ but to linear combinations
of these (in Table~\ref{detailed}, this is indicated by braces).
In the opposite limit of $\Gamma / \Delta_L \to \infty$
at $\varepsilon_d = 0$, one has
$\varepsilon_{1,1} \simeq 
 \Delta_L({1 \over  2} - {\Delta_L \over 4 \pi^2 \Gamma})$.

\subsection{Consistency Checks}
\label{consistency}

Several consistency checks on the above solution are possible.
Firstly, let us check  Eq.~(\ref{Bnormal}):
In the special case that $\varepsilon$ or $\varepsilon'$ is 
$\varepsilon_{0,1} = 0$, Eq.~(\ref{e00}) is easily checked to
be consistent with Eq.~(\ref{Bnormal}). If
$\varepsilon$ and $\varepsilon' \neq 0$, 
one finds by
writing out   Eq.~(\ref{Bnormal}) for $ (\varepsilon = \varepsilon', \nu = -)$
and $(\varepsilon \neq \varepsilon', \nu = \pm)$, respectively,
that [analogously to (\ref{S2a}),(\ref{S2b})]
the following relations must hold: 
\begin{eqnarray}
\label{S22--}
{1 \over \varrho^2 (\varepsilon) \varepsilon^2 } -1 =
\sum_{\bar k \ge 0} {4 V_{\bar k}^2
       \, \bar k^2
       \over (\bar k^2 - \varepsilon^2)^2 }
&\quad &\left[\equiv S_{2-} (\varepsilon)\right],
\\
\nonumber
- { \varepsilon_d^2 \over \varepsilon \varepsilon' }
=
 \sum_{\bar k \ge 0} {4 V_{\bar k}^2
       \, \varepsilon \varepsilon'
       \over (\bar k^2 - \varepsilon^2)(\bar k^2 - {\varepsilon'}^2) }
&\quad& \left[\equiv S_{3+}  (\varepsilon, \varepsilon')\right],
\\
\nonumber
-1 = 
\sum_{\bar k \ge 0} {4 V_{\bar k}^2
       \, \bar k^2
       \over (\bar k^2 - \varepsilon^2)(\bar k^2 - {\varepsilon'}^2) } 
&\quad &  \left[ \equiv S_{3-} (\varepsilon, \varepsilon') \right].   
\end{eqnarray}
One can verify that indeed they do, 
by noting from Eqs.~(\ref{S1a}), (\ref{S2b}) that the sums defined above can be
rewritten as 
$S_{2-} = S_1 + S_{2+}$ and 
$$
\nonumber
S_{3\pm}  (\varepsilon, \varepsilon') =
\frac12 \left[
  {\varepsilon S_1 (\varepsilon) - \varepsilon' S_1 (\varepsilon') 
    \over \varepsilon - \varepsilon'}
  \mp  
  {\varepsilon S_1 (\varepsilon) + \varepsilon' S_1 (\varepsilon') 
    \over \varepsilon + \varepsilon'}
  \right]
$$
and simplifying these using 
Eqs.~(\ref{eigen0}) and~(\ref{S2a}). 

Secondly, one can verify explicitly that our transformation does
indeed diagonalize $H_x$: 
insert the inverse  Bogoliubov transformation
(\ref{invertBogolubov}) and  
Eqs.~(\ref{allBeqs}) 
for the coefficients $B_{\varepsilon n \nu}$
into the original form (\ref{Hxmaj}) for $H_x$ and 
express the resulting $\sum_{\bar k}$ sums in terms of 
$S_1$, $S_{2-}$ and $S_{3-}$: 
\begin{eqnarray}
\nonumber
\lefteqn{  H_x =  |\varepsilon_d|/2 + \sum_{k > 0}  k/2  
   + \sum_{\bar k > 0}\bar k \beta_{\bar k}^\dagger \beta_{\bar k} }
\\
& & 
+   \sum_{\varepsilon ,\varepsilon' > 0} 
  i \tilde \gamma_{\varepsilon +} \tilde \gamma_{\varepsilon' -}
  \, \varrho (\varepsilon) \varrho (\varepsilon') \, \varepsilon'
  \Biggl[
   \varepsilon_d^2
    - \varepsilon^2 S_1 (\varepsilon) 
\Biggr.
\\ && 
\nonumber
        \Biggl.
 + \, \varepsilon^2 \biggl(\delta_{\varepsilon \varepsilon'}
    S_{2-} (\varepsilon) + (1 - \delta_{\varepsilon \varepsilon'})
    S_{3-} ( \varepsilon, \varepsilon')     \biggr)
    \Biggr].
\end{eqnarray}
(The terms with $\varepsilon$ or $\varepsilon' = 0$
can be checked to be zero.) 
Evaluating this
using Eq.~(\ref{eigen0}) and the equations for
$S_{2-}$ and $S_{3-}$ readily yields the sought-after diagonal form
(\ref{Hdiagmaj}) for $H_x$ and confirms Eq.~(\ref{2ck:EG})
for $\delta E_G$.

\subsection{Ground State Energy Shift  $\delta E_G$}
\label{2ck:calcEG}

We now show how to calculate the ground state energy shift 
$\delta E_G \equiv
 \delta E_G^0 + P \delta E_G^P$ of
(\ref{2ck:EG}), for $\Gamma /\Delta_L \gg 1$ and both $T_h=0$ and
$T_h / \Delta_L \gg 1$, i.e.\ we derive Eqs.~(\ref{sumsPP})
and (\ref{EG0h=0}). 
As explained in Section~\ref{groundstateshift},
the coefficient of the $P$-dependent term,  $\delta E_G^P$, 
must be  extracted  with care to obtain the correct
finite-size spectrum. 

In the notation of Eq.~(\ref{2ck:epsappr}), 
Eq.~(\ref{2ck:EG}) becomes
\begin{eqnarray}
\nonumber
\delta 
E_G &=& {1 \over 2} \left[
 \sum_{j=1}^{N_{\rm max}} 
\Delta_L 
 \Bigl( j - {1 \over 2} -{ P \over 2} \Bigr) + 
|\varepsilon_d| 
  - \sum_{j=0}^{N_{\rm max}} 
\varepsilon_{j,P}  \right] 
\\
&=& {1 \over 2} \left[ |\varepsilon_d| - \varepsilon_{0,P} - \Delta_L
  \sum_{j=1}^{N_{\rm max}}  \delta_{j,P} \right]
\label{sumf}
\end{eqnarray}
where we introduced a ``band cutoff'' $N_{\rm max} \equiv D/\Delta_L $
to regularize the sum (with $D \sim 1/a$). The task at hand
is to perform the sum on $j$ sufficiently carefully to extract
its leading $P$-dependence.

\subsubsection{Zero magnetic field}
 
We first  consider the case $\Gamma / \Delta_L \gg 1$ and
$T_h = 0$.
To isolate the $P$-dependence of $\delta_{j,P}$, we write 
Eq.~(\ref{delta-shift}) as
\begin{eqnarray}
\label{shift-g}
\delta_{j,P} &=& {1 /2} + g(j - {P / 2} + \delta_{j,P}) , 
\\
  \label{eq:gdefine}
   g(x) & \equiv &  - {1 \over \pi} \arctan 
   \left[{ \Delta_L (x - 1/2 ) \over 4 \pi \Gamma } \right] \; ,
\end{eqnarray}
and solve  (\ref{shift-g}) for $\delta_{j,P}$ by
expanding its r.h.s.\ in the small parameter
$ (\delta_{j,P} - P/2) \Delta_L / \Gamma \ll 1$, finding 
\begin{equation}
\delta_{j,P}  =  {1/2 + g(j) \over 1 - g^\prime (j)}
-  g^\prime (j) {P \over 2} 
+ {\cal O}\left({\Delta_L^2 \over \Gamma^2 }\right)\; .
\label{app:fappr}
\end{equation}
The first term is $P$-independent and gives the leading
contribution to $\delta E_G^0$.
The second term  is
${\cal O}\left({\Delta_L / \Gamma }\right) $, contains the
full $P$-dependence of $\delta_{j,P}$ to this order and
contributes to $\delta E_G^P$.  Inserting
(\ref{app:fappr}) into (\ref{sumf}) gives
\begin{eqnarray}
\label{EG0-explicit}
\delta E_G^0 & = &   - {\Delta_L \over 2}
        \sum_{j=1}^{N_{\rm max}}
{1/2 + g(j) \over 1 - g^\prime (j)}  
\\
\nonumber 
& = & 
-  \Gamma \left[ \int_0^{D/4 \pi \Gamma} \!\! d y \, 
 [\pi - 2 \, \mbox{arctan} \, y ] + {\cal O} (\Delta_L / \Gamma)\right]
\\
&=& 
-2 \Gamma \Bigl[ \ln(D/4 \pi \Gamma) + 1 + 
{\cal O}\left(\Delta_L/ \Gamma, {\Gamma/D}\right) \Bigr] \; ,
\end{eqnarray}
while the $P$-dependent part,  $\delta E_G^P$ is equal to 
\begin{eqnarray}
& &  -   \varepsilon_{0,P} /2
+ (\Delta_L  / 4) \sum_{j=1}^{N_{\rm max}}  g'(j) 
\\ & \approx  & 
- \varepsilon_{0,P} / 2 + (\Delta_L  /4) \left[ g(N_{\rm max} ) - g(1)
        \right] 
\label{app:2ck:EG-semifinal}
\\
 & \approx & - \Delta_L \left[ 1 / 8 + {\cal O} (\Delta_L / \Gamma)
        \right] \; ,
\label{app:2ck:EG}
\end{eqnarray}
where for (\ref{app:2ck:EG}) we used $\varepsilon_{0,P} = 0$
for $T_h=0$ [by (\ref{eq:zeromodes})]. 

\subsubsection{Large magnetic field} 

Next we consider the case $\Gamma / \Delta_L \gg 1$ and
$T_h / \Delta_L \gg 1$ (for arbitrary  $ T_h / \Gamma$).
This can be treated analogously, except  that now 
(\ref{eq:gdefine}) must be replaced by [from (\ref{delta-shift})]
\begin{eqnarray}
\nonumber
   g (x) & \equiv &   {1 \over \pi} \arctan 
   {1 \over 4 \pi } \left[ {T_h \over   \Delta_L (x - 1/2)} 
- { \Delta_L (x - 1/2 ) \over  \Gamma } \right]  
\end{eqnarray}
[thus $g' (x)$ is of order 
$ {\cal O} (\Delta_L / \Gamma, \Delta_L /T_h) $ for all  $x \ge 1$]. 
Since  now $\varepsilon_{0,P} = \Delta_L (1 - P)/2$
[by (\ref{eq:zeromodes}) and (\ref{eq:e0asympt=1/2})],  
Eq.~(\ref{app:2ck:EG-semifinal}) now yields
$ \Delta_L [1/2 +  {\cal O} (\Delta_L / \Gamma,
\Delta_L/ T_h)] \; ,
$
which is $P$-{\em in}\/dependent. This implies that 
$\delta E_G^P = 0$ to this order, in other words that
for $T_h, \Gamma \gg \Delta_L$ (and independent
of the ratio $T_h/\Gamma$) the ground
state energy shift $\delta E_G$ is no longer $P$-dependent. 

The $P$-independent part of the shift, $\delta E_G^0$,  can be obtained from
(\ref{EG0-explicit}), plus the $|\varepsilon_d|/2$ of (\ref{sumf}):
\widetext
\begin{eqnarray}
\delta E_G^0 & = &  
\nonumber 
{|\varepsilon_d| \over 2}
-  \Gamma \left[ \int_0^{D/4 \pi \Gamma} \!\! d y \, 
 \left[\pi - 2 \, \mbox{arctan} \! \left(y - {T_h \over
16 \pi^2 \Gamma y}\right) \right] + {\cal O} (\Delta_L / \Gamma, \Delta_L/T_h)
\right]
\\
&=& 
\label{EG0-hhh}
-2 \Gamma \Bigl[ \ln(D/ |\varepsilon_d| ) + 1 + 
{\cal O}\left(\Delta_L/ \Gamma,  \Delta_L/T_h, {\Gamma/D}, 
T_h / D , \Gamma / T_h \right) \Bigr] \; .
\end{eqnarray}
Although the integral can be evaluated for arbitrary 
values of its parameters using 
\begin{eqnarray}
%\nonumber
  \int d y \,  \mbox{arctan} (y - b/y) &=&
    y  \, \mbox{arctan}  (y - b/y) 
 - {1 \over 4} \ln (y^4 + y^2 -2y^2 b + b^2) 
%\\ & & 
\, - \, 
{1 \over 2} \sqrt{4b-1} \,  \mbox{arctan} \!
\left({1 + 2y^2 - 2b \over \sqrt{4 b -1}} \right) \;  ,
  \label{eq:hintegral-explicit}
\end{eqnarray}
\narrowtext
\noindent
we gave in (\ref{EG0-hhh}) only the limit of large magnetic
fields, $T_h / \Gamma  \gg 1$.

\subsection{Fluctuations in ${\cal N}_x$}
\label{app:Nxfluctuations}

The results for 
$\langle \hat {\cal N}_x \rangle$ 
and  $\langle \hat  {\cal N}_x^2 \rangle$ 
discussed 
and interpreted in Section~\ref{physicalffs}
are  obtained as follows. We consider only  $\varepsilon_d = 0$
and the states  $|\tilde 0 \rangle_{{\cal S}_{\rm ext}}$ and 
$\tilde \alpha^\dagger_{\varepsilon_{0,P} }|\tilde 0 \rangle_{{\cal S}_{\rm ext}}$,
which represent,  respectively, the physical ground states in sectors with
excitation parity ${\cal P}_{\tilde E} = 0$ and 1.
We express
$\hat {\cal N}_x$ through  $\alpha_{\bar k}$ and  $\beta_{\bar k}$, 
using (\ref{xN_k}) and (\ref{ddeeffalphabeta}):
\begin{equation}
  \label{eq:Nxalphabeta}
  \hat {\cal N}_x =  \sum_{\bar k>0} 
  i (\alpha^\dagger_{\bar k} \beta_{\bar k}
  - \beta^\dagger_{\bar k} \alpha_{\bar k} ) +
  P (1/2 - \alpha_0^\dagger \alpha_0 ) \; .
\end{equation}
Now, in a  $P=1$ sector, we have [analogously to (\ref{eq:deltaEmin})]
\begin{eqnarray}
\nonumber
  1/2 - \langle \alpha_0^\dagger \alpha_0 \rangle
& = & { 1 \over 2} \Bigl[
   \pm B_{\varepsilon_{0,1} 0+} B_{\varepsilon_{0,1} 0-} + 
 \sum_{\varepsilon > \varepsilon_{0,1}  }
   B_{\varepsilon 0 +} B_{\varepsilon 0 -} \Bigr] \\
& = &  0 ,  
\label{eq:Nxexpect} 
\end{eqnarray}
for all $\Gamma/\Delta_L$ [using (\ref{allBeqs}) and (\ref{e00})].
(Here and below the upper or lower signs in $\pm$ (and $\mp$)  refer to  
${\cal P}_{\tilde E} = \langle \tilde \alpha^\dagger_{\varepsilon_{0,P}} 
\tilde \alpha_{\varepsilon_{0,P}} \rangle
= 0$ or 1.) 
Since moreover $\beta_{\bar k} | \tilde 0 \rangle_{{\cal S}_{\rm ext}} = 0$,
we conclude from (\ref{eq:Nxalphabeta}) and (\ref{eq:Nxexpect}) that 
$ \langle   \hat {\cal N}_x \rangle = 0$ for both $P=0$ and 1
and  all $\Gamma/\Delta_L$. 

The calculation of $\langle \hat {\cal N}_x^2 \rangle $
is more involved:
\begin{eqnarray}
\nonumber
\lefteqn{  \langle \hat {\cal N}_x^2 \rangle + P/4 
  = \sum_{\bar k \ge 0}
  \langle \alpha^\dagger_{\bar k} \alpha_{\bar k} \rangle }
\\
\nonumber 
 &=& { 1 \over 2}
 \sum_{\bar k \ge 0} \Bigl[ 1 \mp  
   B_{\varepsilon_{0,P} \bar k +}   B_{\varepsilon_{0,P} \bar k -}
   - \sum_{j > 0}
   B_{\varepsilon_{j,P} \bar k +}   B_{\varepsilon_{j,P} \bar k -} \Bigr]
\\
\nonumber
 &=&   { 1 \over 2}
 \sum_{\bar k \ge 0} \Bigl[ 1 - \sum_{\varepsilon > 0}
 \rho^2(\varepsilon)  \varepsilon^2 
{ 4 V_{\bar k}^2 \varepsilon \bar k \over
  (\varepsilon^2 - \bar k^2 )^2} \Bigr] 
\\
&=&
 \label{eq:DeltaNx-4}
  \sum_{\varepsilon > 0}
 \rho^2(\varepsilon)  \varepsilon^2 
\Bigl[ { 1 \over 4} + \sum_{\bar k \ge 0}
{  V_{\bar k}^2 \over
  (\bar k + \varepsilon )^2} \Bigr]
+ {1 \over 2} \Bigl[ \sum_{\bar k \ge 0} -  \sum_{\varepsilon > 0} \Bigr] .
\end{eqnarray}
For the first equation we used (\ref{eq:Nxalphabeta}), 
(\ref{eq:Nxexpect}); for the second
(\ref{ddeeffalphabeta}),
(\ref{gamman}), (\ref{invertBogolubov}), (\ref{gammae});
for  the third 
(\ref{allBeqs}), (\ref{eq:p=0ed=0})
or (\ref{e00}); we tamed the ``divergence'' at $\varepsilon \simeq
\bar k$ using 
\begin{eqnarray}
\nonumber 
  {\varepsilon \bar k \over (\varepsilon^2 - \bar k^2)^2}
= {\bar k^2 \over (\bar k^2 - \varepsilon^2)^2}
- {1 \over 2 (\bar k^2 - \varepsilon^2)}
- {1 \over 2 (\bar k + \varepsilon)^2 }
\end{eqnarray}
and performing  the $\sum_{\bar k}$ sums over the first two terms
using (\ref{eigen0}), (\ref{S22--}) and  (\ref{S1a}), thus obtaining  
(\ref{eq:DeltaNx-4}), in which the last two 
(diverging) terms cancel exactly.

The limit $\Gamma/\Delta_L \to 0$ of (\ref{eq:DeltaNx-4})
yields $\langle \hat {\cal N}_x^2 \rangle \to P/4 $
[for $P=1$ the mode $\varepsilon_{1,1} \to 2 \sqrt {\Gamma \Delta_L}$
(compare Section~\ref{2ck:zeromodes}) 
makes a non-zero contribution]. To obtain the
leading behavior of (\ref{eq:DeltaNx-4}) in the opposite limit
$\Gamma/\Delta_L \gg 1$, we evaluate the sums as integrals:
\begin{eqnarray}
\nonumber
\langle \hat {\cal N}_x^2 \rangle & \approx &
\int^\infty_{\Delta_L} \! d \varepsilon
{2 \Gamma \over {1 \over 4} \varepsilon^2 + 4 \pi^2 \Gamma^2}
\left[ {1 \over 4}  + 
\int^\infty_0 \! d \bar k
 {2 \Gamma \over (\bar k+ \varepsilon)^2} \right]
\\
& \approx &
{1 \over \pi^2} \ln (\Gamma L) \; + \; {\cal O} [(\Delta_L/ \Gamma)^0]  \; .
\end{eqnarray}

\subsection{Perturbing around EK line by $\delta H'_z$}
\label{app:pertEK}

In this section we determine the
scaling dimension $\gamma_{\delta \lambda_z}$
of the operator  $\delta H'_z$ of (\ref{dev:EKline}),
which  arises as soon as one leaves the EK line,
i.e.\ when $\lambda_z = 1 + \delta \lambda_z$.
To this end we perturbatively calculate,
 in a given
subspace ${\cal S}_{\rm phys} (S_T, {\cal N}_c,  {\cal N}_f)$
at $\varepsilon_d = 0$, 
the level shifts $\delta \tilde {\cal E} (L)$ [see (\ref{deltauniversal})]
induced by $\delta H'_z$. 
As first step, we express  $\delta H'_z$
in terms of operators  that diagonalize $H'(\delta \lambda_z = 0)$:
\begin{eqnarray}
\label{partphisq}
\partial_x \varphi_s (x) &= &
\Delta_L \sum_{n_q >0 }
\sqrt{n_q}\,  i \,  (b_{qs} - b_{qs}^\dagger) \; ,
\\
\label{NsSzvare}
 \hat {\cal N}_s S_z & = & S_T S_z - 1/4 \; , 
\\
\label{SzatNFL}
S_z  &=& 
   {  1 \over 2} \sum_{\varepsilon \varepsilon'}
    B_{\varepsilon d +} B_{\varepsilon' d -}  
   ( \tilde\alpha_\varepsilon^\dagger +  \tilde\alpha_\varepsilon)
   ( \tilde\alpha_{\varepsilon'}^\dagger -
   \tilde\alpha_{\varepsilon'}) \; .
\end{eqnarray}
[Eq.~(\ref{partphisq}) follows from  (\ref{bosonfieldsa}) and
(\ref{2ck:varphidef}),
Eq.~(\ref{NsSzvare}) from (\ref{2ck:tildeNS}),
and Eq.~(\ref{SzatNFL}) from  (\ref{definec_d}), (\ref{defalphad}),
(\ref{gamman}), (\ref{invertBogolubov}) and (\ref{gammae}).]
These relations show that although the NFL fixed point spectrum
is highly degenerate, there is no need for 
degenerate perturbation theory, because
 $\langle \tilde E| \delta H'_z | \tilde E'
\rangle = 0$ whenever $| \tilde E \rangle$ and $|\tilde E' \rangle$ 
are degenerate but distinct eigenstates of $H'$
[by inspection; compare Table~\ref{2ckbigtable}]. 
To first order in $\delta \lambda_z$, the dimensionless energy shift 
$\delta \tilde {\cal E} (L)$ of (\ref{deltauniversal}) due
to $\delta H'_z$ is thus simply given by 
\begin{eqnarray}
\nonumber
\delta \tilde {\cal E} (L) &=&
\left[ \langle \tilde E | \delta H'_z | \tilde E \rangle 
- { \langle \tilde E}_{\rm min} | \delta H'_z | \tilde E_{\rm min}
 \rangle \right] / \Delta_L
\\
  \label{eq:shiftHz}
&=& \delta \lambda_z S_T \left[
\langle \tilde E | S_z| \tilde E \rangle 
- { \langle \tilde E}_{\rm min} | S_z | \tilde E_{\rm min}
 \rangle \right] .
\end{eqnarray}
(Note that since $\partial_x \varphi_s$ [by  (\ref{partphisq})]
 is linear in $b_{qs}$ and $b^\dagger_{qs}$, 
in (\ref{eq:shiftHz}) only in the second, ``zero-mode term'' of
$\delta H'_z$ 
contributes, which does not
occur in the continuum limit considered in Ref.~\onlinecite{EK92}.)
Now, $ \tilde E_{\rm min}$ corresponds to the physical
ground states in the two sectors ${\cal S}_{\rm phys} (S_T =  \pm
1/2,{\cal N}_c =  0, {\cal N}_f = 0)$
[with ${\cal E}_{\rm NFL} = 0$ in Table~\ref{2ckbigtable}], namely
\begin{eqnarray}
  \label{eq:groundstates}
  | \tilde 0 \rangle_{{\cal S}_{\rm ext}(1/2,0,0)} \; , 
\qquad
 \tilde\alpha^\dagger_{\varepsilon_{0,0}}
 | \tilde 0 \rangle_{{\cal S}_{\rm ext}(-1/2,0,0)} \; .
\end{eqnarray}
From (\ref{SzatNFL}), one readily finds for these, respectively,
\begin{equation}
  \label{eq:deltaEmin}
S_T \langle S_z \rangle  = \pm  { 1 \over 4}\Bigl(\pm 
     B_{\varepsilon_{0,0} d +}     B_{\varepsilon_{0,0} d -} +
     \sum_{j > 0}
   B_{\varepsilon_{j,0} d +}     B_{\varepsilon_{j,0} d -} \Bigr) .
\end{equation}
Evaluating this using (\ref{B_ed-}) and (\ref{eq:p=0ed=0}),
the first term yields  $  { 1 \over 4}
   [1 + 4 \pi^2 \Gamma/\Delta_L]^{-1/2} $ for both cases,
and the second term vanishes for $\varepsilon_d = 0$.
Thus the two-fold ground state degeneracy
persists, in keeping with the fact
that  $\delta H'_z$ respects spin reversal symmetry. 

The first four excited
states in Table~\ref{2ckbigtable}, with $ = {1
  \over 8}$, all have  $S_T = 0$, hence 
$\langle S_T S_z \rangle = 0$. 
They are thus not shifted
by $\delta H'_z$ themselves. Nevertheless, their 
{\em relative shift}\/ w.r.t.\ the ${\cal E}_{\rm NFL} = 0$  
physical ground states just discussed 
 is non-zero, since the ground states were shifted upwards:
by  (\ref{eq:shiftHz}), it is 
\begin{eqnarray}
  \label{eq:shift1/8}
\delta \tilde {\cal E} (L) = - \, \delta \lambda_z 
{\textstyle  { 1 \over 4}}
   [1 + 4 \pi^2 \Gamma/\Delta_L]^{-1/2}  \sim L^{-1/2} .
\end{eqnarray}
It follows [from (\ref{deltauniversal})] that the sought-after scaling
dimension of $\delta H'_z$ is $\gamma_{\delta \lambda_z} = 1/2$.
Thus this perturbation is irrelevant, and the RG flow in
the vicinity of the EK-line is always towards it.
[It is easy to check that all  10 of
the next-higher excited states in Table~\ref{2ckbigtable},
with ${\cal E}_{\rm NFL} = {1 \over 2}$, have $\delta \tilde {\cal E}
(L) =  - \delta \lambda_z  {\textstyle  { 1 \over 2}}
   [1 + 4 \pi^2 \Gamma/\Delta_L]^{-1/2}$, which
again is $ \sim L^{-1/2} $, as expected.]

Let us now turn
on a local magnetic field $\varepsilon_d = h_i$,
in which case the second term of  (\ref{eq:deltaEmin}) is
non-zero and contributes to lifting the ground state
degeneracy. In the continuum limit $L\to \infty$ 
(so that  $\Delta_L \ll h_i, \Gamma$),  it in fact gives
a much larger contribution than the first term of (\ref{eq:deltaEmin}), namely
$  \delta \tilde {\cal E} (L)  = \pm {\delta \lambda_z \over 4} I
(\bar h)$, where $\bar h \equiv |h_i|/(2 \pi \Gamma)$, 
and $I(\bar h)$ is given, after the substitution
$x = \varepsilon^2_{j,0}/\varepsilon_d^2$,  by the following integral:
\begin{eqnarray}
  \label{eq:EKperthh}
I(\bar h) &=&   {\bar h \over \pi}
  \int_0^\infty  dx {1 \over 
      \bar h^2 (x - 1)^2 / 2  + 2x }  
\\ 
\nonumber
& = &  \left\{ \!\! 
\begin{array}{ll} 
{\displaystyle
\quad \stackrel{\bar h \to 0}{\longrightarrow}
  \vspace*{4mm} - {2 \over \pi} \bar h \ln(\bar h/2) \; ; } 
\phantom{\rule[-5mm]{0mm}{4mm}}
\\
{\displaystyle \quad \stackrel{\bar h \to \infty}{\longrightarrow}
 1 - {2 \over \pi \bar h} \; . } 
\end{array} 
\right. 
\end{eqnarray}

\section{The single-channel Kondo model}
\label{app:1ck}

This Appendix deals with the
 anisotropic {\em single}\/-channel Kondo
(1CK) model, which is of interest not only as the most
basic Kondo model, but also since it is equivalent
to a dynamic two-state system coupled to an ohmic
environment.\cite{Costi} 
 We shall solve the 1CK model 
 along the so-called Toulouse line,\cite{Toulouse,EKrev}
the 1CK analog of the EK line, calculating the crossover 
of the finite-size spectrum from the free Fermi liquid fixed
point to the strong-coupling Fermi liquid fixed point, well-known from
Wilson's NRG calculations.\cite{Wilson} 
Since the 1CK calculation is a straightforward
adaption of that developed above for the 2CK case,
it will be presented in less detail than the latter, though technical
differences will be pointed out.

\subsection{Conserved Quantum Numbers}

The 1CK model is defined by Eqs.~(\ref{hkinetic}) or (\ref{H0a}) for
$H_0$, (\ref{H_int1D}) for $H_{\rm int}$ and (\ref{Hh}) for $H_h$, the
only difference being that the channel index only has the value $j=1$
and hence can be dropped throughout.
To exploit the fact that the total charge is conserved,
we transform from the $\alpha = (\uparrow, \downarrow)$ basis to a 
$y = (c, s)$ basis by writing:
\begin{mathletters}
\label{EK-1cktransform}
\begin{eqnarray}
b^\dagger_{ qc/s} & \equiv & ( b^\dagger_{q \uparrow } \pm  
b^\dagger_{q \downarrow})/\sqrt{2}\;,
\label{1ck:transformationb}
\\
\varphi_{c/s}(x) &\equiv & 
(\phi_{\uparrow}(x) \pm \phi_{\downarrow}(x) )
/ \sqrt 2  \; .
\label{1ck:varphi_c,s}
\\
\label{1ck:transformationN}
\hat{\cal N}_{c/s} &\equiv & ({\hat N}_\uparrow \pm
{\hat N}_\downarrow) /2 \; ,
\label{1ck:tr}
\end{eqnarray}
\end{mathletters}
Note that  the  normalization constants
in Eqs.~(\ref{1ck:transformationb}) and {\ref{1ck:varphi_c,s})
differ from that of (\ref{1ck:transformationN})
[this contrasts with the 2CK case, and affects many
of the equations below]:
the $1/\sqrt 2$ in the former ensures that the transformations for
$b_{qy}^\dagger$ and $\varphi_y$ are unitary,
so that these operators satisfy commutation relations analogous
to those of  $b_{q \alpha}$ and $\phi_\alpha$
[namely (\ref{bqcom}) and (\ref{[phi,dxphi]})];  
the $1/2$ in (\ref{1ck:transformationN}) ensures that 
$\hat {\cal N}_c$  and
$\hat {\cal N}_s$ can be interpreted as  half the total charge and the
total electron spin, whose eigenvalues are
are either  both integers or both half-integers
(whereas a $1/\sqrt 2$ in (\ref{1ck:transformationN})
would have yielded irrational eigenvalues):
\begin{equation}
\vec {\cal N}\equiv ({\cal N}_c,{\cal N}_s ) \in ({\mathbb Z} + P/2)^2 \; . 
\label{1CKgluing}
\end{equation}
Here the {\em parity index}\/ $P$ equals 
0 or 1 if the  total number of
electrons is even or odd, respectively. 
Eq.~(\ref{1CKgluing}) is the
{\em free gluing condition}\/ for the 1CK model. 

Evidently, the total charge ${\cal N}_c$ and the total spin,
\begin{equation}
\label{deftildeNs}
S_T =   {\cal N}_s + S_z \; ,
\end{equation}
are conserved, where  (\ref{deftildeNs}) will be called the
{\em spin-conservation condition.}\/ Hence 
we can restrict our attention to the invariant subspace 
\begin{equation}
{\cal S}_{\rm phys} (S_T,{\cal N}_c )  \equiv 
\{ | {\cal N}_{c}, S_T - {\textstyle {1\over2}}; \Uparrow \rangle 
  \oplus  
  | {\cal N}_{c}, S_T + {\textstyle {1\over2}} ; \Downarrow \rangle\} \; 
  \; . 
\label{1ck:S}
\end{equation}
The difference between  Eqs.~(\ref{1ck:S}) and
(\ref{2ck:S})  makes explicit a major difference 
between the 1CK and 2CK models: though for
both the quantum number ${\cal N}_s$ fluctuates ``mildly''
between $S_T \mp 1/2$,
the {\em 1CK model lacks a ``wildly'' fluctuating quantum number}\/
such as ${\cal N}_x$; this is the ``deep reason''
why it also lacks NFL behavior. 

In the new charge-spin basis, the bosonized form of
the 1CK Hamiltonian takes the following simple form:
\begin{eqnarray}
\nonumber
 H_0 &=& \Delta_L \left[
         \hat {\cal N}_c (1 - P_0) +
    \hat {\cal N}_c^2 +     \hat {\cal N}_s^2 \right] 
\\
& & + \sum_{ q>0} q \, (b^\dagger_{q c} b_{q c} 
+ b^\dagger_{q s} b_{q s} ) \; ,
\label{1ck:H0y}
\label{1ck:H0bos}
\\
\label{1ck:bosonh}
  H_z &= & \lambda_z    \left[
  \partial_x \varphi_s (0) / \sqrt 2   \, + \,  \Delta_L  \hat {\cal N}_s 
 \right] \, S_z \; , 
\\
  H_\perp  &=& {\lambda_\perp \over 2a} \biggl[ 
e^{- i \sqrt{2} \varphi_s (0)} S_+ F^\dagger_{\downarrow }
F_{\uparrow} + \mbox{h.c.} \biggr] \; . 
\label{1ck:Hperpbos}
\end{eqnarray}

\subsection{EK transformation}

To simplify $H_z$, we use the same Emery-Kivelson
transformation $U(\lambda) = e^{i \gamma S_z \varphi_s (0)}$
 of (\ref{1ck:ektr}) as for the 2CK model,
 but now with $\varphi_s(x)$ given by
Eq.~(\ref{1ck:varphi_c,s}) instead of 
(\ref{2ck:varphidef}). 
The impurity spin operators $S_\pm$ and the spin field
$\varphi_s$ transform  according to 
Eqs.~(\ref{S^pm'}) and (\ref{varphi_s'}), just as in the 2CK case, 
but in contrast to the latter, 
\begin{eqnarray}
& & U (H_0 + H_z) U^{-1}
  = H_0 + (\lambda_z / \sqrt 2  - \gamma) 
\partial_x\varphi_s(0) S_z\nonumber \\
&&+  \lambda_z  \Delta_L  
        {\hat {\cal N}}_s\; S_z
+ \mbox{const}\; ,
\label{H_z'}
\\
 \label{eq:EK-psi-1ck}
& &  \psi_{\alpha } (x) \to   \psi_{\alpha } (x)
e^{i \sqrt 2 \alpha \gamma S_z  \arctan (x/a)} \;, \quad (|x| \ll L) \; .
\end{eqnarray}
Moreover, since in the 1CK case
the spin density is $\partial_x\varphi_s(x) / (2\pi
\sqrt 2) $, a spin $- \gamma S_z / \sqrt 2$ from the
conduction band is tied to the impurity.  To eliminate the $S_z
\partial_x \varphi_s$ term, we choose $ \gamma \equiv
\lambda_z/\sqrt{2}$ (in contrast to $\gamma \equiv \lambda_z$ for the
2CK case).  Then 
\begin{eqnarray}
\label{1ck:lambda=0}
\lefteqn{\hspace{-5mm} H'(\lambda_\perp = 0)  = \Delta_L \left[
  \hat {\cal N}_c( {\cal N}_c + 1 - P_0) + 
\hat {\cal N}_s^2 +  \lambda_z \hat N_s S_z \right] }
  \nonumber \\
        &+& \sum_{  q>0} q \, (  b^\dagger_{q c} b_{q c} + 
b^\dagger_{q s} b_{q s}) + 
        H_h + \mbox{const}\; ,
\label{1ck:Hgamma=0}
\end{eqnarray}
and $H'_\perp$ contains the 
factors $e^{\pm i (\sqrt 2 - \lambda_z/ \sqrt2) \varphi_s
(0)}$.  

These factors simplify for two special values of
$\lambda_z$. The first case,  $\lambda_z=2$ (i.e.\ $\gamma=\sqrt{2}$), is
called the {\it decoupling point,}\/ since the $\varphi_s$-dependence
drops out completely:
\begin{equation}
H_\perp^\prime 
= {\textstyle \lambda_\perp \over 2a}\bigl(S_+ F_\downarrow^\dagger
F_\uparrow + \mbox{h.c.}\bigr)\;.
\label{decoupling}
\end{equation}
In this case, the spin $- \gamma S_z / \sqrt 2$
from the conduction band that is tied to the impurity
is precisely $-S_z$, thus we have perfect screening.
Indeed, by (\ref{eq:EK-psi-1ck})
the phase shift $\delta$ in  $ \psi_{ \alpha } (0^-) \equiv
 e^{i 2 \delta} \psi_{\alpha } (0^+)  $ is $|\delta|  = \pi / 2$, 
corresponding to the unitarity limit. The dynamics of the 
electron-hole excitations described by the $\varphi_{s,c}$
fields evidently decouples from $S_z$ [by (\ref{decoupling})]. 
Thus it is trivial to find the spectrum,
which turns out to 
coincide with the fixed-point spectrum shown 
in the strong-coupling limit of Fig.\ref{1ck:scal}.
Note, incidentally, that at the decoupling point
the model can be mapped to
a two-level system without dissipation.\cite{Leggett,kotliarsi}

The other solvable point is the {\it Toulouse  point}, with
\begin{eqnarray}
\lambda_z^\ast &\equiv & 2-\sqrt{2} \;, \qquad
\gamma^\ast \equiv \sqrt{2}-1 \; , 
\\
\label{HperpToulouse1}
H_\perp^\prime 
&=& { \lambda_\perp \over 2a}\bigl(S_+ F_\downarrow^\dagger
F_\uparrow e^{-i \varphi_s(0)} + \mbox{h.c.}\bigr)\;.
\end{eqnarray}
We henceforth focus on this point,
which is the analog of the EK line 
in the 2CK context, since the factors $e^{i \varphi_s}$ 
can be treated by refermionization, as shown below.
Note, though, that the spin $- \gamma^\ast S_z / \sqrt 2$
from the conduction band that is tied to the impurity
does not fully screen the latter. 

\subsection{Refermionization}

To ensure proper anticommutation relations for the pseudofermions
to be defined  below, it is convenient to make one more unitary
transformation with the operator $U_2=e^{i\pi{\hat {\cal N}}_s
\, S_z}$, which changes the phases of the Klein factors and the
spin operators (and of the basis states in Fock space):
\begin{eqnarray}
U_2 F_\downarrow^\dagger F_\uparrow U_2^{-1} &= & e^{-i\pi S_z}
F_\downarrow^\dagger F_\uparrow \;,\\
U_2 S_\pm U_2^{-1} &= &  e^{\pm i \pi {\hat {\cal N}}_s} S_\pm\; .
\end{eqnarray}
Then  $H'_\perp$ of Eq.~(\ref{HperpToulouse1})
takes the very simple  form
\begin{equation}
\label{finalHperp1cK}
U_2 H_\perp^\prime  U_2^{-1} 
= {\textstyle \lambda_\perp \over 2\sqrt{a}}
\left(c_d^\dagger \psi_s(0) + \psi_s^\dagger (0) c_d \right)\;,
\end{equation}
where we introduced the following pseudofermions,
\begin{eqnarray}
\label{defd}
c_d^\dagger  \equiv && S_+ e^{i\pi ({\hat {\cal N}}_s - S_z)} \;,
\qquad c^\dagger_d c_d = S_z + 1/2 \; , 
\\
\label{referms2}
\psi_s(x) \equiv && 
\; {F_\downarrow^\dagger F_\uparrow \over a^{1/2}} 
e^{-i({\hat {\cal N}}_s- {\rm sgn}(S_T)[1+P]/2 ) 2 \pi x/L -i 
\varphi_s(x)} \;
 \\ 
\equiv && \sqrt{2\pi / L}
        \sum_{\bar k} c_{\bar k s}
e^{-i \bar k x}\; , 
\label{referm}
\end{eqnarray}
with ${\rm sgn}(S_T=0) \equiv 1$ .
By including the factor $e^{ i [1+ {\rm sgn}(S_T)P] \pi x/L} $ in 
the definition (\ref{referms2}) of $\psi_s$,
we purposefully ensured that $\psi_s$ has
the {\em same} boundary conditions (namely periodic)
for both $P = 0$ and 1, in order not to have to distinguish 
between these two sectors
(the reason for the
$\mbox{sgn}(S_T)$ factor is explained below). 
As a consequence, the $\bar k$'s in
Eq.~(\ref{referm}) must be of the form 
\begin{equation}
\bar k= \Delta_L 
        \mbox{[} n_{\bar k} - {\rm sgn}(S_T) 1/2 \mbox{]} ,
\qquad n_{\bar k}\in{\mathbb Z}.
\end{equation}
[i.e.\ the periodicity parameter $P_0$
of (\ref{k_free}) here equals 1].

By Eqs.~(\ref{Fajcomms-y}), these pseudofermions  
have the properties
\begin{eqnarray}
\{c_{\bar k s},c^\dagger_{\bar k' s}\} & = & \delta_{\bar k \bar
k'}\;,\\ 
 \{c_d,c^\dagger_d\}&=&1\;,\\
\{c_d,c_{\bar k s}^\dagger\} &=& \{c_d,c_{\bar k s}\} = 
0 \;,\\
\mbox{[}c_d , {\hat {\cal N}}_s\mbox{]} &=& c_d \; .
\label{[N_s,c_d]}
\label{newanticom}
\end{eqnarray}
Note in particular that the anticommutation of $c_d$ and $c_{\bar k s}$
is ensured by the factor $e^{-i\pi {\hat
{\cal N}}_s}$ in the definition (\ref{defd}) of $c_d$. 
Note further that the individual action of both 
$c_d$ and  $\psi_s$  violates
the conservation (\ref{deftildeNs}) of the total spin $S_T = {\cal N}_s + S_z$. 
When  diagonalizing $H'$, we shall therefore work not
in the  physical subspace ${\cal S}_{\rm phys} (S_T,{\cal N}_c)$
of (\ref{1ck:S}), but in a correspondingly
extended subspace ${\cal S}_{\rm ext}
({\cal N}_c)$,
in which  ${\cal N}_s$ is unrestricted and not linked to $S_z$.
At the end of the calculation we shall retain only 
the physical states in ${\cal S}_{\rm ext}({\cal N}_c) $, 
which we identify using
a {\em generalized spin-conservation condition}\/ to be derived from 
(\ref{deftildeNs}).

Let $|0 \rangle_{{\cal S}_{\rm ext}}$ be a free reference ground state
in ${\cal S}_{\rm ext}$ defined as in
(\ref{eq:defnormorderbark}), and let $: \, :$ denote normal
ordering w.r.t.\ it. Then 
$: \! c^\dagger_d c_d \! : \, = 
c_d^\dagger c_d - {n_d^{(0)}}$. Moreover
 $\hat {\bar N}_s \equiv \sum_{\bar k} 
: \! c^\dagger _{\bar k s}
c_{\bar k s} \! :$, which  counts the number of $s$-pseudofermions,
is related to $\hat {\cal N}_s$ by [compare (\ref{xN_k})]
\begin{equation}
{\hat {\bar N}}_{s}  = {\hat {\cal N}}_s -  \mbox{sgn}(S_T) P/ 2\; .
\label{N_k}
\end{equation}
The ${\rm sgn}(S_T)$ factors above are needed because we
purposefully included one in the refermionization relation (\ref{referms2}); 
we did this to
ensure that the spin reversal
transformation $(S_z,{\cal N}_s, \varphi_s) \to
(-S_z, -{\cal N}_s, - \varphi_s)$ can also be simply implemented 
in terms of the new pseudo-fermions, for which it implies
\begin{equation}
  \label{spin-reverse}
(S_T,c_d, \psi_s, \bar N_s) \to 
(-S_T,c_d^\dagger, \psi_s^\dagger,
 - \bar N_s).
\end{equation}

The pseudofermions' kinetic energy is [compare (\ref{kinxferm})]
\begin{equation}
\label{kinsferm}
\sum_{\bar k} \bar k : \! c^\dagger _{\bar k s}
c_{\bar k s} \! : \, 
= \Delta_L
        {\hat {\bar { N}}}_s^2 /2 
        \;+\; \sum_q q\; b_{qs}^\dagger b_{qs}\;.
\end{equation}
This result can be used to rewrite $H_0$ in  (\ref{1ck:H0y})
in terms of the new pseudofermions $c_{\bar k s}$ and $c_d$.
Though (\ref{kinsferm}) differs from  (\ref{1ck:H0y}) by terms in both 
$\hat {\cal N}^2_s$ and $\hat {\cal N}_s$,
the difference can be expressed in terms of
$ c_d^\dagger c_d $ using the spin-conservation condition 
(\ref{deftildeNs}), namely $\hat {\cal N}_s = S_T + {1 \over 2} -
c_d^\dagger c_d$  (those states in 
${\cal S}_{\rm ext}$ for which (\ref{deftildeNs}) does not hold will
be discarded at the end anyway). 
In this way, the EK-transformed Hamiltonian of Eqs.~(\ref{1ck:H0y}),
(\ref{H_z'}})  and (\ref{finalHperp1cK})
can be brought into the following refermionized form: 
\begin{eqnarray}
\label{Hprime1ck}
&& U_2 H' U_2^{-1} = H_c + H_s + E_G + const. , 
\\
&& H_{c}=
\sum_q q\; b_{qc}^\dagger b_{qc} \; ,
\\
 &&       H_s = \sum_{\bar k} \bar k :\! c^\dagger _{\bar k s} c_{\bar
        k s} \! : \; + \; \varepsilon_d 
      : \! c_d^\dagger c_d \! :
\nonumber \\ 
&& \qquad + 
         \sqrt{\Delta_L \Gamma}
\sum_{\bar k} (c_{\bar ks}^\dagger c_d + c_d^\dagger c_{\bar ks}) \; , 
\label{1ck:resonant}
\\
&& E_G =\Delta_L
       \mbox{$\Bigl[ \Bigr. $} {\cal N}_c ({\cal N}_c + 1 - P_0)  \;+\; 
  {\textstyle {1 \over 2}} S_T  
[S_T +  {\rm sgn}(S_T)P ]
  \nonumber \\
&&   \qquad    +   {\textstyle {1-P \over 8}}
 - {\textstyle { \lambda_z^\ast  \over 4}}
\mbox{$\Bigl. \Bigr]$}
+ S_T h_e + \varepsilon_d \left[ {n_d^{(0)}} - {\textstyle {1
      \over 2}}  \right]  , 
\label{1ck:EGbare} 
\\
&& \varepsilon_d = \Delta_L {\cal E}_{d,0} + h_i - h_e \, , 
\\ 
&& {\cal E}_{d,0}  =          ( \lambda_z^\ast -1 ) S_T 
        - {\rm sgn}(S_T) {\textstyle  {P / 2}}  \; . 
\label{1ck:varepsilon_d}
\end{eqnarray}
Evidently, the charge sector decouples
completely. In the spin sector, $H_s$ 
corresponds to a quadratic resonant level model,
whose ``resonant level'' has energy $\varepsilon_d$ and 
width $\Gamma \equiv \lambda_\perp^2/4a$, 
and $E_G$ is the ``free ground state energy''
 of the spin sector in the extended subspace 
${\cal S }_{\rm ext}$, in the presence of magnetic fields. 
Note that in the absence of magnetic fields,
i.e.\ $h_i = h_e =  0$,  we also
have ${\rm sgn}(\varepsilon_d) = 
- {\rm sgn}(S_T)$; thus it follows by inspection that in this case
$H_s$ and $E_G$ are invariant under the spin reversal
 transformation (\ref{spin-reverse}), as they should be. 
As for the 2CK case, we henceforth set $h_e=0$, 
the generalization to $h_e \neq 0$ being straightforward. 

\subsection{Diagonalization of $H_s$}

We wish to bring $H_s$ into the diagonal form
\begin{eqnarray}
\label{1ck:diagonalH}
&& H_s = \sum_{\varepsilon} \varepsilon
:\! \tilde c^\dagger_\varepsilon \tilde c_\varepsilon \! : \; + \; 
\delta E_G  \; .
\end{eqnarray}
Here $: \, :$ denotes normal ordering of the
$\tilde c_\varepsilon$ operators
with respect to an exact reference  ground
state  $|\tilde 0\rangle_{{\cal S}_{\rm ext}}$
of the subspace  ${\cal S}_{\rm ext}$, 
defined by
the conditions $\tilde c_\varepsilon|\tilde 0\rangle_{{\cal S}_{\rm ext}}=0$ for
$\varepsilon>0$ and $\tilde c^\dagger_\varepsilon|\tilde 0\rangle_{{\cal
    S}_{\rm ext}}=0$ for $\varepsilon\le0$. 
This may be accomplished by a unitary 
transformation of the form 
\begin{eqnarray}
&&\tilde c^\dagger_\varepsilon = \sum_{n=\{\bar k, d\}}B_{\varepsilon n}
c^\dagger_n\; . 
\label{1ck:diag1}
\end{eqnarray}
Analogously to Appendix~\ref{app:2ck}, the $B_{\varepsilon n}$'s
and $\varepsilon$'s can be determined starting from the 
relations 
\begin{equation}
\bigl[H_s , \tilde c^\dagger_\varepsilon \bigr]= \varepsilon\;
\tilde c^\dagger_\varepsilon, \qquad
\{ \tilde c_\varepsilon, \tilde c_{\varepsilon'}^\dagger \} =
\delta_{\varepsilon \varepsilon'} \; .
\label{1ck:diag2}
\end{equation}
One  readily finds the following results:
\begin{eqnarray}
        {\pi \Gamma \over \varepsilon-\varepsilon_d}
         &=&  - {\rm \cot}\left(\pi \varepsilon
        / \Delta_L \right)\; , 
\label{1ck:eigenv}
\\
B_{\varepsilon d} &=& \left[ {\Gamma\; \Delta_L 
        \over \Gamma^2\pi^2 +
\Gamma \Delta_L + (\varepsilon -\varepsilon_d)^2} \right]^{1/2} \;,\\
B_{\varepsilon \bar k} &=& \sqrt{\Delta_L  \Gamma} \; {1\over
\varepsilon - \bar k} \; B_{\varepsilon d}\; , 
\\
\delta E_G &=&
 \sum_{\varepsilon <0} \varepsilon - \sum_{\bar k <0} 
\bar k - \varepsilon_d {n_d^{(0)}}  \;.
\label{shift1ck}
\end{eqnarray}
The eigenenergies $\varepsilon$ are the roots of 
Eq.~(\ref{1ck:eigenv}). Their general behavior as functions of
$\Gamma$ and $\varepsilon_d$ can be determined graphically, similarly to 
Fig.~\ref{fig:graphical}. 
To identify the crossover scales of the
problem, we write a general solution, in analogy to (\ref{2ck-ejp}),
as 
\begin{mathletters}
  \label{eq:1ck:ejandshift}
\begin{eqnarray}
  \label{eq:1ck:ej}
  \varepsilon_j &=& \Delta_L (j - 1/2 + \delta_j ) \, ,
\\
  \label{eq:1ck:ej=shift}
   \delta_j &=& {1 \over \pi} \arctan \left[
   { \pi \Gamma \over \varepsilon_j - \varepsilon_d }
   \right] \; .
\end{eqnarray}
\end{mathletters}
Evidently, all solutions with  $|\varepsilon_j - \varepsilon_d| \ll \Gamma$
are strongly perturbed, with upward or downward shifts
$\delta_j \simeq \pm 1/2$ for $\varepsilon_j >$ or 
$< \varepsilon_d$, whereas those with $|\varepsilon_j- \varepsilon_d| 
\gg \Gamma$ are only weakly perturbed, with $\delta_j \simeq 0$.
(The solution $\varepsilon_j$ closest
to $\varepsilon_d$ can be associated with the 
$d$ level, which, measured in units of $\Delta_L$,
is pushed to the integer closest to it as  $\Gamma/ \Delta_L \to \infty$.)

This implies the following crossover scales: \\
(i) Without magnetic fields ($h_i = h_e = 0$,
so that  $\varepsilon_d =$ $ \Delta_L  {\cal E}_{d,0}$) and 
in the limit $\Gamma \gg  \Delta_L$ 
(i.e.\ also $\Gamma  \gg |\varepsilon_d| $), 
 the crossover scale separating the strongly and weakly
perturbed spectral regimes is $\Gamma$,
which can thus again be associated with 
the Kondo temperature, i.e.\  $T_K \simeq \Gamma$. \\
(ii) For  a large local magnetic field   $h_i \gg \Delta_L$
(so that $\varepsilon_d \simeq h_i$),
the $h_i$-induced shifts in
the lowest-lying levels with $|\varepsilon_j|  \ll \Gamma$
become large ($\simeq 1/2$) roughly when  $h_i$
 reaches the crossover field $h_c \simeq \Gamma$.
In other words, for $h_i \gg h_c$, the local magnetic field
is strong enough to effectively erase the effects
of spin-flip scattering 
from the lowest-lying part of spectrum. 
Note the contrast to the 2CK case, where the crossover field 
is smaller, namely $h_c \simeq \sqrt{\Gamma \Delta_L}$.

For $\Gamma
\gg \varepsilon_d, \Delta_L$, the ground state energy shift $\delta
E_G$, calculated similarly to Appendix~\ref{2ck:calcEG}, turns out to be
$\delta E_G =   \delta E_G^0 +  \delta E_G^d$, 
where $\delta E_G^0 \approx -{\textstyle \Gamma }\ln
(\textstyle D / \Gamma )$ is the ``binding energy'',
and 
\begin{equation}
\label{1ck:deltaEG}
\delta E_G^d = \varepsilon_d [1/2 - {n_d^{(0)}} ] \; 
+ \dots 
\end{equation}
where the dots represent terms that
are either independent of $P$ and $\varepsilon_d$, or
of order  ${\cal O} (\varepsilon_d^2 / \Gamma,
\Delta_L^2/ \Gamma, \Delta_L \varepsilon_d / \Gamma)$.
Note that $\delta E_G^d$ precisely cancels the
$\varepsilon_d$ term in $E_G$ of 
(\ref{1ck:EGbare}), thus  the fixed point spectrum
at $\Gamma/\Delta_L = \infty$ satisfyingly 
does not depend on the  parameter $\lambda_z^\ast = 2 - \sqrt 2$
occurring in $\varepsilon_d$.

\subsection{Generalized Spin-Conservation Conditions}
\label{1CK.gluing}

To identify and discard 
all states in ${\cal S}_{\rm ext}$ that violate the total-spin-conservation 
condition (\ref{deftildeNs}), we now derive a {\em
generalized spin-conservation condition.}\/ The argument is analogous
to that for the 2CK generalized gluing condition in
Section~\ref{sec:generalEstate}, but more straightforward, since $H_s$
conserves the {\em number}\/ of $c_n$ excitations (not only their number
parity).

The number of excitations of a general eigenstate
$|\tilde E \rangle$ of $H_s$ relative to ${\cal S}_{\rm ext}$
is 
\begin{equation}
\label{1ck:GGC1}
{\cal N}_{\tilde E} = \langle \tilde E | 
\sum_{\varepsilon} : \! \tilde c_\varepsilon^\dagger \tilde c_\varepsilon
\! : | \tilde E \rangle \; .
\end{equation}
When $\Gamma$ is turned off adiabatically and 
$|\tilde E \rangle$ reduces to 
$| E \rangle = \lim_{\Gamma \to 0} |\tilde E \rangle$,
its excitation number  ${\cal N}_{\tilde E}$ remains fixed.
It hence equals ${\cal N}_{\tilde E} (\Gamma \to 0)$, which can be written
as 
\begin{eqnarray}
\nonumber 
&& \langle  E | 
 \sum_{\bar k} :  \!  c_{\bar k s}^\dagger  c_{\bar k s}  \! :
+ : \! c_d^\dagger c_d \! : | E \rangle \; 
\\
\nonumber
& = & 
\langle  E | \Bigl[ 
\hat { \bar { N}_s } 
+ S_z + 1/2  - {n_d^{(0)}} \Bigr]
| E \rangle \; .
\end{eqnarray}
Using  (\ref{N_k}) for $\hat { \bar {N}_s }$
and imposing  the condition that any physical $|E \rangle$ must
satisfy the total-spin-conservation condition 
(\ref{deftildeNs}), we obtain 
\begin{equation}
\label{generalspingluing}
{\cal N}_{\tilde E} = \left\{
\begin{array}{ll}
S_T- \mbox{[} \mbox{sgn}(S_T) P-1 \mbox{]} /2  
& \phantom{mm} (\varepsilon_d>0) \\
S_T - \mbox{[}\mbox{sgn}(S_T) P+1 \mbox{]}/2 
& \phantom{mm} (\varepsilon_d\le 0)
\end{array}\right.\; .
\end{equation}
This {\em generalized spin-conservation condition}\/
specifies which of all the possible states in 
${\cal S}_{\rm ext}$ are physical;
it supplements the free gluing condition 
(\ref{1CKgluing}),
which stipulates 
that $S_T \mp 1/2$ must
be integer (half-integer) if ${\cal N}_c$
is integer (half-integer).

\subsection{Finite-Size Spectrum}

\begin{figure}
\epsfxsize=8.5cm
\epsfbox{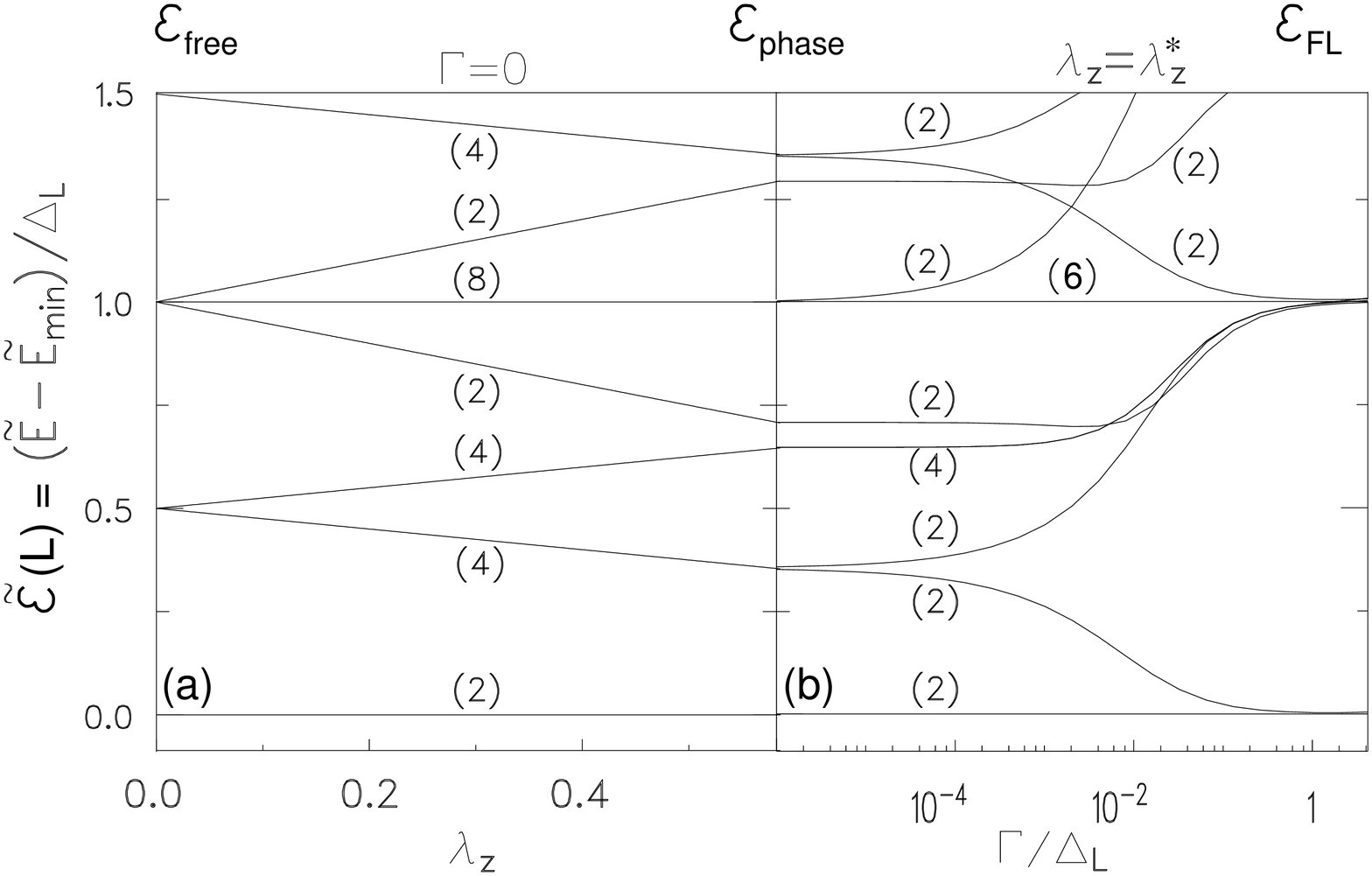}%{1ck.eps}
\vskip0.7truecm
\caption[1CK-eigenspectrum]{
\label{1ck:scal} Evolution of the many-body finite-size spectrum 
of the 1CK model, for anti-periodic boundary
conditions ($P_0 = 1$), from the free Fermi liquid
fixed point to the strong-coupling
Fermi liquid fixed point. 
All eigenstates of $H'$ of Eq.~(\ref{Hprime1ck}) 
are shown for which ${\cal E}_{\rm FL} \le 1$, 
 as well as some higher-lying states,
with degeneracies given in brackets.
(a)  When $\lambda_z$ is tuned from 0 to its
Toulouse-point  value $\lambda^\ast = 2 - \sqrt 2$, 
with  $\lambda_\perp = \varepsilon_d= 
0$, the free Fermi-liquid spectrum ${\cal E}_{\rm free}$ at
$\lambda_z=0$ evolves smoothly into a simple phase-shifted 
spectrum ${\cal E}_{\rm phase}$ at $\lambda_z=\lambda_z^\ast$.
(b) When $\Gamma/ \Delta_L = \lambda_\perp^2/(4 a \Delta_L)$  
is tuned from 0 to $\infty$
at the Toulouse point,  i.e.\ with 
$\lambda_z=\lambda_z^\ast$ and $\varepsilon_d = 0$, 
the spectrum crosses over from ${\cal E}_{\rm phase}$ to the
strong-coupling Fermi liquid spectrum ${\cal E}_{\rm FL}$
at $\Gamma/ \Delta_L = \infty$.
The latter is identical to the free
Fermi-liquid spectrum ($\lambda_z = \lambda_\perp = \varepsilon_d = 0$)
for periodic boundary conditions $(P_0 = 0)$,
in agreement with Wilson's NRG results. 
}
\end{figure}

We consider here only the case $P_0 = 1$ of anti-periodic boundary
conditions ($P_0 = 0$ is analogous),
and zero  magnetic fields, $h_i = h_e = 0$.
The construction of the finite-size spectrum
is entirely analogous to the
2CK case of Section~\ref{sec:ffs}, but 
a little more cumbersome, since  
${\cal E}_{d,0}$ of (\ref{1ck:varepsilon_d}) and hence also $\varepsilon_d$ 
is not equal to zero;
instead it depends on $S_T$, 
i.e.\ changes from one sector ${\cal S}_{\rm ext}$ to the next.
The results are summarized in Fig.~\ref{1ck:scal} and
Table~\ref{1cktable}. The latter's caption also 
summarizes the technical details of the construction.

(i) {\em Phase-Shifted Spectrum:---}\/ 
The evolution of the phase-shifted
spectrum ${\cal E}_{\rm phase}$ for $\lambda_z \in [0, \lambda_z^\ast]$ at
$\lambda_\perp=0$ 
is given by $H'(\lambda_\perp = 0)$ of (\ref{1ck:lambda=0});
it evolves linearly with increasing $\lambda_z$,
from ${\cal E}_{\rm free}$ at $\lambda_z = 0$ to 
${\cal E}_{\rm phase}$ at $\lambda_z = \lambda^\ast
=2 -  \sqrt 2$, as shown in Fig.~\ref{1ck:scal}(a). 

(ii) {\em Crossover Spectrum:---}\/ The crossover spectrum as function of
$\Gamma / \Delta_L \in [0, \infty]$ at the Toulouse point $\lambda_z =
\lambda_z^\ast$ is shown in Fig.~\ref{1ck:scal}(b).  The spectrum evolves
continuously from the phase-shifted values ${\cal E}_{\rm phase}$ at $\Gamma =
0$ to a Fermi liquid fixed-point spectrum ${\cal E}_{\rm FL}$ 
at $\Gamma /
\Delta_L = \infty $, which is constructed analytically in
Table~\ref{1cktable}.  The fixed-point spectrum corresponds precisely to the
Fermi-liquid spectrum of free fermions [of (\ref{1ck:H0y})] obeying the
periodic boundary condition $P_0 = 0$.  This agrees with the standard results
of Wilson's
numerical renormalization group calculations,\cite{Wilson} 
and is expected, because at
$\Gamma = \infty $ one electron is bound so tightly to the impurity that the
total number of free electrons effectively changes by one, and hence the
chemical potential shifts by $\Delta_L / 2$.\cite{Wilson}

 The fact that all ${\cal E}_{\rm FL}$'s are integers or
half-integers is a very direct sign of Fermi liquid
 physics, since it implies the absence of non-fermionic operators.

\begin{figure}
\epsfxsize=8.5cm
\epsfbox{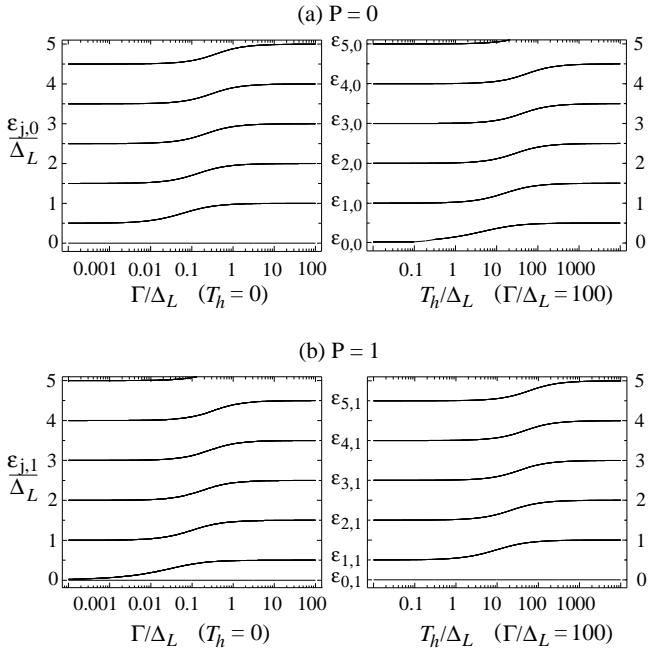}%{combinedroots.eps}
\vspace{5mm}
\caption[Evolution-ov-e]{\label{onepscaling}
(a)  Evolution of the excitation energies 
$\varepsilon_{j,P}$, found 
by numerically solving 
the eigenvalue equation (\ref{2ck:eigenenergies})
(or by a graphical analysis as in Fig.~\ref{fig:graphical}). 
On the left the evolution is shown as function of $\Gamma/\Delta_L \in
[0,\infty)$ at $T_h = 0$,
and on the right as function of $T_h/ \Delta_L
\in [0,\infty)$ at fixed $\Gamma / \Delta_L \gg 1$,
for (a) $P=0$ and (b) $P = 1$. These excitation
energies are combined in Table~\ref{2ckbigtable} 
with excitations in the charge, spin and flavor sectors
to obtain the evolution  of the full finite-size spectrum shown
in Fig.~\ref{2ckscaling}.} 
\end{figure}

\begin{figure}
\epsfxsize=8.5cm
\epsfbox{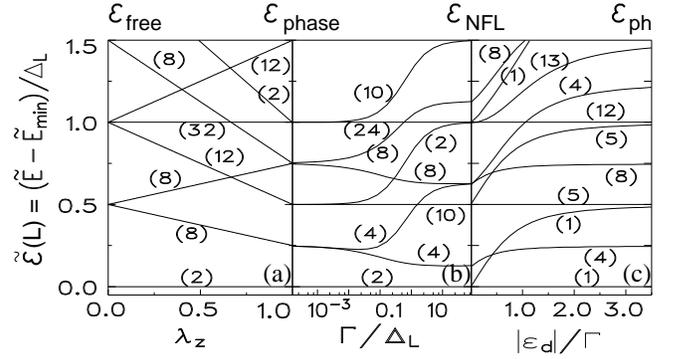}%{prl2fig.eps}
\vskip0.7truecm
\caption[2CK-eigenspectrum]{\label{2ckscaling}\label{fig:scaling}   
Evolution of the many-body finite-size spectrum 
of the 2CK model, for antiperiodic boundary conditions $(P_0 = 1)$,
from the free Fermi liquid fixed point to the NFL fixed point,
and the additional crossover induced by a local magnetic field
to a phase-shifted Fermi liquid fixed point.
All eigenstates of $H'$ of Eq.~(\ref{fullH'}) 
are shown for which 
 ${\cal E}_{\rm NFL} \le 1$ , as well as some higher-lying states,
with degeneracies given in brackets (in Ref.~\cite{vDZF},
the degenaries for ${\cal E}_{\rm NFL} = 1$ were incorrect). 
(a)  When $\lambda_z$ is tuned from 0 to its
Emery-Kivelson value $\lambda_z=1$, with  $\lambda_\perp = \varepsilon_d= 
0$, the free Fermi-liquid spectrum ${\cal E}_{\rm free}$ at
$\lambda_z=0$ evolves smoothly into a simple phase-shifted 
spectrum ${\cal E}_{\rm phase}$ at $\lambda_z=1$.
(b) When $\Gamma/ \Delta_L = \lambda_\perp^2/(4 a \Delta_L)$
 is tuned from 0 to $\infty$
along the EK line, i.e.\ with 
$\lambda_z=1$ and $\varepsilon_d = 0$, 
the spectrum crosses over from ${\cal E}_{\rm phase}$ to the
non-Fermi liquid spectrum ${\cal E}_{\rm NFL} $
at $\Gamma/ \Delta_L = \infty$, which 
agrees with NRG and CFT results. 
(c) Turning on a local magnetic field $\varepsilon_d = {h_i}$ 
(with $h_e = 0$) by 
tuning $|\varepsilon_d| /\Gamma$ from 0 to $\infty$
with $\lambda_z=1$, $\Gamma\gg \Delta_L$ fixed,
then induces a further crossover from ${\cal E}_{\rm NFL}$
to ${\cal E}_{\rm ph}$. 
For the lowest levels this crossover occurs when
$|\varepsilon_d| /\Gamma \gtrsim 1$, since then
the crossover parameter used in  Fig.~\ref{onepscaling},
namely  $T_h / \Delta_L = (\varepsilon_d/\Gamma)^2 (\Gamma/ \Delta_L)$,
is $\gg 1$. 
The ${\cal E}_{\rm ph}$ spectrum is identical to 
the phase-shifted spectrum  ${\cal E}_{\rm phase}$
 of $\lambda_z=1$ and 
$\lambda_\perp =\varepsilon_d = 0$,  
apart from a degeneracy factor of two due to the lack
of spin reversal symmetry. 
}
\end{figure}

Finally, it is instructive to  deduce another well-known fact, namely
that a local  magnetic field $h_i$ is a marginal perturbation, 
from the deviations from ${\cal E}_{\rm FL}$
which it produces:  For the lowest-lying levels, the value
of the shift $\delta_j$ of (\ref{eq:1ck:ejandshift}) at the
Fermi liquid fixed point ($h_i = 0$,  $\Gamma \gg \Delta_L$)
is $|(\delta_j)_{\rm FL}| = 1/2$.  For a 
 small local field $h_i\ll \Delta_L$,
an expansion in powers of $h_i/\Delta_L$ and $\Delta_L / \Gamma$ yields
\begin{eqnarray}
\nonumber 
  \delta_j - (\delta_j)_{\rm FL} = 
 {h_i \over  \pi^2 \Gamma} \left[
  1 -  {( j - 1/2 + \delta_j - {\cal E}_{0,d})^2
 \Delta_L^2 \over \pi^2 \Gamma^2}
 \right]  .
\end{eqnarray}
Since the $L$-dependence of the leading term is $\sim h_i L^0$, the
local magnetic field has scaling dimension $\gamma_{h_i} = 0$ [cf.\ 
(\ref{deltauniversal})] and hence is a {\em marginal}\/ perturbation
(in contrast to the 2CK case, where it is relevant with $\gamma_{h_i}
= -1/2$, cf.\ Section~\ref{ss:finsize}). A marginal perturbation
always implies the existence of a line of fixed points, parameterized
by a {\em non-universal quantity.}\/ Indeed, for $h_i$ non-zero, the
fixed point spectrum obtained for $L\to \infty$ {\em is}\/
non-universal, since in this limit (\ref{eq:1ck:ej=shift}) yields
non-universal shifts for the lowest-lying levels, namely $ \delta_j
\to - {1 \over \pi} \arctan (\pi \Gamma / h_i)$. In contrast, for the
2CK case with $\Gamma, h_i \neq 0$, the limit $L \to \infty$
necessarily implies $\Gamma , T_h \gg \Delta_L$; hence
(\ref{delta-shift}) yields universal shifts $\delta_{j,P} \to 1$ [see
Fig.~\ref{onepscaling}], which is why along the EK line the
phase-shifted fixed point spectrum ${\cal E}_{\rm ph}$ of
Fig.~\ref{fig:scaling}(c) is independent of $\Gamma$ and $h_i$ too
(though away from the EK line it does acquire a a dependence on
$h_i/\Gamma$, see Section~\ref{ss:finsize}).

\newpage

\newpage

%% The tables I and III in the text are so big that they
%  do not fit on one page when printed in preprint format.
%  They are only correctly formatted in two-column format,
%  for which, instead of the \documentstyle command in the first line,
%  the alternative command 
%
%\documentstyle[aps,prb,amsfonts,epsf]{revtex}
%
% in the preamble of this .tex file should be activated.
%
\widetext
\begin{table}
\begin{tabular}{ccc||ccc|c|cc|l|clc|cc}
$S_T$ &${\cal N}_c$ & ${\cal N}_f$ & ${\cal N}_s$ & 
${\cal N}_x$ & $S_z$ & 
$\begin{array}{c} {\cal E}_{\rm free} \\ 
{\rm Eq.\,(\protect\ref{2ck:H0y})} \end{array}$ &
\multicolumn{2}{c|}{
$\begin{array}{c}  {\cal E}_{\rm phase} 
\\ {\rm Eq.\,(\protect\ref{2ck:lambda=0})} \end{array}$}
     & & 
$\begin{array}{c} {\cal E}_G \\ {\rm Eq.\,(\protect\ref{EGfree})}
\hspace{-1mm} \end{array}$ &
$\begin{array}{c}\mbox{excitations w.r.t.\
     $|\tilde 0 \rangle_{{\cal S}_{\rm ext}}$ }
\\ \mbox{$ {\cal E}_{ex}(0) \to
 {\cal E}_{ex}  (\infty)$ } \end{array}$& 
$\begin{array}{c} \delta  {\cal E}_G^P \\ 
\hspace{-1mm} {\rm Eq.\,(\protect\ref{sumsPP})} \hspace{-1mm}
\end{array}$ &
     \multicolumn{2}{c}{${\cal E}_{\rm NFL}$ }  \\
%1a%   
\hline
& $\phantom{-}0$  & $\phantom{-}0$ &  
$ \phantom{-}0$ &  $\phantom{-} 0$ & $\Uparrow$  
& 0 & 0 &\hspace{-0.002mm}(1) &   &
 &
\hspace{17.22mm} \mbox{---} & 
 & %%EP
0 & \hspace{-0.002mm}(1) 
\\
%11%
&  $\begin{array}{c} \phantom{-}0 \\ \phantom{-}0 \end{array}$ &
$\begin{array}{c} \phantom{-}0 \\ \phantom{-}0 \end{array}$ &
$\begin{array}{c} \phantom{-} 1 \\ \phantom{-} 1 \end{array}$ & 
$\begin{array}{r} 1 \\ -1 \end{array}$ & 
$\begin{array}{c} \Downarrow \\ \Downarrow \end{array}$ 
& $\begin{array}{c}  1 \\ 1 \end{array}$ & 
$\begin{array}{c} 1/2 \\ 1/2 \end{array}$ & 
$\begin{array}{c} \hspace{-0.002mm}(1) \\ \hspace{-0.002mm}(1) \end{array}$ &
\hspace{-3mm}$\left.{
\mbox{ \phantom{\rule[-3pt]{1pt}{12pt}} } } \right \}
\hspace{-1mm}  \left\{
\mbox{ \phantom{\rule[-3pt]{1pt}{12pt}} } \hspace{-3mm} \right.
$
   & 
$\begin{array}{c}  \phantom{0} \\ \phantom{0}  \end{array}$ & 
$\begin{array}{l} \varepsilon_{0,0} \! :0 \to 0 \; ,\;
 \;  \bar k_\beta =  1/2 \\ \varepsilon_{0,0} \! : 0 \to 0 \; , \;  
\; \varepsilon_{1,0} \! : 1/2  \to 1 \phantom{/2} \end{array}$ & 
 & %%EP
$\begin{array}{c} 1/2 \\ 1 \end{array}$ &
$\begin{array}{c} \hspace{-0.002mm}(1) \\ \hspace{-0.002mm}(1) \end{array}$
\\
$\phantom{-}1/2$ 
&$\phantom{-}0$ & $\phantom{-}0$ & $\phantom{-}0$  
& $\phantom{-'}0'$ & $\Uparrow$  
& 1 & 1 &\hspace{-0.002mm}(1) &   &
0 &
$\; \varepsilon_{1,0} \! : 1/2 \to 1 \; , \; \; \bar k_{\beta}  =  1/2  $
& 
0  & %%EP
3/2 & \hspace{-0.002mm}(1) 
\\
&$\phantom{-}0$ & $\phantom{-}0$ & $\phantom{-'}0'$  
& $\phantom{-} 0$ & $\Uparrow$  
& 1 & 1 &\hspace{-0.002mm}(1) &   &
 &
\hspace{14.33mm}
$q_s = 1$ & 
 & %%EP
1 & \hspace{-0.002mm}(1) 
\\
&$\phantom{-}0$ & $\phantom{-'}0'$ & $\phantom{-}0$  
& $\phantom{-} 0$ & $\Uparrow$  
& 1 & 1 &\hspace{-0.002mm}(1) &   &
 &
\hspace{14.33mm}
$q_f = 1$ & 
 & %%EP
1 & \hspace{-0.002mm}(1) 
\\
&$\phantom{-'}0'$ & $\phantom{-}0$ & $\phantom{-}0$  
& $\phantom{-} 0$ & $\Uparrow$  
& 1 & 1 &\hspace{-0.002mm}(1) &   &
 & %%\hspace{3.123mm} 
\hspace{14.33mm}
$q_c = 1$ & 
 & %%EP
1 & \hspace{-0.002mm}(1) 
\\
%3a,b%
\hline
%% 1& 
$\phantom{-}0$ &
$\pm 1/2$ & %%$\begin{array}{r} \pm 1/2 \\ \pm 1/2  \end{array}$ &
$\mp 1/2$ & %%$\begin{array}{r} \mp 1/2 \\ \mp 1/2  \end{array}$  &
$\begin{array}{r} -1/2 \\ 1/2  \end{array}$ &
$\begin{array}{r} 1/2 \\  -1/2  \end{array}$ &
$\begin{array}{r} \Uparrow \\ \Downarrow  \end{array}$ 
& $1/2$ & $1/4$ & \hspace{-0.002mm}(4) &
\hspace{-3mm}$\left.{
\mbox{ \phantom{\rule[-3pt]{1pt}{12pt}} } } \right \}
\hspace{-1mm}  \left\{
\mbox{ \phantom{\rule[-3pt]{1pt}{12pt}} } \hspace{-3mm} \right.
$
 & 
1/4 & %%\hspace{3.123mm} & 
$\begin{array}{c}  \mbox{---} \\ \varepsilon_{0,1} \! : 0 \to 0 \; , \; \;  
\varepsilon_{1,1} \! : 0 \to 1/2 \end{array}$ 
&
$-1/8$ & %%EP
$\begin{array}{r}1/8 \\ 5/8  \end{array}$ &
$\begin{array}{r} \hspace{-0.002mm}(2) \\  \hspace{-0.002mm}(2) \end{array}$ 
\\
%4a,b%
\hline
$\phantom{-} 1$ &
$\pm 1/2$ & 
$\pm 1/2$ & 
$\begin{array}{r} \phantom{-} 1/2 \\ 3/2 \end{array}$ &
$\begin{array}{r} 1/2 \\  -1/2  \end{array}$ &
$\begin{array}{r} \Uparrow \\ \Downarrow  \end{array}$ 
& 
$\begin{array}{r} 1/2 \\ 3/2  \end{array}$ &
$3/4$ &
$\hspace{-0.002mm}(4)$ &
\hspace{-3mm}$\left.{
\mbox{ \phantom{\rule[-3pt]{1pt}{12pt}} } } \right \}
\hspace{-1mm}  \left\{
\mbox{ \phantom{\rule[-3pt]{1pt}{12pt}} } \hspace{-3mm} \right.
$
 & 
3/4 & %%\hspace{3.123mm} 
$\begin{array}{c}  \mbox{---}
 \\ \varepsilon_{0,1} \! : 0 \to 0 \; , \; \;  
\varepsilon_{1,1} \! : 0 \to 1/2 \end{array}$ 
&
$-1/8$ & %%EP
$\begin{array}{r} 5/8 \\  9/8  \end{array}$ &
$\begin{array}{r} \hspace{-0.002mm}(2) \\  \hspace{-0.002mm}(2) \end{array}$ 
\\
%5a,b%
\hline
$ - 1 $&
$\pm 1/2$ & 
$\pm 1/2$ & 
$\begin{array}{r} -3/2 \\ -1/2  \end{array}$ &
$\begin{array}{r} 1/2 \\ - 1/2  \end{array}$ &
$\begin{array}{r} \Uparrow \\ \Downarrow  \end{array}$ & 
$\begin{array}{c} 3/2 \\ 1/2 \end{array}$ & 
$3/4$ & \hspace{-0.002mm}(4) &
\hspace{-3mm}$\left.{
\mbox{ \phantom{\rule[-3pt]{1pt}{12pt}} } } \right \}
\hspace{-1mm}  \left\{
\mbox{ \phantom{\rule[-3pt]{1pt}{12pt}} } \hspace{-3mm} \right.
$
 & 
3/4 & %%\hspace{3.123mm} 
$\begin{array}{c} \mbox{---}
   \\  \varepsilon_{0,1} \! : 0 \to 0 \;  , \; \;  
\varepsilon_{1,1} \! : 0 \to 1/2 \end{array}$ 
&
$-1/8$ & %%EP
$\begin{array}{r} 5/8 \\ 9/8  \end{array}$ &
$\begin{array}{r} \hspace{-0.002mm}(2) \\  \hspace{-0.002mm}(2) \end{array}$ 
\\
%8a%
\hline
& & & $-1$ & \phantom{-} 0 & $\Uparrow$ & & $1/2$ & \hspace{-0.002mm}(2) & &
 & \hspace{17.22mm} \mbox{---} & & $1/2$ & \hspace{-0.002mm}(2)  
\\
$\begin{array}{c} -1/2 \\ \phantom{-1/2 } \end{array}$ &
$\begin{array}{c}  \pm 1   \\  \phantom{-0} \end{array}$  & 
$\begin{array}{c}  \phantom{-}0 \\  \phantom{-0} \end{array}$ & 
$\begin{array}{c} \phantom{-}0 \\ \phantom{-}0 \end{array}$ & 
$\begin{array}{c} \phantom{-} 1 \\ -1 \end{array}$ & 
$\begin{array}{c} \Downarrow \\ \Downarrow \end{array}$ &
$\begin{array}{c} 1 \\ \phantom{-0} \end{array}$ & 
$\begin{array}{c} 1 \\ 1 \end{array}$& 
$\begin{array}{c} \hspace{-0.002mm}(2) \\ \hspace{-0.002mm}(2) 
\end{array}$ & 
$
 \hspace{-3mm}
\left.{
\mbox{ \phantom{\rule[-3pt]{1pt}{12pt}} } } \right \}
\hspace{-1mm}  \left\{
\mbox{ \phantom{\rule[-3pt]{1pt}{12pt}} } \hspace{-3mm} \right.
$ & 
$\begin{array}{c} 1/2 %%\hspace{3.123mm} 
\\  \phantom{-0} \end{array}$ &
$ \begin{array}{l} 
\varepsilon_{0,0} \! : 0 \to 0 \; , \; \;  \; \bar k_\beta =  1/2  \\
\varepsilon_{0,0} \! : 0 \to 0 \; , \; \;  
\varepsilon_{1,0} \! : 1/2 \to 1 
 \end{array}$ &
$\begin{array}{c} 0 \\  \phantom{0} \end{array}$ & %%EP 
$\begin{array}{c} 1 \\ 3/2  \end{array} $ & 
$\begin{array}{c} \hspace{-0.002mm}(2) \\ \hspace{-0.002mm}(2)  \end{array}$
\\
\hline
& & & $-1$ & \phantom{-} 0 & $\Uparrow$ & & $1/2$ & \hspace{-0.002mm}(2) & &
 & \hspace{17.22mm} \mbox{---} & & $1/2$ & \hspace{-0.002mm}(2)  
\\
$\begin{array}{c} -1/2 \\ \phantom{-1/2 } \end{array}$ &
$\begin{array}{c} \phantom{-}0  \\  \phantom{-0} \end{array}$  & 
$\begin{array}{c} \pm 1  \\  \phantom{-0} \end{array}$ & 
$\begin{array}{c} \phantom{-}0 \\ \phantom{-}0 \end{array}$ & 
$\begin{array}{c} \phantom{-} 1 \\ -1 \end{array}$ & 
$\begin{array}{c} \Downarrow \\ \Downarrow \end{array}$ &
$\begin{array}{c} 1 \\ \phantom{-0} \end{array}$ & 
$\begin{array}{c} 1 \\ 1 \end{array}$& 
$\begin{array}{c} \hspace{-0.002mm}(2) \\ \hspace{-0.002mm}(2) 
\end{array}$ & 
$
 \hspace{-3mm}
\left.{
\mbox{ \phantom{\rule[-3pt]{1pt}{12pt}} } } \right \}
\hspace{-1mm}  \left\{
\mbox{ \phantom{\rule[-3pt]{1pt}{12pt}} } \hspace{-3mm} \right.
$ & 
$\begin{array}{c} 1/2 %%\hspace{3.123mm} 
\\  \phantom{-0} \end{array}$ &
$ \begin{array}{l} 
\varepsilon_{0,0} \! : 0 \to 0 \; , \;  \; \; 
\bar k_\beta =  1/2  \\
\varepsilon_{0,0} \! : 0 \to 0 \; , \; \;  
\varepsilon_{1,0} \! : 1/2 \to 1 
 \end{array}$ &
$\begin{array}{c} 0 \\  \phantom{0} \end{array}$ & %%EP 
$\begin{array}{c} 1 \\ 3/2  \end{array} $ & 
$\begin{array}{c} \hspace{-0.002mm}(2) \\ \hspace{-0.002mm}(2)  \end{array}$
\\
%9a%
\hline
$\phantom{-}1/2$ & $\pm1$ & $\pm1$  &  
$\phantom{-}0$ & $\phantom{-}0$ & $\Uparrow$ 
& 1  & 1 & \hspace{-0.002mm}(2) & &
1 & %%\hspace{3.123mm} 
\hspace{17.22mm} \mbox{---} & 
0 & %%EP
1 & \hspace{-0.002mm}(2) 
\\
\hline
$\phantom{-}1/2$ & $\pm1$ & $\mp1$  &  
$\phantom{-}0$ & $\phantom{-}0$ & $\Uparrow$ 
& 1  & 1 & \hspace{-0.002mm}(2) & &
1 & %%\hspace{3.123mm} 
\hspace{17.22mm} \mbox{---} & 
0 & %%EP
1 & \hspace{-0.002mm}(2) 
\\
\hline
$-3/2$ & $\phantom{-}0$ & $\phantom{-}0$ &  $-2 $ 
& $\phantom{-}0$ & $\Uparrow$ 
& 2  & 1 & \hspace{-0.002mm}(1) & &
1 & %%\hspace{3.123mm}
\hspace{17.22mm} \mbox{---} & 
0 & %%EP
1 & \hspace{-0.002mm}(1) 
\end{tabular}\vspace*{4mm}
\caption[Table-1]{
\label{detailed}  \label{2ckbigtable} 
Construction of the 2CK model's finite-size spectrum  
for $P_0=1$, corresponding to 
Fig.~\protect\ref{2ckscaling}.  
The table shows all
states that have 
 excitation parity ${\cal P}_{\tilde E}= 0$ [see (\protect\ref{GGC})]
and a NFL fixed-point energy ${\cal E}_{\rm NFL}$
that is $ \le 1$, as well as some higher-lying states.
(The states with   $P_{\tilde E} = 1$ double 
the degeneracies of those with $P_{\tilde E} = 0$ listed here,
as explained below.)
All energies are given in units of $\Delta_L$, e.g.\
$ {\cal E}_G \equiv E_G / \Delta_L$, with degeneracies in brackets.
States in the same sector ${\cal S}_{\rm phys} (S_T, {\cal N}_c, 
{\cal N}_f)$ are grouped together between a pair of horizontal lines
and have the same $ {\cal E}_G$ and $\delta  {\cal E}_G^P$.
(i) The construction of the 
{\em phase-shifted spectrum}\/ for $\lambda_\perp = 0$
and   $\varepsilon_d = 0$
is shown  to the left of the brace column:
in each  sector, we list 
the lowest-lying free eigenstates of $H'(\lambda_\perp = 0)$,
some of which are
illustrated in the left and middle columns of Fig.~\ref{manystate}. Each
such state is   labeled by the further quantum numbers
$({\cal N}_s,{\cal N}_x,S_z)$, 
 satisfies the free gluing conditions (\protect\ref{gluingall}),
and has
energy ${\cal E}_{\rm free}$ or ${\cal E}_{\rm phase}$
for $\lambda_z = 0$ or 1, respectively,
as shown in Fig.~\protect\ref{2ckscaling}(a). 
(${\cal N}_y = 0'$ here denotes a particle-hole excitation
with ${\cal N}_y=0$, $\langle b_{1 y}^\dagger b_{1 y} \rangle = 1$
and dimensionless energy $q_y = 1$.) 
(ii) The construction of the
{\em crossover spectrum}\/ for $\lambda_\perp \neq 0$ 
at $\lambda_z = 1$ and   $\varepsilon_d = 0$ is 
shown to the right of the brace column:
in each sector,  we list 
the lowest-lying physical  eigenstates $|\tilde E\rangle$ 
 of the full $H'(\lambda_\perp \neq 0)$,
some of which are
illustrated in the right column of Fig.~\ref{manystate}. Each 
such $|\tilde E\rangle$ is 
characterized by the excitation energies
$ {\cal E}_{ex} = \varepsilon_{j,P}$, $\bar k_\beta$ 
or $q_y$ of those excitations $\tilde
\alpha_{\varepsilon_{j,P}}^\dagger$, 
$\beta^\dagger_{\bar k}$ or  $b^\dagger_{qy}$ which it contains 
relative to the reference  state $| \tilde 0 \rangle_{{\cal
S}_{\rm ext}}$ in ${\cal S}_{\rm ext}(S_T, {\cal N}_c, 
{\cal N}_f)$ [see (\protect\ref{NFLreferenceS})],
and  satisfies the generalized gluing condition (\protect\ref{GGC}).
As $\Gamma / \Delta_L$ increases from 0 to $\infty$,
the excitation energies evolve from $ {\cal E}_{ex} (0) \to  {\cal E}_{ex} 
(\infty)$
[as can be read off from Fig.~\protect\ref{onepscaling}(a)];
correspondingly, the energy of each eigenstate $|\tilde E \rangle$ evolves from
${\cal E}_{\rm phase} =  {\cal E}_G + \sum  {\cal E}_{ex} (0)$ to
$ {\cal E}_{\rm NFL} =  {\cal E}_G  + \sum  {\cal E}_{ex} (\infty) + 
\delta  {\cal E}_G^P$
(the sum goes over all excitations listed), as shown in 
Fig.~\protect\ref{2ckscaling}(b). 
When $\Gamma$ is turned off, each
$|\tilde E \rangle$ reduces to the free state on its left
in the same row, unless they are separated by braces, in
which case $| \tilde E \rangle$ reduces to a linear combination
of the two degenerate free states grouped within  the braces to its left 
(as illustrated in Fig.~\protect\ref{manystate} and
Section~\ref{2ck:zeromodes}).
By spin reversal  symmetry, for each 
state shown here (all have ${\cal P}_{\tilde E}= 0$)
there exists  a degenerate
partner with ${\cal P}_{\tilde E}= 1$, 
obtained by setting $(S_T,{\cal N}_f) \to (-S_T,-{\cal N}_f)$ and
either, for $\lambda_\perp=0$,
$({\cal N}_s,{\cal N}_x,S_z) \to (-{\cal N}_s,-{\cal N}_x,-S_z) $,
 or, for $\lambda_\perp \neq 0$,  by adding (or removing) a 
$\varepsilon_{0,P}=0$ excitation if it was absent (or present),
as illustrated by the first and fifth rows of Fig.~\ref{manystate}(a). 
The degeneracies 
in Fig.~\protect\ref{2ckscaling}(a) and (b), and
summarized in Table~\ref{concise},  are thus twice those
listed here. (iii) A {\em crossover induced by a local magnetic
field}\/ $\varepsilon_d = h_i$  (with $h_e = 0$) 
occurs as  $T_h/\Delta_L = \varepsilon_d^2/ (\Gamma \Delta_L)$ increases from
0 to $\infty$ (at fixed $\Gamma / \Delta_L \gg 1$),
as shown in Fig.~\protect\ref{2ckscaling}(c). It
 results from the further evolution 
of the listed excitations' energies $ {\cal E}_{ex}$ 
with $T_h/\Delta_L$
[as can be read off from Fig.~\protect\ref{onepscaling}(b)].
The fixed-point
spectrum at $T_h/\Delta_L = \infty$ is
given by ${\cal E}_{\rm ph}=  {\cal E}_G  + \sum  {\cal E}_{ex} 
(\Gamma/\Delta_L=\infty,
T_h/\Delta_L = \infty)$ 
(note  from (\protect\ref{eq:deltaEG-P-h})
that  $\delta  {\cal E}_G^P = 0$ for $T_h/\Delta_L \gg 1$).
}
\end{table}

\vspace*{5cm}
\newpage   
\phantom{.} 
\narrowtext

\begin{table}
\begin{tabular}{cc|cc|cc|cc}
\multicolumn{2}{c|}{$ {\cal E}_{\rm free}$} & 
\multicolumn{2}{c|}{$ {\cal E}_{\rm phase}$} &
\multicolumn{2}{c}{${\cal E}_{\rm NFL}$} &
\multicolumn{2}{c}{${\cal E}_{\rm ph}$}
\\
\hline
0 & (2) &0 & (2) & 0 & (2) & 0 & (1)
\\
1/2 & (16) &1/4 & (8) & 1/8 & (4) & 1/4 & (4)
\\
1 & (54) & 1/2 & (12) & 1/2 & (10) & 1/2 & (6)
\\
 &  &3/4 & (16) & 5/8 & (12) & 3/4 & (8) 
\\
 & & 1 & (34) & 1 & (26) & 1 & (17) 
\end{tabular}
\vspace*{4mm}
\caption[Degeneracies]{\label{concise} \label{2cksmalltable} 
Summary of the finite-size spectrum of  Fig.~\ref{fig:scaling}
for the 2CK model,   at the four  points
$\lambda_z=\lambda_\perp = \varepsilon_d=0$ (${\cal E}_{\rm free}$);
$\lambda_z=1$, $\lambda_\perp = \varepsilon_d=0$ (${\cal E}_{\rm phase}$);
$\lambda_z=1$, $\Gamma/\Delta_L = \infty$, $ \varepsilon_d = 0$ 
(${\cal E}_{\rm NFL}$); and 
$\lambda_z=1$, $\Gamma/\Delta_L = \infty$, $T_h /\Delta_L = \infty$ 
(${\cal E}_{\rm ph}$).
We list  all  energies ${\cal E} \le 1$ (in
units of $\Delta_L$)  and give their total degeneracies in brackets. }
\end{table}

\widetext
%\onecolumn 
\begin{table}
\begin{tabular}{cc||cclc|c|cc|ccc|cc}
$S_T$ &
${\cal N}_c$ & 
${\cal N}_s$ \hspace{-4.099mm}
 & 
$S_z$ & 
$\begin{array}{c} 
{\cal E}_{d,0}  \\ (\protect\ref{1ck:varepsilon_d}) \end{array}$ 
\hspace{6mm}& 
\hspace{-6mm} $\begin{array}{r} {\cal N}_{\tilde E} \\ 
(\protect\ref{generalspingluing}) 
\end{array}$ & 
$\begin{array}{c} 
\hspace{-.50101mm} {\cal E}_{\rm free} \hspace{-.50101mm}
\\ (\protect\protect\ref{1ck:H0y}) \end{array}$ &
\multicolumn{2}{c|}{
$\begin{array}{c} {\cal E}_{\rm phase} 
\\ (\protect\ref{1ck:lambda=0}) \end{array}$} & 
$\begin{array}{c}  {\cal E}_G \\ (\protect\ref{1ck:EGbare}) \end{array}$
\hspace{-.03mm} &
$\begin{array}{l}\mbox{excitations w.r.t.\
     $|\tilde 0 \rangle_{{\cal S}_{\rm ext}}$}% \hspace{-4mm} }
\\ \mbox{$ {\cal E}_{ex} (0) \to
 {\cal E}_{ex} (\infty)$ } \end{array}$& 
$\begin{array}{c} \hspace{-2mm} \delta  {\cal E}_G^d - 
{1- \lambda^\ast \over  4 }
\hspace{-1mm} \\ 
(\protect\ref{1ck:deltaEG}) \end{array}$ &
\multicolumn{2}{c}{${\cal E}_{\rm FL}$ }  \\
\hline
0                                 &%% S_T
$
\begin{array}{c} \hspace{-2mm}\pm 1/2 \\ \\ \hspace{-2mm} 
\pm 1/2'  \end{array}$
%\\ \begin{array}{c} \phantom{'}0' \\ 0 \end{array}\end{array}$
                                    &%% {\cal N}_c
$
\begin{array}{c} 
\begin{array}{c}  \hspace{-2mm} -1/2 \\  1/2  \end{array}
\\ \begin{array}{c}\hspace{-2mm} - 1/2 \\ 1/ 2 \end{array}\end{array}$
\hspace{-4.099mm}                          &%% {\cal N}_s 
$\begin{array}{c} 
\begin{array}{c} \Uparrow\\ \Downarrow \end{array}
\\
\begin{array}{c} \Uparrow \\ \Downarrow \end{array}
\end{array}$   
                                    &%% S_z 
$-1/2$                                &%% \varepsilon_d (0)
$-1$                                  &%% {\cal N}_\varepsilon
$\begin{array}{c} 1/2 \\ \\ 3/2 \end{array}$
                                    &%% \Delta E_{\rm free} 
$\begin{array}{c} 
1/2 -\lambda^\ast \! / 4 
%%{2-\lambda^\ast \over 4} 
\\ \\ 
3/2 -\lambda^\ast \! / 4 
%%{6-\lambda^\ast \over 4} 
\end{array}$ 
                                    &%% \Delta E_{\rm phase} 
\hspace{-2.398mm}
$\begin{array}{c} (4) \\ \\ (4)  \end{array}$
                                    &%% phase degeneracy
$\hspace{-8mm} \begin{array}{l}
\begin{array}{l} \left.{
\mbox{ \phantom{\rule[-3pt]{1pt}{13pt}} } } \right \}
\hspace{-1mm}  \left\{
\mbox{ \phantom{\rule[-3pt]{1pt}{13pt}} } \right. \vspace*{1mm}
\\
\left.{
\mbox{ \phantom{\rule[-3pt]{1pt}{13pt}} } } \right \}
\hspace{-1mm}  \left\{
\mbox{ \phantom{\rule[-3pt]{1pt}{13pt}} }  \right.
\end{array} 
\end{array}$
 \hspace{-6mm}
$- \lambda^\ast \! / 4$  \hspace{-2mm}
                                    &%% E_G 
$\begin{array}{l} 
|{\rm -}1/2| \to |0| \\ 
|{\rm -}1/2| \to |{\rm -}1| \\ 
|{\rm -}1/2| \to |0| \; , \; \; \quad  \; \; q_c = 1 \\ 
|{\rm -}1/2| \to |{\rm -}1| \; , \; \; \; q_c = 1 
\end{array}$
                                    &%% excitations 
$ \lambda^\ast \! / 4 $
                                    &%% \delta E_G^d 
$\begin{array}{c} 0 \\ 1 \\ 1 \\ 2 \end{array}$
                                    &%% \Delta \tilde E^\ast
\hspace{-2.398mm} 
$\begin{array}{c} (2) \\ (2) \\ (2) \\ (2) \end{array}$
                                     %% E^\ast degeneracy
\\
\hline
1/2                                 &%% S_T
$\begin{array}{c} 0 \\ 0 \\  \phantom{'}0' \\ 0 \end{array}$
                                    &%% {\cal N}_c
$\begin{array}{c} 0 \\ 1 \\ 0 \\ \phantom{'}0' \end{array}$
\hspace{-4.099mm}                   &%% {\cal N}_s 
$\begin{array}{c} \Uparrow\\ \Downarrow \\ \Uparrow \\ \Uparrow \end{array}$   
                                    &%% S_z 
$\lambda^\ast \! /2 - 1/ 2$
                                    &%% \varepsilon_d
$\phantom{-}0$                      &%% {\cal N}_\varepsilon
$\begin{array}{c} 0 \\ 1 \\ 1 \\ 1 \end{array}$
                                    &%% \Delta E_{\rm free} 
$\begin{array}{c} 0 \\ 
1 -\lambda^\ast \! /2
\\ 1 \\  1 \end{array}$ 
                                    &%% \Delta E_{\rm phase} 
\hspace{-2.398mm}
$\begin{array}{c} (1) \\ (1) \\ (1) \\ (1) \end{array}$
                                    &%% phase degeneracy
0                                   &%% E_G 
$\begin{array}{l} 
\hspace{11.22mm}  \mbox{---} \\
|\varepsilon_d| \to |0| \, , \; \;   1/2 \to 1  \\
 \hspace{8.33mm} q_c = 1 \\
|{\rm  -}1/2| \to |{\rm -}1| \, , \; \; 1/2 \to 1 
\end{array}$
                                    &%% excitations 
0                                   &%% \delta E_G^d 
$\begin{array}{c} 0 \\ 1 \\ 1 \\ 2 \end{array}$
                                    &%% \Delta \tilde E^\ast
\hspace{-2.398mm} 
$\begin{array}{c} (1) \\ (1) \\ (1) \\ (1) \end{array}$
                                    %% E^\ast degeneracy
\\
\hline
1/2                                 &%% S_T
$\hspace{-2mm} \pm 1 $              &%% {\cal N}_c
$0 $
\hspace{-4.099mm}                   &%% {\cal N}_s 
$\Uparrow $     
                                    &%% S_z 
$\lambda^\ast \! / 2 - 1/ 2$              
                                    &%% \varepsilon_d
$\phantom{-}0$                        &%% {\cal N}_\varepsilon
$1$     
                                    &%% \Delta E_{\rm free} 
$ 1$     
                                    &%% \Delta E_{\rm phase} 
\hspace{-2.398mm}        
$ (2) $
                                    &%% phase degeneracy
1                                   &%% E_G 
 \mbox{---}   \hspace{10.22mm}
                                    &%% excitations 
0                                   &%% \delta E_G^d 
$1$     
                                    &%% \Delta \tilde E^\ast
\hspace{-2.398mm} 
$(2) $
                                    %% E^\ast degeneracy
\\
\hline
1                                   &%% S_T
$\hspace{-2mm} \pm 1/2$             &%% {\cal N}_c
$1/2$ \hspace{-4.099mm}             &%% {\cal N}_s 
$\Uparrow $                         &%% S_z 
$\lambda^\ast  - 3 /2$              &%% \varepsilon_d
$\phantom{-}0$                        &%% {\cal N}_\varepsilon
$ 1/2 $                             &%% \Delta E_{\rm free} 
$1/2 + \lambda^\ast \! /4$          &%% \Delta E_{\rm phase} 
\hspace{-2.398mm} 
$ (2)$
                                    &%% phase degeneracy
$1/2 + \lambda^\ast \! / 4$         &%% E_G 
  \mbox{---}    \hspace{10.22mm}     & %% excitations 
$1/2 - \lambda^\ast \! / 4$         &%% \delta E_G^d 
$ 1 $                               &%% \Delta \tilde E^\ast
\hspace{-2.398mm} 
$ (2) $                             %% E^\ast degeneracy
\\
\hline
$3/2$                               &%% S_T
$0$                                 &%% {\cal N}_c
$1$ \hspace{-4.099mm}               &%% {\cal N}_s 
$\Uparrow $                         &%% S_z 
$(\lambda^\ast   - 1) 3/2$ \hspace{-2.5mm}             
                                    &%% \varepsilon_d
$\phantom{-}1$                        &%% {\cal N}_\varepsilon
$1 $                                &%% \Delta E_{\rm free} 
$1+ \lambda^\ast \! / 2$            &%% \Delta E_{\rm phase} 
\hspace{-2.398mm} 
$ (1)$
                                    &%% phase degeneracy
$1/2 + \lambda^\ast \! / 2$         &%% E_G 
 $1/2 \to 1$ \hspace{10.22mm}       &%% excitations
$1/2 - \lambda^\ast \! / 2$         &%% \delta E_G^d 
$ 2 $                               &%% \Delta \tilde E^\ast
\hspace{-2.398mm} 
$ (1) $                             %% E^\ast degeneracy
\end{tabular}
\vspace*{4mm}
\caption[Table-2]{ \label{1cktable} \label{1ck-detailed}  \label{1ckbigtable} 
Construction of the 1CK model's finite-size spectrum  for
 $h_i=h_e = 0$ and $P_0 = 1$, corresponding to Fig.~\protect\ref{1ck:scal}. 
The table shows all states with $S_T \ge 0 $ that have 
a strong-coupling Fermi liquid
fixed-point energy ${\cal E}_{\rm FL}$ that is $\le 1$, 
as well as some higher-lying states. 
(The states with $S_T < 0$ double the degeneracies of
those with $S_T > 0$ listed here, as explained below.) All energies 
are given in units of $\Delta_L$, e.g.\ $ {\cal E}_G \equiv E_G / \Delta_L$, 
with degeneracies in brackets.
States in the same sector ${\cal S}_{\rm phys} (S_T, {\cal N}_c)$ 
are grouped together between a pair of horizontal lines
(and have the same $\varepsilon_d$, ${\cal N}_{\tilde E}$,
$ {\cal E}_G$ and $\delta  {\cal E}_G^d$).
(i) The construction of the {\em phase-shifted spectrum}\/ 
for $\lambda_\perp = 0$ is shown to the left of the braces:
in each sector, we list 
the lowest-lying free eigenstates of $H'(\lambda_\perp = 0)$; each is  
labeled by the further quantum numbers
$({\cal N}_s,S_z)$, satisfies the free gluing 
condition~(\protect\ref{1CKgluing}) and spin-conservation
condition~(\ref{deftildeNs}), and has 
energy ${\cal E}_{\rm free}$ or ${\cal E}_{\rm phase}$
for $\lambda_z = 0$ or $\lambda_z= \lambda_z^\ast = 2- \sqrt 2$, 
respectively. 
(${\cal N}_y = 0'$ 
and the braces here have
 the same meaning as in Table~\protect\ref{2ckbigtable}.) 
(ii) The construction of the {\em crossover spectrum}\/ 
for $\lambda_\perp \neq 0$ at $\lambda_z = \lambda_z^\ast$ is shown
to the  right of the braces: in each sector, we list 
the lowest-lying physical eigenstates $|\tilde E\rangle$ 
 of the full $H'(\lambda_\perp \neq 0)$ in that sector; each 
such $|\tilde E \rangle$ is 
characterized by the excitation energies
$ {\cal E}_{ex} = \varepsilon$, 
$|\varepsilon|$ or $q_y$ of the excitations $\tilde
c_{\varepsilon > 0}^\dagger$, 
$\tilde c_{\varepsilon \le 0}$ or $b^\dagger_{qy}$ which 
it contains relative to the reference  state $| \tilde 0 \rangle_{{\cal
S}_{\rm ext}}$ in ${\cal S}_{\rm ext}(S_T, {\cal N}_c)$,
and satisfies the generalized spin-conservation condition 
(\protect\ref{generalspingluing}). 
(For $\varepsilon \le 0$, $|\varepsilon|$ denotes the hole
excitation $\tilde c_\varepsilon | \tilde 0 \rangle_{{\cal
S}_{\rm ext}}$.) As $\Gamma / \Delta_L$ increases from 0 to $\infty$,
the excitation energies evolve from 
$ {\cal E}_{ex} (0) \to  {\cal E}_{ex} (\infty)$;
correspondingly, the energy of each eigenstate $|\tilde E \rangle$
evolves from ${\cal E}_{\rm phase} =  {\cal E}_G + \sum  {\cal E}_{ex} (0)$ to
${\cal E}_{\rm FL} =  {\cal E}_G + \sum  {\cal E}_{ex} (\infty) + 
\delta  {\cal E}_G^d - 
{1 - \lambda^\ast \over 4}$ (the latter constant corresponds
to subtracting $ {\cal E}_{\rm min}$, 
the sum goes over all excitations listed),
as shown in Fig.~\protect\ref{1ck:scal}. 
By spin reversal symmetry, each $S_T > 0$ state shown
here has a degenerate
partner with total spin $-S_T$, 
obtained, for $\Gamma=0$, by setting
$({\cal N}_s,S_z) \to (-{\cal N}_s,-S_z) $,
 or, for $\Gamma \neq 0$,  by setting
$(\varepsilon_d ,{\cal N}_{\tilde E} ) \to (-\varepsilon_d,
-{\cal N}_{\tilde E})$ and interchanging particle- and
hole-excitations, $|\varepsilon| \leftrightarrow \varepsilon$
[cf.\ (\protect\ref{spin-reverse})].
For all $S_T \neq 0$ levels, the degeneracies 
in Fig.~\protect\ref{1ck:scal}
are thus twice those listed here. 
}  
\end{table}
\narrowtext

\widetext

\begin{thebibliography}{WWW}
\bibitem{a-to-0} The limit $a\to0$ must be taken first (before $x_0
  \to 0$), since the bosonization and refermionization
identities (\ref{fermiboson}) and (\ref{newpsixgeneral}) 
are exact only for $a=0$,
  as discussed in Appendix~\ref{app-bos-cutoff}.  
A detailed evaluation of such point-split
products is given in  Appendix~G.3 of
  Ref.~\onlinecite{jvdschoeller}, where,
however, it was mistakenly stated
 (and even emphasized) that
  the ``point-splitting'' parameter $x_0$ should be set equal to 
  the bosonization cutoff parameter $a$.
  This is incorrect.  The calculations in Appendix~G.3 of
  Ref.~\onlinecite{jvdschoeller} can be corrected by 
 first replacing the explicit occurence
  of $a$ by $x_o$, then setting $a=0$ in the boson fields $\phi_\eta$,
 and then taking $x_0 \to 0$.
\end{thebibliography}
\end{document}